\documentclass[aps,prb,10pt,twocolumn,showpacs,letterpaper,superscriptaddress]{revtex4-1}
\usepackage{graphicx}
\graphicspath{{./figures/}}
\usepackage{hyperref}
\hypersetup{
  colorlinks=true,
  allcolors=blue
}

\usepackage{amsmath}
\usepackage{chemformula}

\newcommand{\tio}[0]{\ch{TiO2}}
\newcommand{\tios}[0]{\ch{TiO2}(110)}
\newcommand{\wf}[0]{\texttt{PolFlow}}
\newcommand{\vbml}[0]{ConfML}
\newcommand{\cVO}[0]{$c_{\rm VO}$}
\newcommand{\cNb}[0]{$c_{\rm Nb}$}
\newcommand{\EPOL}[0]{$E_{\rm POL}$}
\newcommand{\eadsG}[0]{$E_{\rm ads}^{GS}$}
\newcommand{\eadsF}[0]{$E_{\rm ads}^{F}$}
\newcommand{\tiu}[0]{$\rm Ti_{\rm 5c}$}

\newcommand{\eg}{\textit{e.g.},}
\newcommand{\ie}{\textit{i.e.},}

\newcommand{\VO}[0]{$V_{\rm O}$}

\begin{document}

\begin{abstract}
Polarons are widespread in functional materials and are key to device performance in several technological applications.
However, their effective impact on material behavior remains elusive, as condensed matter studies struggle to capture their intricate interplay with atomic defects in the crystal. In this work, we present an automated workflow for modeling polarons within density functional theory (DFT). 
Our approach enables a fully automatic identification of the most favorable polaronic configurations in the system. 
Machine learning techniques accelerate predictions, allowing for an efficient exploration of the defect-polaron configuration space. We apply this methodology to Nb-doped \tios\ surfaces, providing new insights into the role of defects in surface reactivity.
Using CO adsorbates as a probe, we find that Nb doping has minimal impact on reactivity, whereas oxygen vacancies contribute significantly depending on their local arrangement via the stabilization of polarons on the surface atomic layer. 
Our package streamlines the modeling of charge trapping and polaron localization with high efficiency, enabling systematic, large-scale investigations of polaronic effects across complex material systems.
\end{abstract}

\title{Automated Modeling of Polarons: Defects and Reactivity on \tios\ Surfaces}
\author{Firat Yalcin}
\email[]{firat.yalcin@univie.ac.at}
\affiliation{University of Vienna, Faculty of Physics and Center for Computational Materials Science, Kolingasse 14-16, 1090 Vienna, Austria}
\author{Carla Verdi}
\email{c.verdi@uq.edu.au}
\affiliation{School of Mathematics and Physics, The University of Queensland, Brisbane 4072 QLD, Australia}
\author{Viktor C. Birschitzky}
\affiliation{University of Vienna, Faculty of Physics and Center for Computational Materials Science, Kolingasse 14-16, 1090 Vienna, Austria}
\author{Matthias Meier}
\affiliation{University of Vienna, Faculty of Physics and Center for Computational Materials Science, Kolingasse 14-16, 1090 Vienna, Austria}
\author{Michael Wolloch}
\affiliation{University of Vienna, Faculty of Physics and Center for Computational Materials Science, Kolingasse 14-16, 1090 Vienna, Austria}
\affiliation{VASP Software GmbH, Berggasse, 1090 Vienna, Austria}
\author{Michele Reticcioli}
\email[]{michele.reticcioli@univie.ac.at}
\affiliation{University of Vienna, Faculty of Physics and Center for Computational Materials Science, Kolingasse 14-16, 1090 Vienna, Austria}
\affiliation{National Research Council, CNR-SPIN, via Vetoio 42, 67100 L'Aquila, Italy}
\affiliation{University of L'Aquila, via Vetoio 10, 67100 L'Aquila, Italy}

\maketitle

\section{Introduction}

Polarizable materials can localize charge carriers into electronic states known as polarons~\cite{Emin1982, Stoneham2007, Alexandrov2010book, Holstein2000}.
In the limit of strong localization, the charge becomes confined almost entirely to a single lattice site, accompanied by distortions in the surrounding atomic structure~\cite{Holstein2000}.
Such ``small polarons'' play a critical role in determining the electronic, optical, and chemical properties of functional materials, with significant implications for technological applications~\cite{Tanner2022, Natanzon2020, Jupille2015book, Strand2023, Pastor2022polrev, Chi2022polcat, Qian2018, xu2024photoexcIR, Franchini2021, Rousseau2020, Ren2023, Kick2020a, Cheng2023, Wiktor2018}.
The impact of polarons on material properties might vary depending on the localization site of the charge carrier~\cite{Yim2016a, Birschitzky2024, Yoon2015}.
For example, in catalysis, polarons on surface sites of the host material interact with molecular adsorbates, transferring electronic charge into the molecular orbitals and thereby acting as active centers;
conversely, polarons populating subsurface layers tend to show no sizable effect on the adsorbates~\cite{Reticcioli2019c, Cao2017a, Cheng2021, Yim2018}.
It is therefore crucial to carefully account for the polaron configuration space in order to provide an accurate description of polaronic materials.

Density functional theory (DFT) has been successfully used to model small polarons in a wide range of materials, especially transition metal oxides, providing insights on their fundamental properties in good agreement with experimental observations~\cite{Franchini2021, DeLile2021, Sio2019a, Rousseau2020, Pacchioni2015chapter, Ambrosio2025, Gionco2015, Reticcioli2022, Ren2023}.
The supercell approach, where a small polaron is localized within a relatively large unit cell, allows us to inspect how polaron properties and effects depend on the particular charge localization sites, thus shedding light on the role of polarons in technological applications~\cite{Deskins2009, Yoon2015, Yang2014, Cao2017a, Tanner2021, Preda2011, Sombut2022, Lopez-Caballero2020, Lopez-Caballero2020a, Reticcioli2019c, Reticcioli2022, Reticcioli2018a}.
However, polaron modeling poses challenges to DFT simulations, especially when attempting to localize charge at custom (user-defined) lattice sites.
In fact, instead of yielding the target polaronic state, in the DFT simulation the electronic charge might converge towards a delocalized-state solution (populating states in the conduction or valence bands rather than forming a polaronic state), or spontaneously localize on a different lattice site.

Several strategies have been proposed to facilitate control on polaron localization.
These approaches typically involve multiple calculations to obtain the polaronic configuration at target.
An initial step is performed to obtain electronic density and atomic structure through charge constraints on specific electronic states, manual atomic distortions, or the addition of trapping potentials~\cite{Pham2020, Allen2014, Reticcioli2019b, Falletta2022a}.
Subsequently, the constraints are removed, and a standard DFT calculation is performed, using the constrained results of the first run as a starting point to fully relax electronic density and atomic structure.
Careful manual intervention is required to monitor polaron localization in all calculation steps and to adopt corrective measures when localization fails (\eg\ by imposing stricter constraints, resulting in more difficult and demanding convergence processes).
These practical difficulties in modeling polaron localization may discourage careful inspections of the polaronic configuration space, leading to potentially spurious results in DFT studies of polaronic materials.

\begin{figure*}[ht]
\centering
\includegraphics[width=\textwidth]{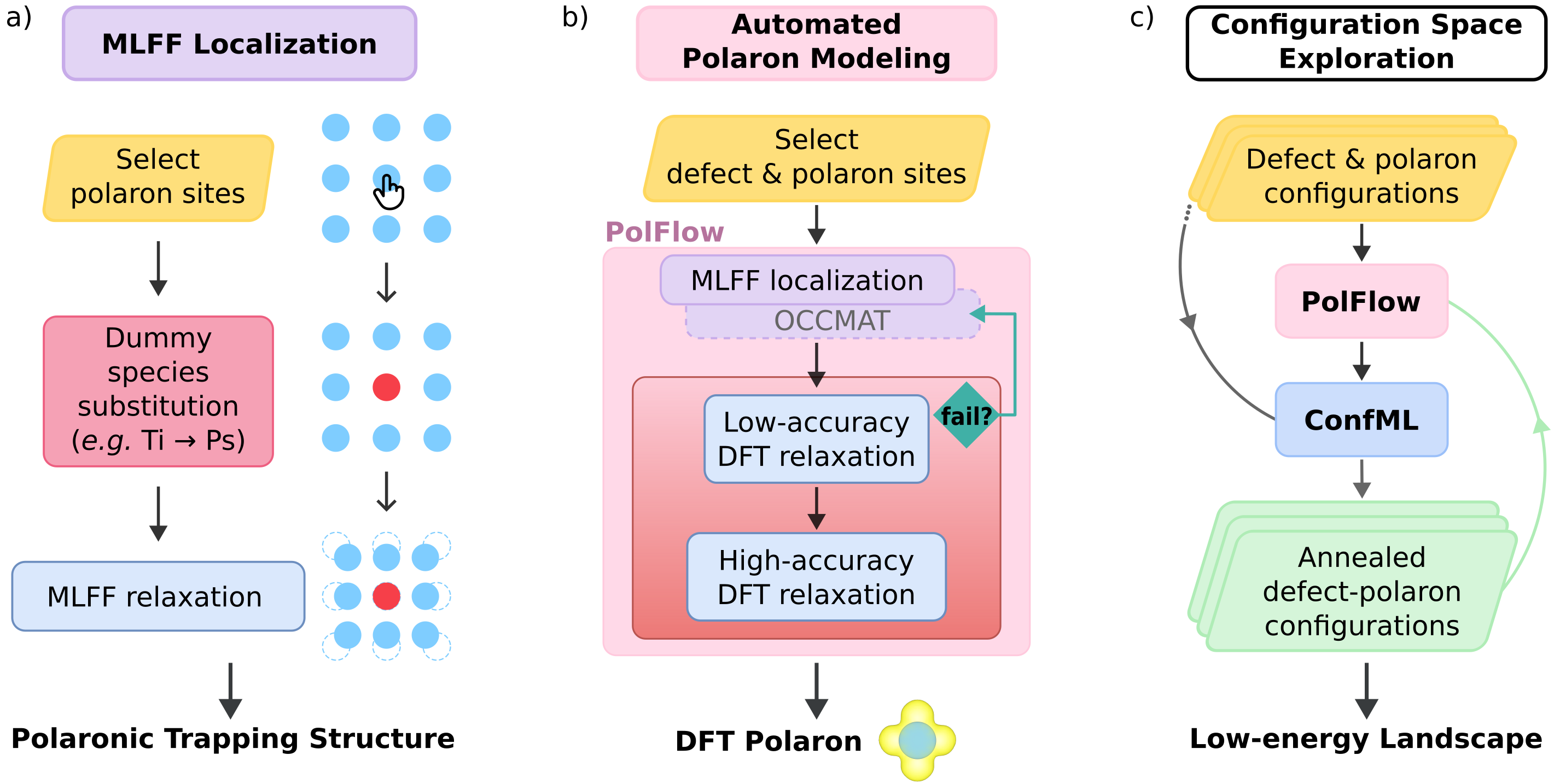}
\caption{Sketch of the computational package components.
(a) Polaron localization via the machine learning force field (MLFF) routine.
A selected atomic site is labeled as polaronic site (Ps), and an MLFF-driven relaxation of internal forces is performed to obtain the polaronic trapping structure.
(b) Flowchart of the \wf\ workflow.
The workflow can automatically perform a polaron localization calculation via the MLFF routine. In the case of a localization failure, stricter approaches are automatically adopted (\eg\ the OCCMAT tool).
\wf\ can progressively increase the accuracy of the calculations to achieve reliable DFT description of polaronic properties, saving computational time.
(c) The \wf\ workflow can automatically scan large numbers of defect-polaron configurations to explore the energy landscape.
The interface with the configuration-space-exploration routine (\vbml) enables an active learning strategy to find low-energy configurations via simulated annealing.}
\label{fig:polaron_flowchart}
\end{figure*}

In addition to the intrinsic challenges of modeling polaron localization, the presence of atomic impurities such as point defects (\eg\ vacancies, interstitials, substitutional dopants) further increases the complexity of simulations~\cite{Chen2023, Ma2025MLdefects, Smart2017}:
Defects and polarons interact through a complex interplay that proves intricate to fully unravel in computational modeling.
On one hand, defects alter polaronic properties such as the polaron stability and the spatial distribution of the localized charge carriers in the material.
In turn, polarons can impact the formation of defects and their arrangement in the material, potentially leading to structural phase transitions and drastic impacts on the material functionalities~\cite{Yim2010, Smart2017, Zhang2019, Birschitzky2024, Reticcioli2017d, Yuan2024, Geiger2022, oesterbacka2022bivo4, Kowalski2010, Deskins2009}. 
Understanding the diversity of properties and effects arising from the different defect-polaron arrangements requires careful exploration of the vast configuration space.
Hence, the need for efficient and robust computational methods for polaron modeling.

Here, we present a fully automated approach for polaron localization that can be used to efficiently sample the defect-polaron configuration space.
Our package includes a high-throughput workflow, \wf, that streamlines all steps required to obtain polaron localization and automatically monitors the status of the simulations in real time.
In the case of localization failure, the calculation is immediately interrupted, a different localization strategy is adopted, and the calculation is resumed, with no manual intervention required.
Furthermore, by leveraging machine learning (ML) force fields (MLFFs), we introduce a novel approach to polaron localization:
We show that MLFFs can be adapted for polaron modeling, enabling an accurate description of the ground-state atomic structure in polaronic materials at a fraction of the computational cost of DFT calculations.

The software package featuring the high-throughput workflow \wf, together with the MLFF-powered localization technique, accelerates the calculations of small polarons and provides broader accessibility to users with limited expertise in polaron modeling.
Furthermore, its high efficiency enables careful investigations of the polaronic configuration space~\cite{Birschitzky2022, Li2024clusterex}, as we complemented our workflow with an interface to the active-learning scheme \vbml~\cite{Birschitzky2024}, to sample the large variety of defect and polaron configurations in materials.

We demonstrate the capabilities of our computational machinery by investigating the polaronic properties on the rutile \tios\ surface, modeling the simultaneous presence of intrinsic oxygen-vacancy (\VO) defects and Nb doping atoms~\cite{DeLile2021, Setvin2014, Diebold2002, DeLile2019a}.
The efficient \wf\ exploration allows us to understand key aspects of the most stable defect-polaron configurations on the surface and sets the basis for analyzing the role of polarons in surface reactivity, which we investigate here by using CO as a probe molecule.
Our extensive sampling of the configuration space reveals that Nb dopants have a negligible direct impact on the reactivity and the stabilization of polaronic states.
Oxygen vacancies, instead, play a major role, as specific \VO\ arrangements can promote the localization of polarons on the surface layer and their interaction with CO adsorbates.

This paper is organized as follows.
In Sec.~\ref{sec:Methodology} we present our automated approach to polaron localization and configuration space exploration, including the description of the \wf\ high-throughput workflow, the MLFF-powered localization technique, and the \vbml-powered configuration search.
In Sec.~\ref{sec:Results} we apply our methodology to study the distribution of polarons, Nb dopants and \VO\ defects on \tios\ surfaces and analyze its performance.
In Sec.~\ref{sec:Discussion} we conclude by showing how the defect-polaron distribution, accessible through our efficient computational approach, influences technologically relevant properties such as CO adsorption on the surface.

\section{Methodology}
\label{sec:Methodology}

In this Section, we describe our automated approach to polaron modeling in the supercell-DFT framework.
As sketched in Fig.~\ref{fig:polaron_flowchart}, our software package features three components, addressing different tasks:
\begin{itemize}
\item an MLFF-powered localization technique to model polaron trapping on target sites, \eg\ user-defined lattice sites (see Sec.~\ref{sec:MetMLFF});
\item an efficient workflow (\wf) designed to streamline the series of DFT calculations required to obtain polaron trapping, including an automated on-the-fly control on the status of the simulations (see Sec.~\ref{sec:MetWFLOW});
\item an interface to an active learning scheme (\vbml) for exploring the configuration space of polarons and atomic defects in the system under investigation (see Sec.~\ref{sec:MetCSPACE}).
\end{itemize}

A description of the computational setup specifically adopted in our study case, \ie\ defect-polaron configurations on the (110) surface of rutile \tio\ with oxygen vacancies and Nb doping, is provided in Sec.~\ref{sec:MetSETUP}.

\subsection{Localizing polarons via MLFF}
\label{sec:MetMLFF}

We propose a robust and efficient approach to model small-polaron localization based on machine learning force fields, as illustrated in Fig.~\ref{fig:polaron_flowchart}(a).

As a preliminary step of the MLFF-localization approach, the force field must be trained if not already available for the polaronic system under investigation.
To this aim, short DFT-based molecular dynamics (DFT-MD) simulations using the on-the-fly learning scheme implemented in the Vienna \textit{Ab-initio} Simulation Package (VASP)\cite{Jinnouchi2019a, Jinnouchi2019} are used to efficiently generate a good training dataset.
The finite temperature in DFT-MD simulations normally leads to spontaneous polaron trapping in polaronic systems, although with no control over the specific localization site.
Thermally activated hopping typically causes the polaron to localize on different lattice sites during the DFT-MD.
As a result, the DFT-MD trajectory naturally samples a variety of localized polaron configurations, enriching the training set and improving the quality of the force field.

The local change in atomic valence states caused by polaron localization poses challenges to standard MLFF techniques~\cite{Birschitzky2025MLMD, Trivisonne2019BSc, Trivisonne2024MSc}.
To overcome this limitation, we adapted the MLFF descriptors to account for polaron localization by labeling the trapping sites in the DFT-MD dataset as polaronic species, denoted ``Ps''.
This assignment is performed 
by identifying sizable perturbations of the local magnetic moments, compatible with polaron localization, and labeling the hosting atom as Ps.
The DFT-MD dataset, with the polaronic atoms properly labeled, is then used to train the MLFF model to evaluate forces and energies in the presence of polarons.

Once trained, the MLFF can be used to localize polarons at any site:
by simply labeling as Ps the target sites designated to host the polarons, an MLFF calculation can be performed to obtain the charge-trapping atomic structure, as well as the corresponding energy and forces.
The MLFF results can then be further refined through a standard DFT calculation.

This MLFF-based localization method is powerful and highly efficient.
It provides users with a simple and convenient tool to select trapping sites, and the MLFF calculations are orders of magnitude faster than standard DFT, while maintaining DFT-level accuracy.
A detailed description of the performance is given in Sec.~\ref{sec:result-benchamark}.

Finally, although this MLFF-based technique for polaron localization can be used independently, we have embedded it into our workflow to further facilitate its use (see the following section below for a description of the workflow).

\subsection{Workflow to streamline polaron modeling}
\label{sec:MetWFLOW}

In this Section, we present the automated workflow \wf, designed to streamline the process of polaron localization within supercell DFT simulations. 
Our goal is to make polaron modeling more accessible, reducing the knowledge barrier required to conduct in-depth investigations of polaronic materials.
To this aim, the workflow is constructed to handle the entire process from input file preparation to calculation execution, and it automatically generates a database of results, which serves as a valuable resource for subsequent post-processing and analysis.

Figure~\ref{fig:polaron_flowchart}(b) outlines the main tasks performed by the workflow, namely 
polaron localization calculations, checks on the charge trapping status, and, eventually, exploration of the configuration space.
To perform these tasks, it leverages a suite of software tools, including Fireworks~\cite{CPE:CPE3505} as workflow manager (which automates and coordinates the execution of complex computational tasks), atomate~\cite{MATHEW2017140} (an interface for VASP-specific workflows), pymatgen~\cite{ONG2013314} for structure generation and manipulation, and custodian~\cite{ONG2013314} for on-the-fly error checking and corrections.

As shown in Fig.~\ref{fig:polaron_flowchart}(b), the workflow is initialized using a user-defined structure (optionally obtained from the Materials Project~\cite{Jain2013} database via pymatgen) and a set of indices for the positions of defects and polarons in the system.
It then automatically sets up the DFT calculations required to localize the polarons at the specified sites.
If a pre-trained MLFF model is available (see Sec.~\ref{sec:MetMLFF}), the workflow performs an MLFF relaxation of the atomic coordinates, with the polaronic species Ps at the custom positions.
A DFT calculation then follows, starting from the relaxed structure as obtained via the MLFF (referred to in the following as MLFF+DFT strategy).
During the execution of the DFT calculation, the workflow checks for possible localization issues:
in the case one of the polarons becomes delocalized or re-localizes on a different site, the calculation is interrupted, and alternative methods to obtain polaron localization can be attempted.
Specifically, we implemented an error-checking mechanism that monitors the status of polarons during runtime. 
This mechanism can dynamically initiate new runs or terminate the entire workflow if polaron migration or delocalization is detected, allowing adaptive responses to these issues and saving a significant amount of computational time (see Sec.~\ref{sec:Results} for our analysis on the performance).

In the current implementation of the workflow, if a failure of polaron localization is detected after the MLFF+DFT step, a second localization attempt is performed using the occupation matrix control tool (OCCMAT)~\cite{Allen2014}.
In this approach, the electronic occupation of the orbitals of the site designed to host the polaron is explicitly constrained to a polaronic solution, followed by a structural relaxation.
The resulting structure and wavefunction are then used as input for a standard, unconstrained DFT relaxation (referred to in the following as OCCMAT+DFT). 

If the localization is successful, all relevant information is stored (\eg\ the free energy for the given defect-polaron configuration).
The workflow is also designed to progressively increase the accuracy level of the DFT calculations.
In practice, the localization attempts are initially performed using a faster, lower-accuracy setup; 
once a localized polaronic solution is obtained, the outputs (such as nuclei coordinates, electronic density, etc\ldots) are used to run calculations at higher accuracy (\eg\ finer sampling of the reciprocal space, higher energy cutoff, tighter convergence thresholds, etc\ldots).
This progressive refinement ensures accuracy while saving a considerable amount of computational time.

The workflow can be executed repeatedly in order to investigate the configuration space, inspecting different arrangements of polarons and defects.
An efficient sampling can be obtained by interfacing the workflow with recently developed \textit{ad-hoc} tools for accelerating the exploration of defect-polaron configurations~\cite{Birschitzky2022, Birschitzky2024, Wang2023a}, as described in the following section.

\subsection{Exploring defect-polaron configuration space: Active learning and simulated annealing}
\label{sec:MetCSPACE}

The efficient automatization of polaron localization via \wf\ facilitates systematic exploration of the defect-polaron configuration space.
To accelerate this, we interfaced the workflow with a machine learning model (\vbml) that can be trained to predict the energy of any defect-polaron configuration, based on DFT data~\cite{Birschitzky2024}, sketched in Fig.~\ref{fig:polaron_flowchart}(c).
The model, implemented using the JAX framework~\cite{jax2018github}, consists of polaron- and defect-type-specific feed-forward neural networks, optimized via stochastic gradient descent and backpropagation.

We design the \wf\ workflow to be able to automatically define unexplored and symmetrically inequivalent defect-polaron arrangements, run the localization calculations, and compute the configuration energies required to train the \vbml\ model, thus implementing an active learning scheme.
During the exploration, we use simulated annealing to optimize defect-polaron configurations, seeking to identify the ground-state structures, in line with our previous work~\cite{Birschitzky2024}. 
Starting from a given or random configuration, defects and/or polarons are transferred to different sites.
ML energies are then used to determine the Metropolis acceptance probability of the new configuration, which governs whether the configuration change is accepted or rejected using a probabilistic method.
This approach allows for efficient exploration of the configuration space.
Comparisons between ML predictions and DFT data are automatically performed in order to identify outliers and incorporate them into the training set to improve model accuracy.

To focus on efficiently exploring low-energy configurations rather than searching the entire polaronic configuration space, we propose a multi-phase approach.
In the first phase, we generate a diverse initial dataset by randomly sampling the defect and polaronic configuration space. 
This ensures a broad coverage of the possible configurations, providing a solid foundation for our \vbml\ model.
The random sampling is intended to avoid biases in the training.
In the second phase, the \vbml\ training dataset is expanded with configurations obtained through simulated annealing, together with additional random polaron configurations in low-energy defect arrangements.
Finally, in the third phase, we focus on sampling the low-energy subspace as predicted by the annealing. 
This targeted approach allows us to refine our understanding of the most stable and relevant polaron and defect arrangements.
The low-energy ML predictions are evaluated against explicit DFT calculations of the proposed polaron configurations.

Throughout this active learning process, we iteratively improve reliability and efficiency, thus determining the ground-state and low-energy defect-polaron configurations.

\begin{figure}[ht]
\centering
\includegraphics[width=0.5\textwidth]{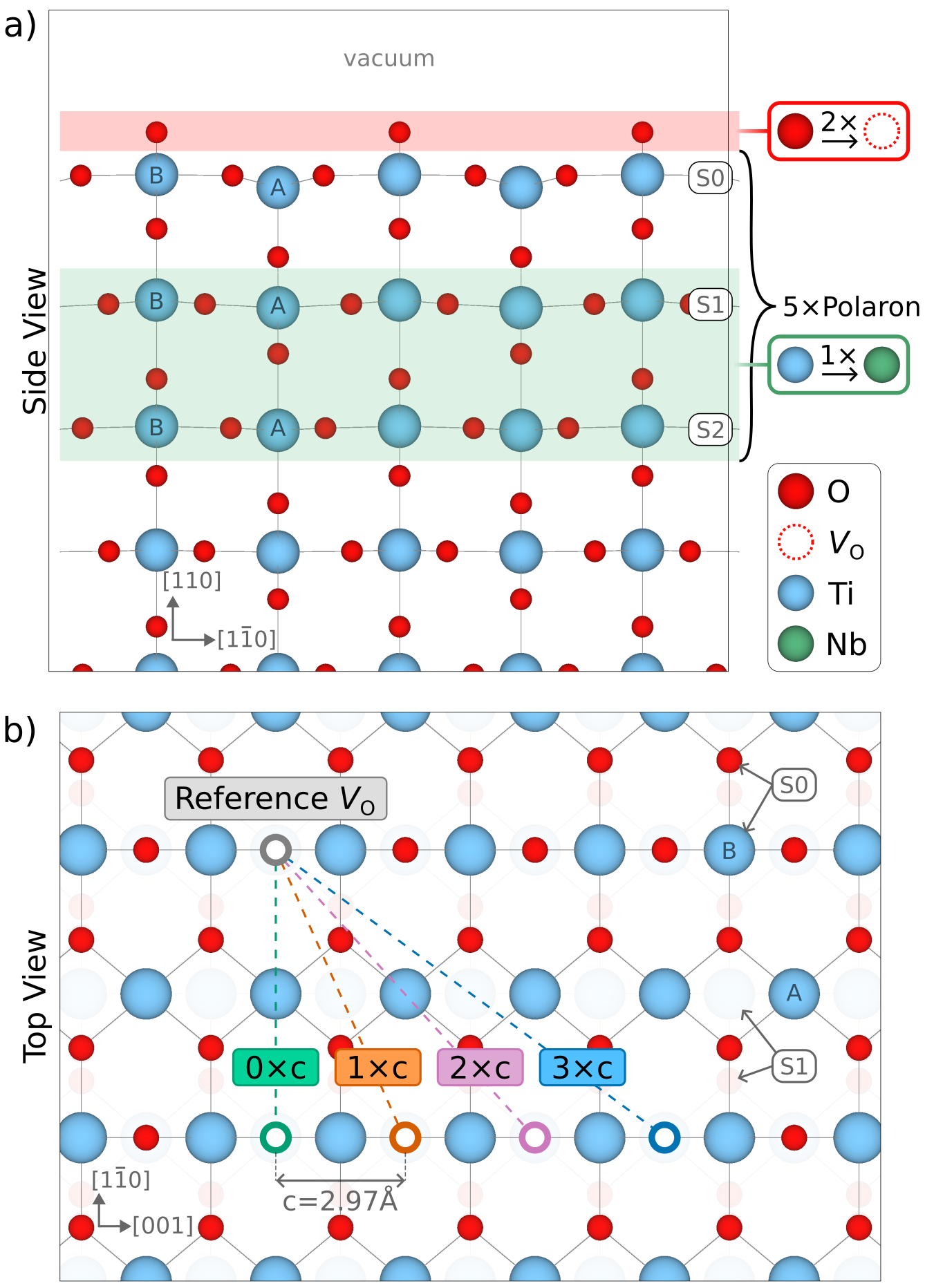}
\caption{Structural Model of \tios. (a) and (b) show the side and top views, respectively.
Two of the surface oxygen bridging atoms are removed to form \VO\ defects (see red box).
Arrangements with the two oxygen vacancies on adjacent [001] O rows are labeled as $0\times$c, $1\times$c, $2\times$c, $3\times$c, which indicate the ideal distance from a reference \VO\ along the c axis.
Polarons can localize on Ti A sites on the surface S0 layer, and on all (A and B) Ti sites of S1 and S2 layers.
The green box indicates the Ti atoms on S1 and S2 layers that can be substituted by the Nb dopant. 
}
\label{fig:multi_phase_gs_evolution}
\end{figure}

\subsection{Computational setup adopted for the \tios\ study case}
\label{sec:MetSETUP}

This Section describes the computational setup adopted for our investigations on the rutile \tios\ surface with Nb dopants and O vacancies.

We performed DFT calculations using VASP~\cite{VASP1,VASP2}.
We adopted the generalized gradient approximation (GGA) within the Perdew, Burke, and Ernzerhof (PBE) parametrization~\cite{Perdew1996a}, with the inclusion of an on-site effective $U$~\cite{Dudarev1998} of $3.9$~eV on the $d$ orbitals of Ti and Nb atoms.
Brillouin-zone sampling was restricted to the $\Gamma$-point for all calculations. 
In the MLFF training, the initial constrained calculations, as well as the subsequent low-accuracy relaxations, a plane-wave energy cutoff of 250~eV was used.
The cutoff was increased to 400~eV in the final calculations.
To train the MLFF, we performed short (15~ps) MD simulations in the NVT ensemble, starting from five different random defect-polaron configurations, as well as the corresponding delocalized configurations.
The time step was set to 1~ps and the temperature to 300~K.
In total, 202 training structures were collected. We used a radial cutoff of 8 and 6~\AA\ for the two- and three-body descriptors, respectively, and 12 and 8 radial basis functions~\cite{Jinnouchi2020}.
The training errors were 0.8~meV/atom for the energy and 0.08~eV/\AA\ for the forces.

Calculations on the adsorption of CO molecules included van der Waals interactions using the non-local SCAN-rVV10 functional~\cite{Peng2016, Sun2016, Sun2015SCAN}.
The spatial extension of the electronic density and scanning tunneling microscopy (STM) images simulated in the Tersoff-Hamann approximation~\cite{Tersoff1985} were calculated by selecting an energy range corresponding to the polaronic eigenvalues only.

Our surface model, shown in Fig.~\ref{fig:multi_phase_gs_evolution}, consists of a $6\times3$ slab with six \tio\ layers (the two bottom stoichiometric layers fixed in their bulk positions).
To simulate the effects of thermal expansion observed during experimental annealing, we used an expanded [001] lattice vector with a constant of 2.968~\AA, corresponding to the high-temperature regime~\cite{Kirby1967rutileT, Hummer2007rutileT}.
We indicate with A the five-fold coordinated Ti atoms (\tiu) on the S0 surface layer and the Ti below, on S1 and S2 layers; B denotes instead all other Ti atoms on S0--S2 layers (see also labels in Fig.~\ref{fig:multi_phase_gs_evolution}).

The slabs contain two surface oxygen vacancies (yielding a vacancy concentration of 11\%) and an Nb atom at a Ti site on the subsurface layers S1 and S2 (resulting in a dopant concentration of approximately 0.5\%)~\cite{Sasahara2013}.
These defects produce five excess electrons that can form polarons (four from the oxygen vacancies and one from the Nb dopant).
Our initial dataset comprises 198 symmetrically distinct configurations.
The A sites on S0, and all the Ti sites on S1 and S2 are considered as possible sites for polaron localization.

\section{Results}
\label{sec:Results}

We used \wf\ to study the most stable defect-polaron configurations of oxygen vacancies and Nb dopants on the \tios\ surface. 
In Sec.~\ref{sec:result-benchamark}, we describe our procedure and comment on the performance of our machinery;
in Sec.~\ref{sec:results-tio2} we focus on the physical outcomes of our test-case study.
The implications for surface reactivity are instead presented in the discussion in Sec.~\ref{sec:Discussion}.




\begin{figure}[ht]
  \centering
  \includegraphics[width=\linewidth]{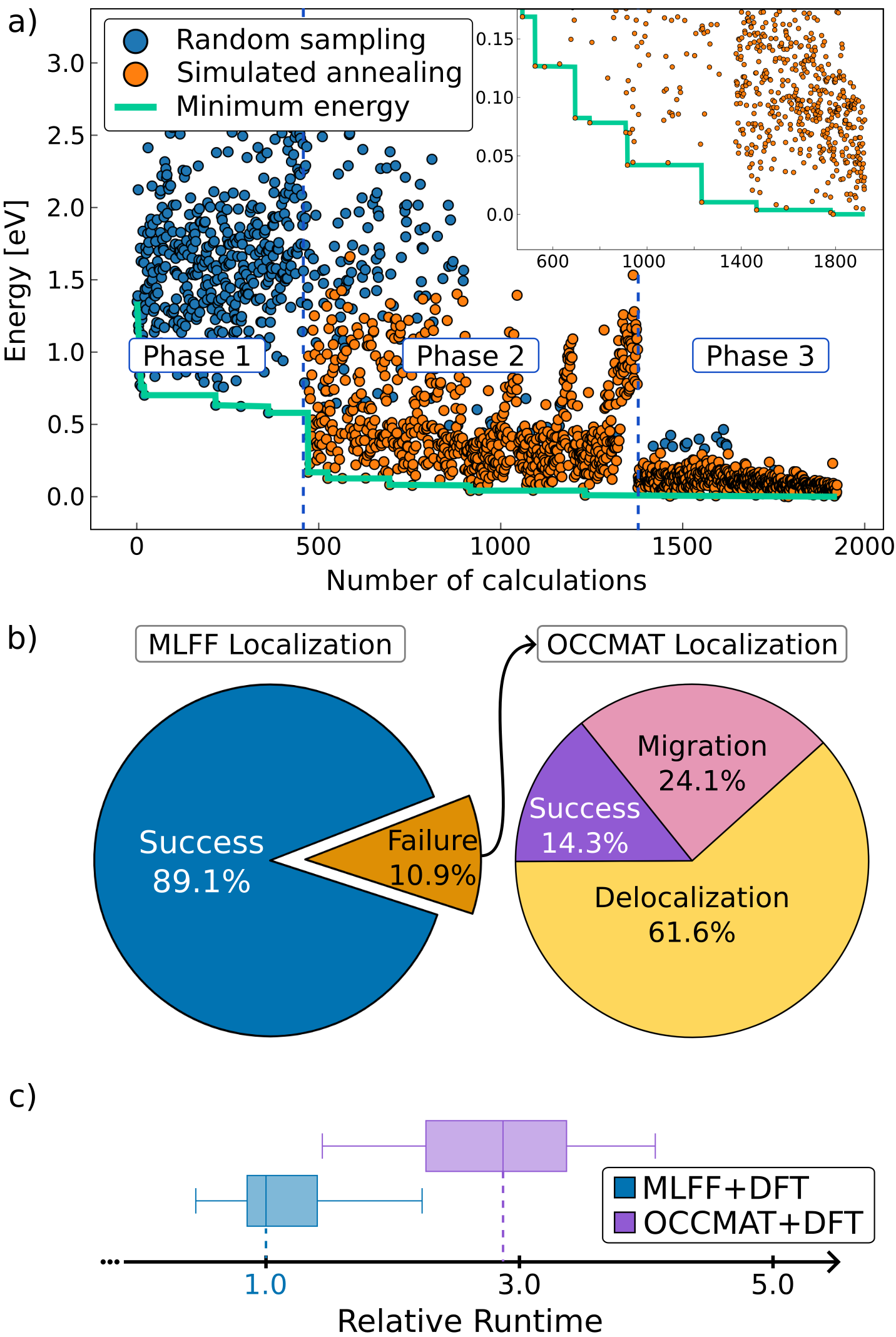}
\caption{Performance analysis of our computational package in the configuration space exploration.
(a) Energy landscape obtained via \wf, using random sampling and simulated annealing.
The inset focuses on the low-energy configurations in Phase~2 and 3 (up to 0.15~eV).
(b) Analysis of polaron localization outcomes using MLFF+DFT (left) and subsequent recovery attempts via OCCMAT (right).
(c) Runtime of the MLFF+DFT and OCCMAT approaches in the low-accuracy runs (boxplots show the distribution of computational times normalized to the MLFF+DFT median).}
  \label{fig:results-main}
\end{figure}

\subsection{Computational package benchmark and performance analysis}
\label{sec:result-benchamark}

Our goal is to assess and analyze the most stable defect-polaron configurations on or near the \tios\ surface.
To this aim, we trained the \vbml\ search tool via the active learning scheme as implemented in the \wf\ package, adopting a multi-phase approach (see Fig.~\ref{fig:results-main}).

Initially, in Phase~1, we modeled the 198 \tios\ $6\times3$ slabs with distinct arrangements of two \VO\ defects and one Nb dopant.
For each slab, we generated three random polaron configurations, resulting in approximately 600 total defect-polaron configurations.
Using \wf, we calculated at the DFT level the energy of all these configurations, taking advantage of the high efficiency of the MLFF-based localization method.
Once we obtained a sufficiently large dataset (approximately 450 valid samples), we used it to train the \vbml\ model.

Next, in Phase~2, the \vbml\ model was iteratively refined by adopting our active learning approach on a mixture of random configurations and annealed samples proposed by the ML model itself.
Finally, in Phase~3, we continued the active-learning training by including only the lowest-energy configurations that we progressively obtained in the annealing.
This strategy allowed us to explore a large number of low-energy configurations [see inset in Fig.~\ref{fig:results-main}(a)].

The three-phase active-learning approach implemented in the \wf\ package enabled efficient training of \vbml\ on a database including in total 1924 symmetrically distinct configurations.
The resulting ML model can predict the energy of any defect-polaron configuration on \tios\ with a mean absolute error (MAE) of 66~meV for our $6\times3$ slab, corresponding to approximately 13~meV per polaron (see Fig.~SF1 in the Supplementary Information).
The accuracy is further improved at low energies due to the annealing-based training in Phase~3 that prioritizes the most stable defect-polaron configurations.
In fact, by considering only low-energy configurations (within 0.15~eV from the minimum, rather than the full range spanning 3~eV), the error is remarkably reduced to 32~meV for the entire slab (6~meV per polaron).

Figures~\ref{fig:results-main}(b) and \ref{fig:results-main}(c) highlight the efficiency of the \wf\ package.
In most cases (89\%), it has been possible to successfully model defect-polaron configurations via an initial MLFF run to localize the polaron at user-defined positions, followed by DFT calculations to refine energy and atomic distortions (\ie\ the MLFF+DFT approach).
In the remaining 11\% of cases, the localization of the polaron failed during the follow-up DFT run [left panel in Fig.~\ref{fig:results-main}(b)].
In response, the \wf\ package automatically employed the occupation matrix control tool to fix the localization failures:
14\% of these OCCMAT calculations resulted in successful localization, while 24\% led to polaron localization on different sites and 62\% to delocalized excess [right panel in Fig.~\ref{fig:results-main}(b)].
Localization failures ought to be expected for some of the sites designed to host polarons, as some arrangements are particularly unstable:
For example, adjacent Ti sites in the \tios\ subsurface typically do not host polarons unless a defect is present in the proximity~\cite{Reticcioli2018a}.
In the cases of re-localization to different sites, if the new configuration is not symmetrically equivalent to any existing entry in the dataset, \wf\ continues the calculation and adds the configuration to the database.
Conversely, if the resulting configuration is symmetrically equivalent to an existing sample, or in the case of electron delocalization, the workflow interrupts the calculation, saving computation time with no need for user intervention. 

As shown in Fig.~\ref{fig:results-main}(c), polaron localization using MLFF is significantly more efficient than alternative approaches.
In our system, low-accuracy polaron localization via the OCCMAT+DFT approach required on average 3~times the computing time compared to the MLFF+DFT strategy (corresponding to $4.4$ and $1.5$ hours per defect-polaron configuration, respectively, when using 64 CPU-cores on our machines).
This gain in efficiency allowed us to save about one-third of the total computational cost in the overall (low- and high-accuracy) polaron-localization process during the exploration of the defect-polaron configuration space in \tios\ (saving in total approximately $300,000$~core hours on our $64$-core machines for calculating $1924$~symmetrically distinct configurations, see also Supplementary Figure~SF2).
Moreover, we estimate that the \wf\ checks on failed polaron localizations ($11$\% of the calculations in the MLFF+DFT, and $86$\% in the OCCMAT+DFT approaches) saved an additional $150,000$~core hours, along with a considerable amount of user time otherwise spent on setup and analysis of localization outcomes.
All considered, the overall computing-time saving amounts to
a reduction of nearly $50$\% in the total computational cost.

Summing up our benchmark analysis, we outline the main advantages of our integrated approach, which combines the MLFF polaron localization routine, the \wf\ automated workflow, and the \vbml\ interface for configuration space exploration.
The MLFF method reduces computational time as compared to state-of-the-art approaches, while maintaining high reliability in polaron localization.
Only $14$\% of the few failed MLFF localization runs were successfully corrected by the OCCMAT approach.
For studies focused on low-energy structures, we find that skipping OCCMAT for failed MLFF cases is generally safe, as these typically correspond to unfavorable configurations, thus resulting in additional overall time saving.
The automated \wf\ workflow and the \vbml\ interface together enable effortless configuration space exploration, even for users with limited experience in polaron modeling.
Their use simplifies the setup and execution of simulations on highly complex systems with entangled interplay between multiple impurities, such as the present case involving polarons, Nb dopants, and oxygen vacancies.
Overall, our computational framework provides a powerful tool for investigating key properties of polaronic materials, as showcased in the following section.

\begin{figure}[ht]
    \centering  \includegraphics[width=\linewidth]{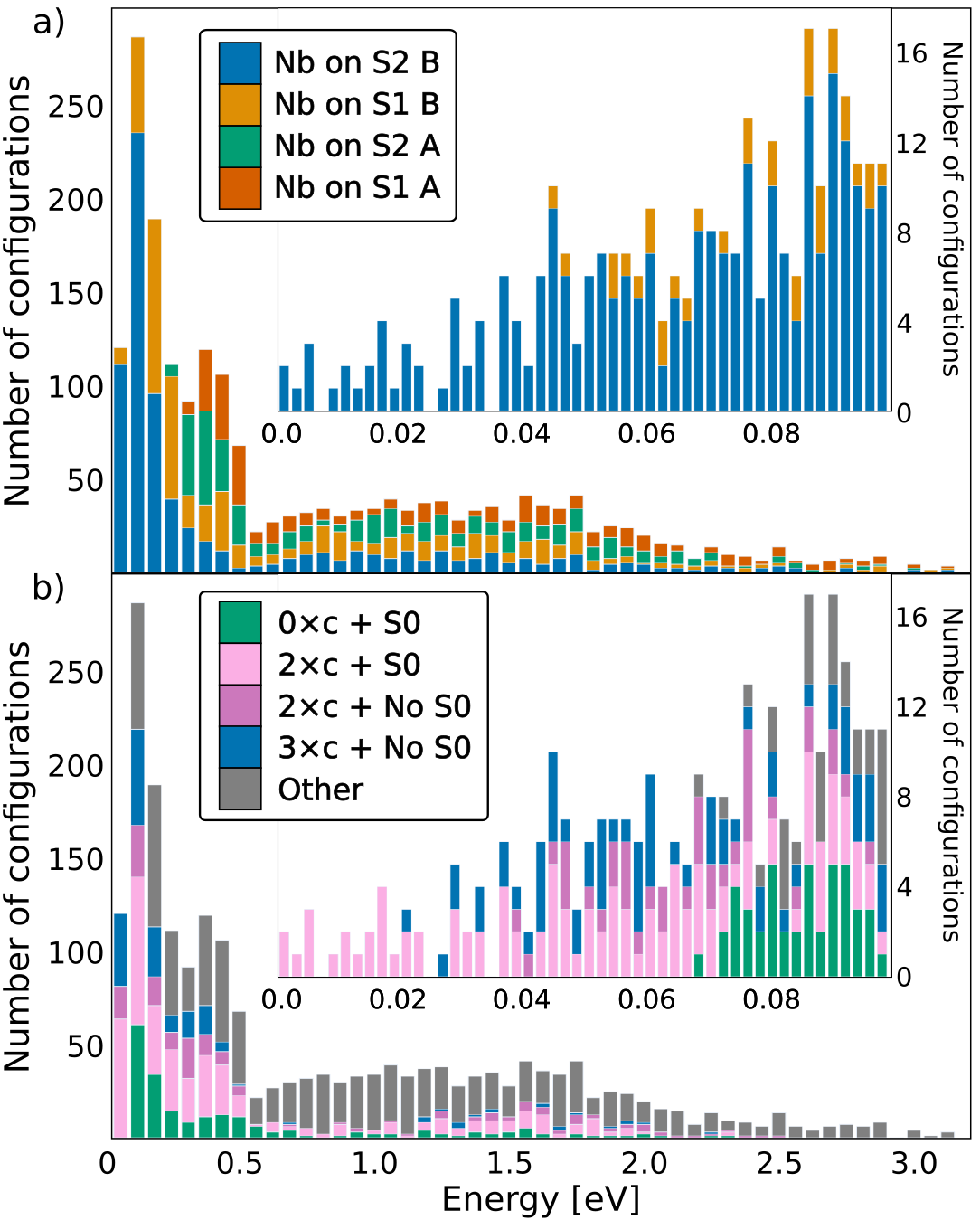}
\caption{Distribution of configuration energies. (a) Histograms are colored according to the Nb position (A or B sites, on the S1 and S2 layers).
(b) Colored histograms represent selected \VO-polaron configurations ($0\times$c, $2\times$c, $3\times$c with/without surface polarons).
Insets in both panels zoom in on the low-energy portion of the configuration space.}    \label{fig:vO_Nb_distribution}
\end{figure}

\begin{figure}[ht]
    \centering
    \includegraphics[width=0.9\linewidth]{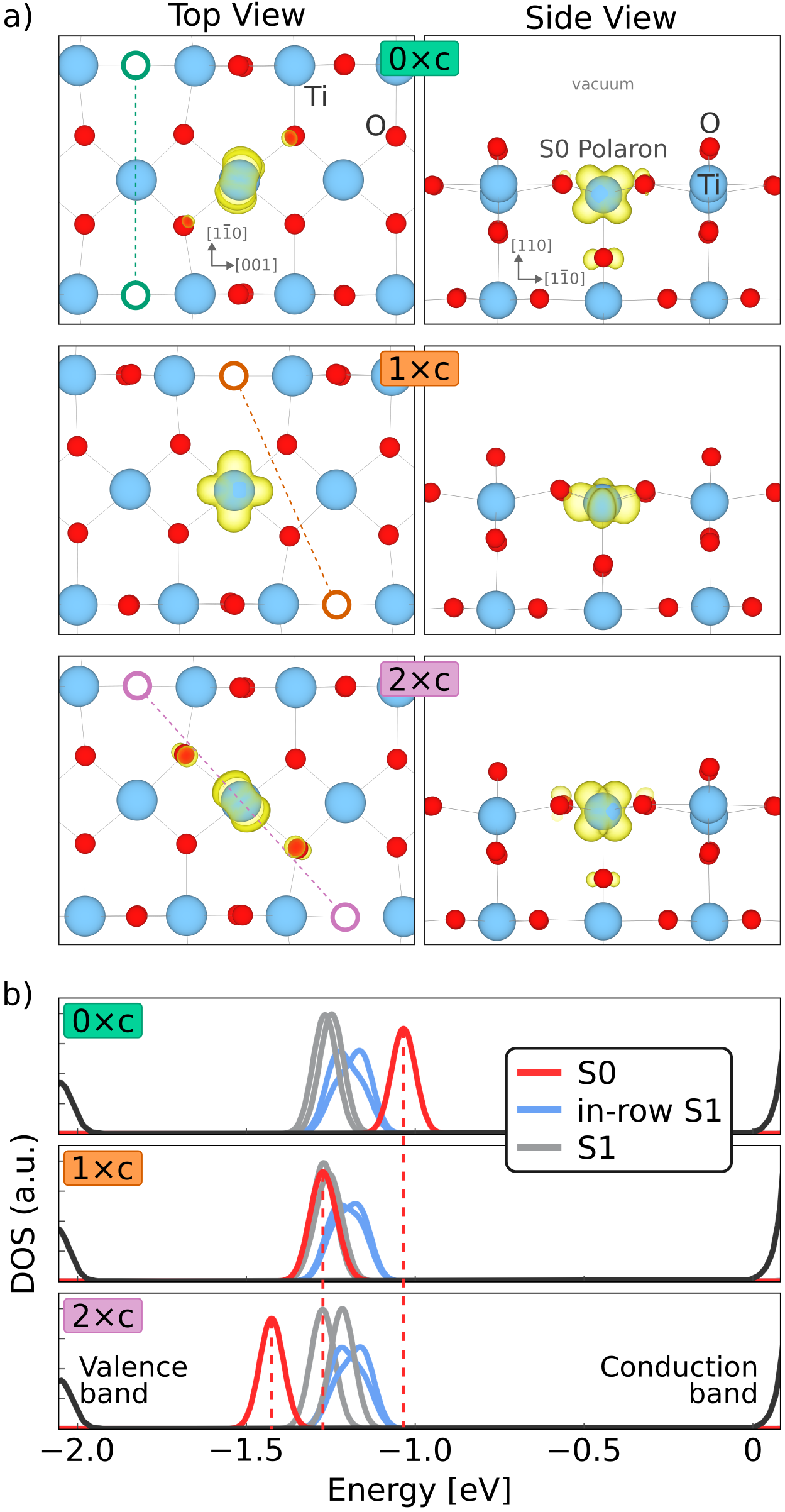}
\caption{Oxygen-vacancies arrangements promoting surface polarons.
(a) Top and side views of S0 polarons in the 0$\times$c, 1$\times$c, and 2$\times$c arrangements (circles indicate \VO\ sites).
(b) Corresponding DOS, projected on S0 polarons (red) and S1 polarons either isolated (gray) or localizing three lattice sites away from a second polaron along the same [001] Ti row (blue).
Dashed lines highlight the shift of the S0 polaron eigenstate due to the different \VO\ arrangements.}
    \label{fig:bestconfmodels}
\end{figure}

\subsection{Distribution of Nb dopants, \VO\ defects, and polarons on \tios}
\label{sec:results-tio2}

The extensive exploration of the Nb-\VO-polaron configuration space described above allows us to identify the most stable defect arrangements on \tios.
Figure~\ref{fig:vO_Nb_distribution} collects our statistical analysis projected onto the Nb and \VO\ contributions (panels (a) and (b), respectively).
The corresponding structural models and polaronic charge densities are shown in Fig.~\ref{fig:bestconfmodels}.

Figure~\ref{fig:vO_Nb_distribution}(a) reveals clear trends for Nb dopants.
The B sites of \tios\ are more favorable substitutional sites compared to the A sites, which are at best 250~meV less stable.
Moreover, there is a clear tendency for Nb to lie on the second subsurface layer:
the inset in panel (a) shows a large number of low-energy S2 configurations, with the most stable one being almost 50~meV more favorable than the best S1 configuration.
We note that the energy instability of Nb in the S1 layer seems to be associated with exceptionally high eigenvalues for S1 polarons in the proximity of the dopant, likely due to unfavorable trapping environment around the defect (see the density of states in Fig.~SF4 in the Supplementary Information).

Moving our attention to the \VO\ defects, Fig.~\ref{fig:vO_Nb_distribution}(b) highlights a remarkable stability for the 2$\times$c arrangement.
All of the most stable configurations up to $20$~meV (see the inset) are formed by a surface S0 polaron in between two 2$\times$c oxygen vacancies [see the model in Fig.~\ref{fig:bestconfmodels}(c)].
This result for the 2$\times$c pattern, obtained here for a \VO\ concentration of 11\% and Nb concentration of 0.5\%, is in agreement with a previous computational study with different levels of impurities (\cVO$=17$\% and \cNb$=0$\%) and with experimental scanning probe images of these patterns in various conditions~\cite{Birschitzky2024, Reticcioli2019c}.

Conversely, the absence of surface polarons favors homogeneous distributions of oxygen vacancies (the 3$\times$c configurations) that are at best $+20$~meV less stable than the ground state (whereas 2$\times$c configurations with no surface polarons appear at higher energies, starting from $+40$~meV).
Considering the lack of such homogeneous patterns in the experimental samples, this result corroborates the hypothesis that polarons play a key role in the distribution of \VO\ defects during the annealing process~\cite{Birschitzky2024}:
Surface S0 polarons likely mediate the vacancy-vacancy repulsion and promote the stabilization of $2\times$c arrangements rather than the homogeneous distribution (which would be expected as dominant pattern arising from a purely repulsive \VO-\VO\ interaction in the absence of intermediate polarons).

Configurations with other oxygen-vacancy arrangements, such as the 0$\times$c and 1$\times$c, appear only at higher energies (above $+70$~meV).
Similarly to the 2$\times$c ground state, also the 0$\times$c and 1$\times$c configurations promote the stabilization of surface polarons.
As shown in Fig.~\ref{fig:bestconfmodels}, the surface polarons exhibit different orbital symmetries depending on the local oxygen-vacancy environment, as evident from the different orientations of the orbital densities in the three panels~\cite{Reticcioli2022}.
Moreover, the local \VO\ environment strongly affects the polaron stability:
The density of states (DOS) reported in Fig.~\ref{fig:bestconfmodels} clearly shows a shift of the S0 polaron eigenvalue, depending on the \VO\ arrangement.
We also note a broadening of the polaronic peaks in the DOS due to the repulsive interaction with other nearby polarons localizing in the same [001] Ti S1 row, three-lattice-site apart (blue curves, labeled as `in-row' in Fig.~\ref{fig:bestconfmodels}, also see Fig.~SF3 in the Supplementary Information for a structural model of 2$\times$c), while isolated polarons show sharper eigenstates (gray curves).
The repulsion between polarons on the same Ti row is indeed known to have a strong effect on the stability of the \tios\ surface itself, leading to reconstruction of the atomic structure at highly reducing conditions~\cite{Reticcioli2017d, Yuan2024, Dohnalek2010, Onishi1994, Li1999a, McCarty2003b, Wang2014a, Mochizuki2016}.
Oxygen vacancies in the proximity might mitigate the polaron-polaron repulsion, reducing the broadening of the polaronic peaks~\cite{Reticcioli2018a}.

\begin{table}[h!]
\renewcommand{\arraystretch}{1.5}
\setlength{\tabcolsep}{12pt}
\centering
\begin{tabular}{|l|c|c|}

\hline
\textbf{Polaron Formation} & Nb on S2 & no Nb \\
\hline
\EPOL(S0) & $-490$~meV & $-530$~meV \\
\EPOL(S1) & $-450$~meV & $-445$~meV \\
\hline
\hline
$\Delta$\EPOL($\rm S0 - S1$) & $-40$~meV & $-85$~meV \\
\hline
\end{tabular}
\caption{Polaron formation energies for surface and subsurface polarons, in the presence and absence of Nb dopants (labeled as `Nb on S2' and `no Nb', respectively). The oxygen vacancies are in the $2\times$c pattern. All polarons have been annealed via \wf, and the best configurations with and without surface polarons are considered in the Table.}
\label{tab:EpolNb}
\end{table}

In contrast to the main role played by \VO\ defects, our configuration space exploration suggests that Nb dopants do not play a primary role in the stabilization of surface polarons, as long as the dopant lies in the S2 layer.
To verify this finding, we performed additional calculations:
By taking advantage of our efficient \wf\ framework, we explored the polaronic landscape on \tios\ surfaces without Nb defects.
In Table~\ref{tab:EpolNb} we compare the polaron formation energy (\EPOL) of different configurations.
In the absence of surface polarons, the optimal $2\times$c configuration with Nb on the S2 layer yields an average \EPOL\ energy of $-450$~meV per S1 polaron, obtained by comparing the total energy of the polaronic system ($E^{\rm loc}$) with respect to a solution with all electrons delocalized ($E^{\rm del}$):
\begin{equation}
E_{\rm POL}(S1) = \left ( E^{\rm loc}-E^{\rm del} \right ) / n_{\rm S1}
\end{equation}
where $n_{\rm S1}$ is the number of subsurface polarons.
Assuming this energy value for S1 polarons unchanged, we can evaluate the formation energy for the S0 polaron 
using the following expression:
\begin{equation}
E_{\rm POL}(S0) = E^{\rm loc}-E^{\rm del} - \left ( n_{\rm S1}\times E_{\rm POL}(S1) \right )~.
\end{equation}
We obtain \EPOL(S0)$= -490$~meV in the $2\times$c configuration with Nb, thus an energy gain $\Delta$\EPOL\ of $-40$~meV upon promotion of a S1 polaron to the surface.

By removing the Nb defect and annealing via \wf\ the polarons in the $2 \times$c configuration, we can calculate the formation energies for surface polarons in the absence of extrinsic dopants:
\EPOL(S1)$=-445$~meV (similar to the Nb case reported above), and \EPOL(S0)$=-530$~meV.
As evident, the energy gain of a polaron localizing on the surface compared to the subsurface increases to $\Delta$\EPOL$=-85$~meV, practically doubling the value obtained in the Nb-doped case.
From this analysis, we conclude that Nb dopants do show an attractive interaction with polarons:
The Nb impurity lying on the S2 layer worsens the stability of the S0 polaron by $45$~meV.
However, this effect is relatively minor when compared to the \EPOL\ energy of the surface polaron in the $2 \times$c complex, which is an order of magnitude larger.
Thus, Nb does not ultimately prevent the stabilization of surface polarons.

As such, doping rutile \tio\ with Nb has been found to be an effective method for introducing additional electrons into the system without altering the surface structure.
This property is particularly advantageous for imaging purposes, such as scanning tunneling microscopy, as it enables conductivity in the bulk, while the preservation of the surface structure might be crucial for applications such as in catalysis, where active sites for adsorption and catalytic reactions remain intact.

\section{Discussion}
\label{sec:Discussion}

Clarifying the distribution of defects and polarons is key to determining the properties of materials and advancing technological applications. 
Here, we use the results of our extensive \wf\ characterization study to shed light on crucial aspects of \tios\ surface reactivity, by modeling CO adsorption on the undercoordinated Ti sites (\tiu)~\cite{Henderson2011, Cao2017a, Cheng2021, Reticcioli2019c, Adachi2021, Wu2020, Petrik2012a, Petrik2018, FarnesiCamellone2011a, Yu2015a, Kunat2009, Dohnalek2006, Mu2017, Zhao2009b, Yoon2015, PratesRamalho2017a}.

\begin{figure}[ht]
  \centering
  \includegraphics[width=0.90\linewidth]{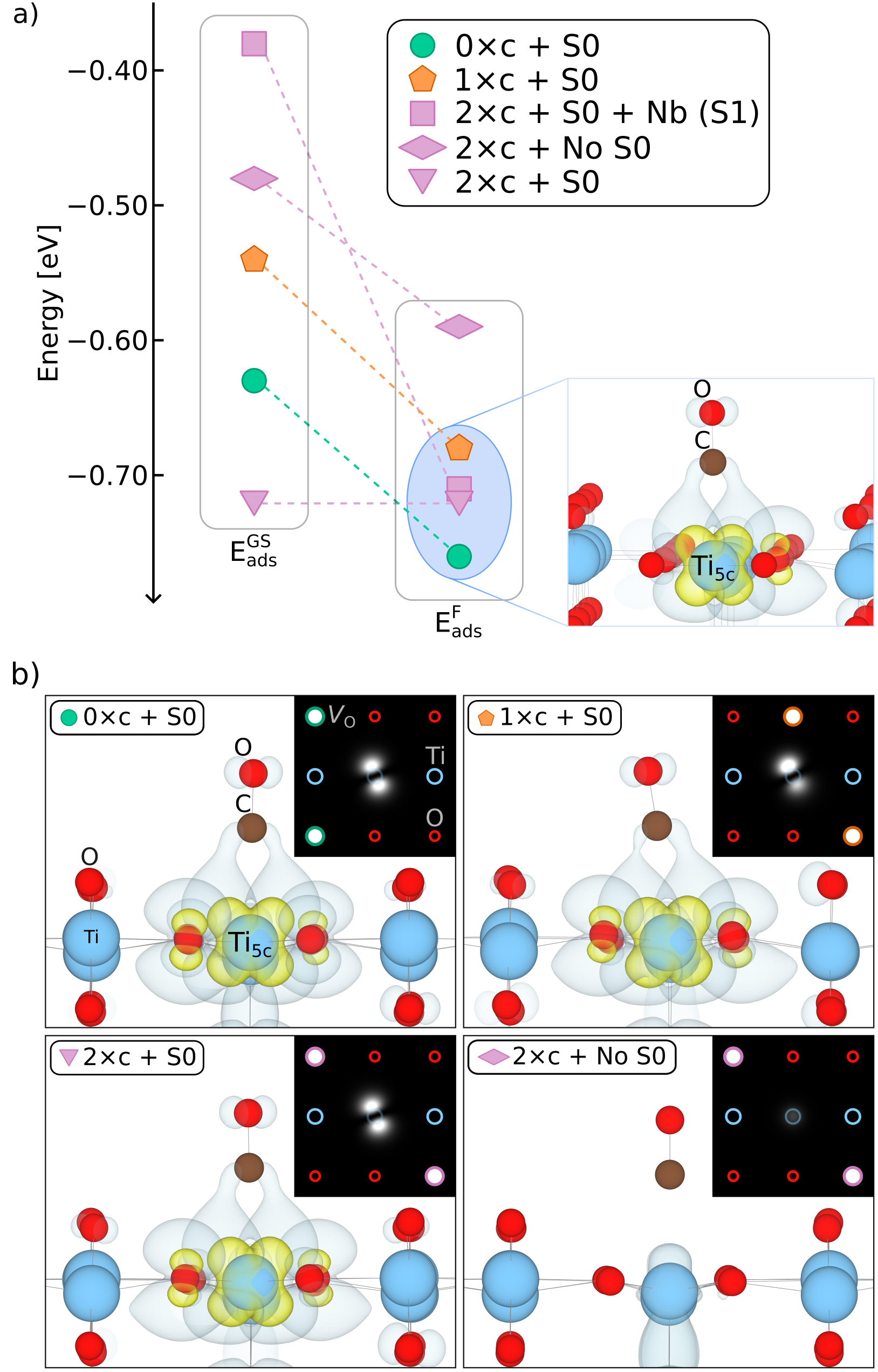}
\caption{CO adsorption on \tiu\ sites.
(a) \eadsG\ (left) and \eadsF\ (right) energies for a CO molecule adsorbing on a \tiu\ site hosting a surface S0 polaron in the proximity of two \VO\ defects in the $0\times$c, $1\times$c and $2\times$c arrangements (circle, pentagon, and triangle symbols, respectively), with the Nb dopant on the S2 layer.
For the $2\times$c pattern, we also show the case of no surface polaron (diamond) and the case of Nb on the subsurface S1 layer (square).
The inset shows the complex forming between CO and the S0 polaron (charge densities with low and high isovalues in light blue and yellow, respectively).
(b) Charge density of selected cases from (a).
Insets show the simulated STM images of the polaronic density (filled circles indicate the position of \VO\ defects).
}
\label{fig:discussion-main}
\end{figure}

Figure~\ref{fig:discussion-main} collects our analysis for the most favorable polaron-Nb arrangements obtained through the \wf-\vbml\ exploration for the \VO\ patterns $0\times$c, $1\times$c, and $2\times$c.
These patterns favor the formation of a polaron on the surface layer, in the proximity of the oxygen vacancies (see Fig.~\ref{fig:bestconfmodels}).
Additionally, for the $2\times$c arrangement, we also considered the most favorable configuration featuring no surface polaron (all five polarons are on the S1 subsurface layer, labeled by a diamond symbol).
Nb is on the S2 subsubsurface layer, except for the configuration labeled by a square symbol, where Nb lies on an S1 site below the S0 polaron (see also model structure in Fig.~SF4).
We modeled CO adsorbing directly above the \tiu\ site hosting the surface polaron (or in the corresponding site in between the two {\VO}s for the diamond case with all S1 polarons).

To inspect the role of polarons in adsorbing CO molecules, we computed two distinct adsorption energies, \eadsG\ and \eadsF, for each defect-polaron configuration $x$:
\begin{equation}
\begin{aligned}
E_{\rm ads}^{\rm GS}(x) &= 
E_{\rm slab+CO}(x)-(E_{\rm slab}(GS)+E_{\rm CO})\\
E_{\rm ads}^{\rm F}(x) &= 
E_{\rm slab+CO}(x)-(E_{\rm slab}(x)+E_{\rm CO})~,
\end{aligned}
\end{equation}
where $E_{\rm slab+CO}(x)$ is the energy of \tios\ with an adsorbed CO molecule in the defect-polaron configuration $x$, $E_{\rm slab}$ is the energy of the \tios\ surface with no adsorbate either in its ground-state defect-polaron configuration [$E_{\rm slab}(GS)$] or fixed in the geometry of configuration $x$ [$E_{\rm slab}(x)$], and $E_{\rm CO}$ is the energy of a CO molecule in the gas phase.
This choice allows us to determine the overall most stable adsorption scenario via \eadsG, as well as the direct contribution of the polaron to the adsorption via \eadsF.

In Fig.~\ref{fig:discussion-main}(a), the left column shows the \eadsG\ values calculated for all the different defect distributions under consideration. 
The most favorable CO adsorption scenario is given by the $2\times$c arrangement, with the molecule adsorbing on a surface polaron trapping in between the two oxygen vacancies (represented by a triangle symbol).
This result is in line with previous DFT calculations modeling selected \VO\ configurations on the Nb-free \tios\ surface~\cite{Reticcioli2019c}, thus suggesting a minor role of Nb dopants in CO adsorption (see also discussion on the effects of Nb at the end of this Section).

The adsorption becomes progressively less favorable for the polaron-CO complexes at the $0\times$c and $1\times$c \VO\ patterns [see \eadsG\ for the circle and pentagon symbols in Fig.~\ref{fig:discussion-main}(a)], and even worse for configurations with no surface polaron (see the diamond symbol, representing the $2\times$c configuration with all polarons in the subsurface layer).
Finally, the configuration with an Nb impurity on the subsurface layer S1 right below the adsorption site yields the most unfavorable \eadsG\ energy (square symbol).
This behavior originates from the intrinsic instability of configurations with shallow Nb defects, rather than from the interaction of the dopant with CO itself, as discussed in the analysis of the \eadsF\ energy below.

The \eadsF\ energy [right column in Fig.~\ref{fig:discussion-main}(a)] clarifies the role of surface polarons in stabilizing the CO adsorption, and the minor contribution of Nb.
In fact, in the expression for \eadsF$(x)$, the reference energy $E_{\rm slab}(x)$ is obtained from a system with no CO adsorbate, while defects and polarons are kept at the same sites as in the slab adsorbing the molecule;
therefore, at variance with \eadsG, the adsorption energy \eadsF\ is not affected by the configuration energy of defects and polarons, and allows us to inspect the direct contributions of local impurities to CO adsorption.
As evident from the right column in Fig.~\ref{fig:discussion-main}(a), a CO molecule adsorbing on a surface polaron shows very similar \eadsF\ energies (within $-760$~meV and $-675$~meV) despite the specific \VO\ arrangement:
The S0 polaron transfers part of the electronic charge into the molecular orbitals, thus stabilizing the adsorbing CO [see inset in Fig.~\ref{fig:discussion-main}(a)].
Figure~\ref{fig:discussion-main}(b) shows the corresponding simulated scanning tunneling microscopy (STM) images.
The double-lobe signal of the CO molecule adsorbing on S0 polarons appears brighter compared to configurations without polarons, which is a clear indication of the sizable transfer of polaronic charge into the molecular orbitals.
We note small differences in the orientation of the double lobes of the polaron-CO complexes and on \eadsF, which are due to the interaction with the oxygen vacancies in the proximity, lying in different arrangements.

The \eadsF\ energy in Fig.~\ref{fig:discussion-main}(a) also allows us to clarify the contribution of Nb.
As mentioned above, the \eadsG\ value obtained for a Nb lying on the S1 layer directly below the polaronic adsorption site is quite unfavorable ($-383$~meV).
This is due to the high configuration energy of subsurface Nb atoms, which is significantly worse compared to that of the reference GS system used to calculate \eadsG.
Conversely, from \eadsF\ we can assess the absence of any sizable role of Nb in the CO adsorption.
In fact, the \eadsF\ value for this case is in the typical range of polaron-CO complexes, despite Nb being in the proximity of the adsorption site (see also structural model in Fig.~SF4 in the Supplementary Information).
To further validate this conclusion, we modeled the CO adsorption on slabs without Nb defects.
The results show that the \eadsF\ adsorption energy is largely unaffected by the removal of Nb (going from $-720$~meV with Nb to $-760$~meV without Nb in the case of S0-polaron driven adsorption, and from $-590$~meV to $-560$~meV in the absence of surface polarons; see also Fig.~SF5).
Therefore, these results further confirm that Nb does not play a direct role in CO adsorption, at variance with the key contribution of oxygen vacancies.

\section{Conclusions}

In this work, we introduced a computational package that enables efficient and fully automated DFT modeling of small polarons in materials.
It provides a powerful framework to uncover the fundamental properties of polarons and defects, and to gain insight into technologically relevant phenomena—such as surface reactivity, which we explored here for \tios\ as a case study.

The MLFF-localization tool included in the package enables rapid and reliable trapping of polarons on user-defined sites.
By leveraging data from finite-temperature DFT molecular dynamics and training the MLFF with polaron-specific labels, our method yields atomic structures with localized charges at target positions, while maintaining DFT accuracy at a fraction of the computational cost.

The \wf\ high-throughput workflow streamlines and automates all steps required for polaron modeling.
It coordinates polaron localization calculations, monitors localization success in real time, and triggers corrective measures when needed.
This ensures robust convergence toward physically meaningful polaronic solutions with minimal manual intervention, making polaron modeling more accessible and reproducible.

Additionally, \wf\ features an interface to active-learning frameworks for exploring the configurational space of defects and polarons.
Using simulated annealing and ML-predicted energies, this module identifies low-energy configurations and iteratively improves its predictions by selecting new, informative ones for training the model.

We applied this package to Nb-doped \tios\ surfaces with oxygen vacancies to investigate how polaron-defect interactions affect surface reactivity.
Our results show that oxygen vacancies play a central role in stabilizing surface polarons, especially in specific vacancy arrangements such as the $2\times$c pattern.
In contrast, Nb dopants contribute by donating excess electrons to the system, but have a limited impact on the stability of surface polarons.

Adsorption energy analysis, using CO as a probe molecule, revealed that surface polarons significantly enhance the adsorption strength on \tiu\ sites by transferring charge to molecular orbitals.
These polaron-CO complexes are practically unaffected by Nb dopants on the S2 layers, and are perturbed only to a minor extent by the local oxygen-vacancy arrangement:
While \VO\ defects play a key role in stabilizing S0 surface polarons, they show minor effects on the polaron-CO complexes, such as slight rotations in the simulated STM signals and small variations in the adsorption energies.

These findings highlight the importance of efficient configuration sampling in first-principles simulations to reveal microscopic links between defect configurations, polaron distributions, and catalytic activity on oxide surfaces.
Our work establishes a robust computational framework for modeling such defect-polaron interplay, opening new research paths for understanding and predicting the role of polarons in functional materials, with potential applications extending from catalysis to electronic and energy technologies.

\section{Acknowledgments}
The computational results have been achieved using the Austrian Scientific Computing (ASC) infrastructure, 
LEONARDO at CINECA (Italy) via an AURELEO (Austrian Users at LEONARDO supercomputer) project, and computational resources provided by the Australian National Computational Infrastructure and Pawsey Supercomputing Research Centre via the National Computational Merit Allocation scheme.
C.V. acknowledges financial support from the Australian Research Council (DE220101147).
M.W. and M.M. acknowledge the Austrian Science Fund (FWF) for financial support through projects P-32711 and 10.55776/PAT2176923, respectively.
M.R. acknowledges Austria's Agency for Education and Internationalisation (OeAD, project RS 05/2024).

\section{Data Availability Statement}

The code developed within this study is openly available via Zenodo repository, see Ref.~\citenum{polflow_repo}.

\section{Competing Financial Interests}

Michael Wolloch declares a competing interest as an employee of the VASP Software GmbH.
All other authors declare no competing financial interest.

\bibliography{mrbib,mybib}

\begin{thebibliography}{100}%
\makeatletter
\providecommand \@ifxundefined [1]{%
 \@ifx{#1\undefined}
}%
\providecommand \@ifnum [1]{%
 \ifnum #1\expandafter \@firstoftwo
 \else \expandafter \@secondoftwo
 \fi
}%
\providecommand \@ifx [1]{%
 \ifx #1\expandafter \@firstoftwo
 \else \expandafter \@secondoftwo
 \fi
}%
\providecommand \natexlab [1]{#1}%
\providecommand \enquote  [1]{``#1''}%
\providecommand \bibnamefont  [1]{#1}%
\providecommand \bibfnamefont [1]{#1}%
\providecommand \citenamefont [1]{#1}%
\providecommand \href@noop [0]{\@secondoftwo}%
\providecommand \href [0]{\begingroup \@sanitize@url \@href}%
\providecommand \@href[1]{\@@startlink{#1}\@@href}%
\providecommand \@@href[1]{\endgroup#1\@@endlink}%
\providecommand \@sanitize@url [0]{\catcode `\\12\catcode `\$12\catcode `\&12\catcode `\#12\catcode `\^12\catcode `\_12\catcode `\%12\relax}%
\providecommand \@@startlink[1]{}%
\providecommand \@@endlink[0]{}%
\providecommand \url  [0]{\begingroup\@sanitize@url \@url }%
\providecommand \@url [1]{\endgroup\@href {#1}{\urlprefix }}%
\providecommand \urlprefix  [0]{URL }%
\providecommand \Eprint [0]{\href }%
\providecommand \doibase [0]{http://dx.doi.org/}%
\providecommand \selectlanguage [0]{\@gobble}%
\providecommand \bibinfo  [0]{\@secondoftwo}%
\providecommand \bibfield  [0]{\@secondoftwo}%
\providecommand \translation [1]{[#1]}%
\providecommand \BibitemOpen [0]{}%
\providecommand \bibitemStop [0]{}%
\providecommand \bibitemNoStop [0]{.\EOS\space}%
\providecommand \EOS [0]{\spacefactor3000\relax}%
\providecommand \BibitemShut  [1]{\csname bibitem#1\endcsname}%
\let\auto@bib@innerbib\@empty
\bibitem [{\citenamefont {Emin}(1982)}]{Emin1982}%
  \BibitemOpen
  \bibfield  {author} {\bibinfo {author} {\bibfnamefont {D.}~\bibnamefont {Emin}},\ }\textsl {\enquote {\bibinfo {title} {{Small polarons}},}\ }\href {\doibase10.1063/1.2938044} {\bibfield  {journal} {\bibinfo  {journal} {Physics Today}\ }\bibinfo {volume} {35},\ \bibinfo {pages} {34} (\bibinfo {year} {1982})}\BibitemShut {NoStop}%
\bibitem [{\citenamefont {Stoneham}\ \emph {et~al.}(2007)\citenamefont {Stoneham}, \citenamefont {Gavartin}, \citenamefont {Shluger}, \citenamefont {Kimmel}, \citenamefont {Mũoz~Ramo}, \citenamefont {R{\o}nnow}, \citenamefont {Aeppli},\ and\ \citenamefont {Renner}}]{Stoneham2007}%
  \BibitemOpen
  \bibfield  {author} {\bibinfo {author} {\bibfnamefont {A.~M.}\ \bibnamefont {Stoneham}}, \bibinfo {author} {\bibfnamefont {J.}~\bibnamefont {Gavartin}}, \bibinfo {author} {\bibfnamefont {A.~L.}\ \bibnamefont {Shluger}}, \bibinfo {author} {\bibfnamefont {A.~V.}\ \bibnamefont {Kimmel}}, \bibinfo {author} {\bibfnamefont {D.}~\bibnamefont {Mũoz~Ramo}}, \bibinfo {author} {\bibfnamefont {H.~M.}\ \bibnamefont {R{\o}nnow}}, \bibinfo {author} {\bibfnamefont {G.}~\bibnamefont {Aeppli}}, \ and\ \bibinfo {author} {\bibfnamefont {C.}~\bibnamefont {Renner}},\ }\textsl {\enquote {\bibinfo {title} {{Trapping, self-trapping and the polaron family}},}\ }\href {\doibase10.1088/0953-8984/19/25/255208} {\bibfield  {journal} {\bibinfo  {journal} {Journal of Physics Condensed Matter}\ }\bibinfo {volume} {19},\ \bibinfo {pages} {255208} (\bibinfo {year} {2007})}\BibitemShut {NoStop}%
\bibitem [{\citenamefont {Alexandrov}\ and\ \citenamefont {Devreese}(2010)}]{Alexandrov2010book}%
  \BibitemOpen
  \bibfield  {author} {\bibinfo {author} {\bibfnamefont {A.~S.}\ \bibnamefont {Alexandrov}}\ and\ \bibinfo {author} {\bibfnamefont {J.~T.}\ \bibnamefont {Devreese}},\ }\href {\doibase10.1007/978-3-642-01896-1} {\emph {\bibinfo {title} {Springer Series in Solid-State Sciences}}},\ \bibinfo {series} {Springer Series in Solid-State Sciences}, Vol.\ \bibinfo {volume} {159}\ (\bibinfo  {publisher} {Springer Berlin Heidelberg},\ \bibinfo {address} {Berlin, Heidelberg},\ \bibinfo {year} {2010})\ pp.\ \bibinfo {pages} {1--163}\BibitemShut {NoStop}%
\bibitem [{\citenamefont {Holstein}(2000)}]{Holstein2000}%
  \BibitemOpen
  \bibfield  {author} {\bibinfo {author} {\bibfnamefont {T.}~\bibnamefont {Holstein}},\ }\textsl {\enquote {\bibinfo {title} {{Studies of polaron motion: Part II. The "small" polaron}},}\ }\href {\doibase10.1006/aphy.2000.6021} {\bibfield  {journal} {\bibinfo  {journal} {Annals of Physics}\ }\bibinfo {volume} {281},\ \bibinfo {pages} {725} (\bibinfo {year} {2000})}\BibitemShut {NoStop}%
\bibitem [{\citenamefont {Tanner}\ and\ \citenamefont {Thornton}(2022)}]{Tanner2022}%
  \BibitemOpen
  \bibfield  {author} {\bibinfo {author} {\bibfnamefont {A.~J.}\ \bibnamefont {Tanner}}\ and\ \bibinfo {author} {\bibfnamefont {G.}~\bibnamefont {Thornton}},\ }\textsl {\enquote {\bibinfo {title} {{TiO2 Polarons in the Time Domain: Implications for Photocatalysis}},}\ }\href {\doibase10.1021/acs.jpclett.1c03677} {\bibfield  {journal} {\bibinfo  {journal} {Journal of Physical Chemistry Letters}\ }\bibinfo {volume} {13},\ \bibinfo {pages} {559} (\bibinfo {year} {2022})}\BibitemShut {NoStop}%
\bibitem [{\citenamefont {Natanzon}\ \emph {et~al.}(2020)\citenamefont {Natanzon}, \citenamefont {Azulay},\ and\ \citenamefont {Amouyal}}]{Natanzon2020}%
  \BibitemOpen
  \bibfield  {author} {\bibinfo {author} {\bibfnamefont {Y.}~\bibnamefont {Natanzon}}, \bibinfo {author} {\bibfnamefont {A.}~\bibnamefont {Azulay}}, \ and\ \bibinfo {author} {\bibfnamefont {Y.}~\bibnamefont {Amouyal}},\ }\textsl {\enquote {\bibinfo {title} {{Evaluation of Polaron Transport in Solids from First-principles}},}\ }\href {\doibase10.1002/ijch.201900101} {\bibfield  {journal} {\bibinfo  {journal} {Israel Journal of Chemistry}\ }\bibinfo {volume} {60},\ \bibinfo {pages} {768} (\bibinfo {year} {2020})}\BibitemShut {NoStop}%
\bibitem [{\citenamefont {Marshall}\ \emph {et~al.}(2015)\citenamefont {Marshall}, \citenamefont {Becerra-toledo}, \citenamefont {Marks},\ and\ \citenamefont {Castell}}]{Jupille2015book}%
  \BibitemOpen
  \bibfield  {author} {\bibinfo {author} {\bibfnamefont {M.~S.~J.}\ \bibnamefont {Marshall}}, \bibinfo {author} {\bibfnamefont {A.~E.}\ \bibnamefont {Becerra-toledo}}, \bibinfo {author} {\bibfnamefont {L.~D.}\ \bibnamefont {Marks}}, \ and\ \bibinfo {author} {\bibfnamefont {M.~R.}\ \bibnamefont {Castell}},\ }\href {\doibase10.1007/978-3-319-14367-5} {\emph {\bibinfo {title} {Defects at Oxide Surfaces}}},\ edited by\ \bibinfo {editor} {\bibfnamefont {J.}~\bibnamefont {Jupille}}\ and\ \bibinfo {editor} {\bibfnamefont {G.}~\bibnamefont {Thornton}},\ \bibinfo {series} {Springer Series in Surface Sciences}, Vol.~\bibinfo {volume} {58}\ (\bibinfo  {publisher} {Springer International Publishing},\ \bibinfo {address} {Cham},\ \bibinfo {year} {2015})\ pp.\ \bibinfo {pages} {327--349}\BibitemShut {NoStop}%
\bibitem [{\citenamefont {Strand}\ and\ \citenamefont {Shluger}(2023)}]{Strand2023}%
  \BibitemOpen
  \bibfield  {author} {\bibinfo {author} {\bibfnamefont {J.}~\bibnamefont {Strand}}\ and\ \bibinfo {author} {\bibfnamefont {A.~L.}\ \bibnamefont {Shluger}},\ }\textsl {\enquote {\bibinfo {title} {{On the Structure of Oxygen Deficient Amorphous Oxide Films}},}\ }\href {\doibase10.1002/advs.202306243} {\bibfield  {journal} {\bibinfo  {journal} {Advanced Science}\ }\bibinfo {volume} {2306243},\ \bibinfo {pages} {1} (\bibinfo {year} {2023})}\BibitemShut {NoStop}%
\bibitem [{\citenamefont {Pastor}\ \emph {et~al.}(2022)\citenamefont {Pastor}, \citenamefont {Sachs}, \citenamefont {Selim}, \citenamefont {Durrant}, \citenamefont {Bakulin},\ and\ \citenamefont {Walsh}}]{Pastor2022polrev}%
  \BibitemOpen
  \bibfield  {author} {\bibinfo {author} {\bibfnamefont {E.}~\bibnamefont {Pastor}}, \bibinfo {author} {\bibfnamefont {M.}~\bibnamefont {Sachs}}, \bibinfo {author} {\bibfnamefont {S.}~\bibnamefont {Selim}}, \bibinfo {author} {\bibfnamefont {J.~R.}\ \bibnamefont {Durrant}}, \bibinfo {author} {\bibfnamefont {A.~A.}\ \bibnamefont {Bakulin}}, \ and\ \bibinfo {author} {\bibfnamefont {A.}~\bibnamefont {Walsh}},\ }\textsl {\enquote {\bibinfo {title} {{Electronic defects in metal oxide photocatalysts}},}\ }\href {\doibase10.1038/s41578-022-00433-0} {\bibfield  {journal} {\bibinfo  {journal} {Nature Reviews Materials}\ }\bibinfo {volume} {7},\ \bibinfo {pages} {503} (\bibinfo {year} {2022})}\BibitemShut {NoStop}%
\bibitem [{\citenamefont {Chi}\ \emph {et~al.}(2022)\citenamefont {Chi}, \citenamefont {Mandal}, \citenamefont {Liu}, \citenamefont {Fauzi}, \citenamefont {Chaudhuri}, \citenamefont {Whitcher}, \citenamefont {Jani}, \citenamefont {Chen}, \citenamefont {Xi}, \citenamefont {Diao}, \citenamefont {Naradipa}, \citenamefont {Yu}, \citenamefont {Yang}, \citenamefont {Castro-Neto}, \citenamefont {Breese}, \citenamefont {Loh}, \citenamefont {Venkatesan},\ and\ \citenamefont {Rusydi}}]{Chi2022polcat}%
  \BibitemOpen
  \bibfield  {author} {\bibinfo {author} {\bibfnamefont {X.}~\bibnamefont {Chi}}, \bibinfo {author} {\bibfnamefont {L.}~\bibnamefont {Mandal}}, \bibinfo {author} {\bibfnamefont {C.}~\bibnamefont {Liu}}, \bibinfo {author} {\bibfnamefont {A.~D.}\ \bibnamefont {Fauzi}}, \bibinfo {author} {\bibfnamefont {A.}~\bibnamefont {Chaudhuri}}, \bibinfo {author} {\bibfnamefont {T.~J.}\ \bibnamefont {Whitcher}}, \bibinfo {author} {\bibfnamefont {H.~K.}\ \bibnamefont {Jani}}, \bibinfo {author} {\bibfnamefont {Z.}~\bibnamefont {Chen}}, \bibinfo {author} {\bibfnamefont {S.}~\bibnamefont {Xi}}, \bibinfo {author} {\bibfnamefont {C.}~\bibnamefont {Diao}}, \bibinfo {author} {\bibfnamefont {M.~A.}\ \bibnamefont {Naradipa}}, \bibinfo {author} {\bibfnamefont {X.}~\bibnamefont {Yu}}, \bibinfo {author} {\bibfnamefont {P.}~\bibnamefont {Yang}}, \bibinfo {author} {\bibfnamefont {A.~H.}\ \bibnamefont {Castro-Neto}}, \bibinfo {author} {\bibfnamefont {M.~B.}\ \bibnamefont {Breese}}, \bibinfo {author} {\bibfnamefont {K.~P.}\ \bibnamefont
  {Loh}}, \bibinfo {author} {\bibfnamefont {T.~V.}\ \bibnamefont {Venkatesan}}, \ and\ \bibinfo {author} {\bibfnamefont {A.}~\bibnamefont {Rusydi}},\ }\textsl {\enquote {\bibinfo {title} {{Unravelling a new many-body large-hole polaron in a transition metal oxide that promotes high photocatalytic activity}},}\ }\href {\doibase10.1038/s41427-022-00364-w} {\bibfield  {journal} {\bibinfo  {journal} {NPG Asia Materials}\ }\bibinfo {volume} {14},\ \bibinfo {pages} {19} (\bibinfo {year} {2022})}\BibitemShut {NoStop}%
\bibitem [{\citenamefont {Qian}\ \emph {et~al.}(2019)\citenamefont {Qian}, \citenamefont {Zong}, \citenamefont {Schneider}, \citenamefont {Zhou}, \citenamefont {Zhao}, \citenamefont {Li}, \citenamefont {Yang}, \citenamefont {Bahnemann},\ and\ \citenamefont {Pan}}]{Qian2018}%
  \BibitemOpen
  \bibfield  {author} {\bibinfo {author} {\bibfnamefont {R.}~\bibnamefont {Qian}}, \bibinfo {author} {\bibfnamefont {H.}~\bibnamefont {Zong}}, \bibinfo {author} {\bibfnamefont {J.}~\bibnamefont {Schneider}}, \bibinfo {author} {\bibfnamefont {G.}~\bibnamefont {Zhou}}, \bibinfo {author} {\bibfnamefont {T.}~\bibnamefont {Zhao}}, \bibinfo {author} {\bibfnamefont {Y.}~\bibnamefont {Li}}, \bibinfo {author} {\bibfnamefont {J.}~\bibnamefont {Yang}}, \bibinfo {author} {\bibfnamefont {D.~W.}\ \bibnamefont {Bahnemann}}, \ and\ \bibinfo {author} {\bibfnamefont {J.~H.}\ \bibnamefont {Pan}},\ }\textsl {\enquote {\bibinfo {title} {{Charge carrier trapping, recombination and transfer during TiO2 photocatalysis: An overview}},}\ }\href {\doibase10.1016/j.cattod.2018.10.053} {\bibfield  {journal} {\bibinfo  {journal} {Catalysis Today}\ }\bibinfo {volume} {335},\ \bibinfo {pages} {78} (\bibinfo {year} {2019})}\BibitemShut {NoStop}%
\bibitem [{\citenamefont {Xu}\ \emph {et~al.}(2024)\citenamefont {Xu}, \citenamefont {Wang}, \citenamefont {Chen},\ and\ \citenamefont {Weng}}]{xu2024photoexcIR}%
  \BibitemOpen
  \bibfield  {author} {\bibinfo {author} {\bibfnamefont {Y.}~\bibnamefont {Xu}}, \bibinfo {author} {\bibfnamefont {Z.}~\bibnamefont {Wang}}, \bibinfo {author} {\bibfnamefont {H.}~\bibnamefont {Chen}}, \ and\ \bibinfo {author} {\bibfnamefont {Y.}~\bibnamefont {Weng}},\ }\textsl {\enquote {\bibinfo {title} {{Photoexcited Electron and Hole Polaron Formation in CdS Single Crystals Revealed by Femtosecond Time-Resolved IR Spectroscopy}},}\ }\href {\doibase10.1021/acs.jpcc.3c06401} {\bibfield  {journal} {\bibinfo  {journal} {The Journal of Physical Chemistry C}\ }\bibinfo {volume} {128},\ \bibinfo {pages} {2096} (\bibinfo {year} {2024})}\BibitemShut {NoStop}%
\bibitem [{\citenamefont {Franchini}\ \emph {et~al.}(2021)\citenamefont {Franchini}, \citenamefont {Reticcioli}, \citenamefont {Setvin},\ and\ \citenamefont {Diebold}}]{Franchini2021}%
  \BibitemOpen
  \bibfield  {author} {\bibinfo {author} {\bibfnamefont {C.}~\bibnamefont {Franchini}}, \bibinfo {author} {\bibfnamefont {M.}~\bibnamefont {Reticcioli}}, \bibinfo {author} {\bibfnamefont {M.}~\bibnamefont {Setvin}}, \ and\ \bibinfo {author} {\bibfnamefont {U.}~\bibnamefont {Diebold}},\ }\textsl {\enquote {\bibinfo {title} {{Polarons in materials}},}\ }\href {\doibase10.1038/s41578-021-00289-w} {\bibfield  {journal} {\bibinfo  {journal} {Nature Reviews Materials}\ }\bibinfo {volume} {6},\ \bibinfo {pages} {560} (\bibinfo {year} {2021})}\BibitemShut {NoStop}%
\bibitem [{\citenamefont {Rousseau}\ \emph {et~al.}(2020)\citenamefont {Rousseau}, \citenamefont {Glezakou},\ and\ \citenamefont {Selloni}}]{Rousseau2020}%
  \BibitemOpen
  \bibfield  {author} {\bibinfo {author} {\bibfnamefont {R.}~\bibnamefont {Rousseau}}, \bibinfo {author} {\bibfnamefont {V.~A.}\ \bibnamefont {Glezakou}}, \ and\ \bibinfo {author} {\bibfnamefont {A.}~\bibnamefont {Selloni}},\ }\textsl {\enquote {\bibinfo {title} {{Theoretical insights into the surface physics and chemistry of redox-active oxides}},}\ }\href {\doibase10.1038/s41578-020-0198-9} {\bibfield  {journal} {\bibinfo  {journal} {Nature Reviews Materials}\ }\bibinfo {volume} {5},\ \bibinfo {pages} {460} (\bibinfo {year} {2020})}\BibitemShut {NoStop}%
\bibitem [{\citenamefont {Ren}\ \emph {et~al.}(2023)\citenamefont {Ren}, \citenamefont {Shi}, \citenamefont {Feng}, \citenamefont {Xu},\ and\ \citenamefont {Hao}}]{Ren2023}%
  \BibitemOpen
  \bibfield  {author} {\bibinfo {author} {\bibfnamefont {Z.}~\bibnamefont {Ren}}, \bibinfo {author} {\bibfnamefont {Z.}~\bibnamefont {Shi}}, \bibinfo {author} {\bibfnamefont {H.}~\bibnamefont {Feng}}, \bibinfo {author} {\bibfnamefont {Z.}~\bibnamefont {Xu}}, \ and\ \bibinfo {author} {\bibfnamefont {W.}~\bibnamefont {Hao}},\ }\textsl {\enquote {\bibinfo {title} {{Recent Progresses of Polarons: Fundamentals and Roles in Photocatalysis and Photoelectrocatalysis}},}\ }\href {\doibase10.1002/advs.202305139} {\bibfield  {journal} {\bibinfo  {journal} {Advanced Science}\ }\bibinfo {volume} {2305139},\ \bibinfo {pages} {1} (\bibinfo {year} {2023})}\BibitemShut {NoStop}%
\bibitem [{\citenamefont {Kick}\ \emph {et~al.}(2020)\citenamefont {Kick}, \citenamefont {Grosu}, \citenamefont {Schuderer}, \citenamefont {Scheurer},\ and\ \citenamefont {Oberhofer}}]{Kick2020a}%
  \BibitemOpen
  \bibfield  {author} {\bibinfo {author} {\bibfnamefont {M.}~\bibnamefont {Kick}}, \bibinfo {author} {\bibfnamefont {C.}~\bibnamefont {Grosu}}, \bibinfo {author} {\bibfnamefont {M.}~\bibnamefont {Schuderer}}, \bibinfo {author} {\bibfnamefont {C.}~\bibnamefont {Scheurer}}, \ and\ \bibinfo {author} {\bibfnamefont {H.}~\bibnamefont {Oberhofer}},\ }\textsl {\enquote {\bibinfo {title} {{Mobile Small Polarons Qualitatively Explain Conductivity in Lithium Titanium Oxide Battery Electrodes}},}\ }\href {\doibase10.1021/acs.jpclett.0c00568} {\bibfield  {journal} {\bibinfo  {journal} {Journal of Physical Chemistry Letters}\ }\bibinfo {volume} {11},\ \bibinfo {pages} {2535} (\bibinfo {year} {2020})}\BibitemShut {NoStop}%
\bibitem [{\citenamefont {Cheng}\ \emph {et~al.}(2023)\citenamefont {Cheng}, \citenamefont {Zhou},\ and\ \citenamefont {Long}}]{Cheng2023}%
  \BibitemOpen
  \bibfield  {author} {\bibinfo {author} {\bibfnamefont {C.}~\bibnamefont {Cheng}}, \bibinfo {author} {\bibfnamefont {Z.}~\bibnamefont {Zhou}}, \ and\ \bibinfo {author} {\bibfnamefont {R.}~\bibnamefont {Long}},\ }\textsl {\enquote {\bibinfo {title} {{Time-Domain View of Polaron Dynamics in Metal Oxide Photocatalysts}},}\ }\href {\doibase10.1021/acs.jpclett.3c02869} {\bibfield  {journal} {\bibinfo  {journal} {The Journal of Physical Chemistry Letters}\ }\bibinfo {volume} {14},\ \bibinfo {pages} {10988} (\bibinfo {year} {2023})}\BibitemShut {NoStop}%
\bibitem [{\citenamefont {Wiktor}\ \emph {et~al.}(2018)\citenamefont {Wiktor}, \citenamefont {Ambrosio},\ and\ \citenamefont {Pasquarello}}]{Wiktor2018}%
  \BibitemOpen
  \bibfield  {author} {\bibinfo {author} {\bibfnamefont {J.}~\bibnamefont {Wiktor}}, \bibinfo {author} {\bibfnamefont {F.}~\bibnamefont {Ambrosio}}, \ and\ \bibinfo {author} {\bibfnamefont {A.}~\bibnamefont {Pasquarello}},\ }\textsl {\enquote {\bibinfo {title} {{Role of Polarons in Water Splitting: The Case of BiVO 4}},}\ }\href {\doibase10.1021/acsenergylett.8b00938} {\bibfield  {journal} {\bibinfo  {journal} {ACS Energy Letters}\ }\bibinfo {volume} {3},\ \bibinfo {pages} {1693} (\bibinfo {year} {2018})}\BibitemShut {NoStop}%
\bibitem [{\citenamefont {Yim}\ \emph {et~al.}(2016)\citenamefont {Yim}, \citenamefont {Watkins}, \citenamefont {Wolf}, \citenamefont {Pang}, \citenamefont {Hermansson},\ and\ \citenamefont {Thornton}}]{Yim2016a}%
  \BibitemOpen
  \bibfield  {author} {\bibinfo {author} {\bibfnamefont {C.~M.}\ \bibnamefont {Yim}}, \bibinfo {author} {\bibfnamefont {M.~B.}\ \bibnamefont {Watkins}}, \bibinfo {author} {\bibfnamefont {M.~J.}\ \bibnamefont {Wolf}}, \bibinfo {author} {\bibfnamefont {C.~L.}\ \bibnamefont {Pang}}, \bibinfo {author} {\bibfnamefont {K.}~\bibnamefont {Hermansson}}, \ and\ \bibinfo {author} {\bibfnamefont {G.}~\bibnamefont {Thornton}},\ }\textsl {\enquote {\bibinfo {title} {{Engineering Polarons at a Metal Oxide Surface}},}\ }\href {\doibase10.1103/PhysRevLett.117.116402} {\bibfield  {journal} {\bibinfo  {journal} {Physical Review Letters}\ }\bibinfo {volume} {117},\ \bibinfo {pages} {116402} (\bibinfo {year} {2016})}\BibitemShut {NoStop}%
\bibitem [{\citenamefont {Birschitzky}\ \emph {et~al.}(2024)\citenamefont {Birschitzky}, \citenamefont {Sokolovi{\'{c}}}, \citenamefont {Prezzi}, \citenamefont {Palot{\'{a}}s}, \citenamefont {Setv{\'{i}}n}, \citenamefont {Diebold}, \citenamefont {Reticcioli},\ and\ \citenamefont {Franchini}}]{Birschitzky2024}%
  \BibitemOpen
  \bibfield  {author} {\bibinfo {author} {\bibfnamefont {V.~C.}\ \bibnamefont {Birschitzky}}, \bibinfo {author} {\bibfnamefont {I.}~\bibnamefont {Sokolovi{\'{c}}}}, \bibinfo {author} {\bibfnamefont {M.}~\bibnamefont {Prezzi}}, \bibinfo {author} {\bibfnamefont {K.}~\bibnamefont {Palot{\'{a}}s}}, \bibinfo {author} {\bibfnamefont {M.}~\bibnamefont {Setv{\'{i}}n}}, \bibinfo {author} {\bibfnamefont {U.}~\bibnamefont {Diebold}}, \bibinfo {author} {\bibfnamefont {M.}~\bibnamefont {Reticcioli}}, \ and\ \bibinfo {author} {\bibfnamefont {C.}~\bibnamefont {Franchini}},\ }\textsl {\enquote {\bibinfo {title} {{Machine learning-based prediction of polaron-vacancy patterns on the TiO2(110) surface}},}\ }\href {\doibase10.1038/s41524-024-01289-4} {\bibfield  {journal} {\bibinfo  {journal} {npj Computational Materials}\ }\bibinfo {volume} {10},\ \bibinfo {pages} {89} (\bibinfo {year} {2024})}\BibitemShut {NoStop}%
\bibitem [{\citenamefont {Yoon}\ \emph {et~al.}(2015)\citenamefont {Yoon}, \citenamefont {Du}, \citenamefont {Garcia}, \citenamefont {Zhu}, \citenamefont {Wang}, \citenamefont {Petrik}, \citenamefont {Kimmel}, \citenamefont {Dohnalek}, \citenamefont {Henderson}, \citenamefont {Rousseau}, \citenamefont {Deskins},\ and\ \citenamefont {Lyubinetsky}}]{Yoon2015}%
  \BibitemOpen
  \bibfield  {author} {\bibinfo {author} {\bibfnamefont {Y.}~\bibnamefont {Yoon}}, \bibinfo {author} {\bibfnamefont {Y.}~\bibnamefont {Du}}, \bibinfo {author} {\bibfnamefont {J.~C.}\ \bibnamefont {Garcia}}, \bibinfo {author} {\bibfnamefont {Z.}~\bibnamefont {Zhu}}, \bibinfo {author} {\bibfnamefont {Z.~T.}\ \bibnamefont {Wang}}, \bibinfo {author} {\bibfnamefont {N.~G.}\ \bibnamefont {Petrik}}, \bibinfo {author} {\bibfnamefont {G.~A.}\ \bibnamefont {Kimmel}}, \bibinfo {author} {\bibfnamefont {Z.}~\bibnamefont {Dohnalek}}, \bibinfo {author} {\bibfnamefont {M.~A.}\ \bibnamefont {Henderson}}, \bibinfo {author} {\bibfnamefont {R.}~\bibnamefont {Rousseau}}, \bibinfo {author} {\bibfnamefont {N.~A.}\ \bibnamefont {Deskins}}, \ and\ \bibinfo {author} {\bibfnamefont {I.}~\bibnamefont {Lyubinetsky}},\ }\textsl {\enquote {\bibinfo {title} {{Anticorrelation between surface and subsurface point defects and the impact on the redox chemistry of TiO2(110)}},}\ }\href {\doibase10.1002/cphc.201402599} {\bibfield  {journal}
  {\bibinfo  {journal} {ChemPhysChem}\ }\bibinfo {volume} {16},\ \bibinfo {pages} {313} (\bibinfo {year} {2015})}\BibitemShut {NoStop}%
\bibitem [{\citenamefont {Reticcioli}\ \emph {et~al.}(2019)\citenamefont {Reticcioli}, \citenamefont {Sokolovi{\'{c}}}, \citenamefont {Schmid}, \citenamefont {Diebold}, \citenamefont {Setvin},\ and\ \citenamefont {Franchini}}]{Reticcioli2019c}%
  \BibitemOpen
  \bibfield  {author} {\bibinfo {author} {\bibfnamefont {M.}~\bibnamefont {Reticcioli}}, \bibinfo {author} {\bibfnamefont {I.}~\bibnamefont {Sokolovi{\'{c}}}}, \bibinfo {author} {\bibfnamefont {M.}~\bibnamefont {Schmid}}, \bibinfo {author} {\bibfnamefont {U.}~\bibnamefont {Diebold}}, \bibinfo {author} {\bibfnamefont {M.}~\bibnamefont {Setvin}}, \ and\ \bibinfo {author} {\bibfnamefont {C.}~\bibnamefont {Franchini}},\ }\textsl {\enquote {\bibinfo {title} {{Interplay between Adsorbates and Polarons: CO on Rutile TiO2(110)}},}\ }\href {\doibase10.1103/PhysRevLett.122.016805} {\bibfield  {journal} {\bibinfo  {journal} {Physical Review Letters}\ }\bibinfo {volume} {122},\ \bibinfo {pages} {016805} (\bibinfo {year} {2019})}\BibitemShut {NoStop}%
\bibitem [{\citenamefont {Cao}\ \emph {et~al.}(2017)\citenamefont {Cao}, \citenamefont {Yu}, \citenamefont {Qi}, \citenamefont {Huang}, \citenamefont {Wang}, \citenamefont {Xu}, \citenamefont {Hu},\ and\ \citenamefont {Yan}}]{Cao2017a}%
  \BibitemOpen
  \bibfield  {author} {\bibinfo {author} {\bibfnamefont {Y.}~\bibnamefont {Cao}}, \bibinfo {author} {\bibfnamefont {M.}~\bibnamefont {Yu}}, \bibinfo {author} {\bibfnamefont {S.}~\bibnamefont {Qi}}, \bibinfo {author} {\bibfnamefont {S.}~\bibnamefont {Huang}}, \bibinfo {author} {\bibfnamefont {T.}~\bibnamefont {Wang}}, \bibinfo {author} {\bibfnamefont {M.}~\bibnamefont {Xu}}, \bibinfo {author} {\bibfnamefont {S.}~\bibnamefont {Hu}}, \ and\ \bibinfo {author} {\bibfnamefont {S.}~\bibnamefont {Yan}},\ }\textsl {\enquote {\bibinfo {title} {{Scenarios of polaron-involved molecular adsorption on reduced TiO2(110) surfaces}},}\ }\href {\doibase10.1038/s41598-017-06557-6} {\bibfield  {journal} {\bibinfo  {journal} {Scientific Reports}\ }\bibinfo {volume} {7},\ \bibinfo {pages} {6148} (\bibinfo {year} {2017})}\BibitemShut {NoStop}%
\bibitem [{\citenamefont {Cheng}\ \emph {et~al.}(2022)\citenamefont {Cheng}, \citenamefont {Zhu}, \citenamefont {Fang}, \citenamefont {Long},\ and\ \citenamefont {Prezhdo}}]{Cheng2021}%
  \BibitemOpen
  \bibfield  {author} {\bibinfo {author} {\bibfnamefont {C.}~\bibnamefont {Cheng}}, \bibinfo {author} {\bibfnamefont {Y.}~\bibnamefont {Zhu}}, \bibinfo {author} {\bibfnamefont {W.~H.}\ \bibnamefont {Fang}}, \bibinfo {author} {\bibfnamefont {R.}~\bibnamefont {Long}}, \ and\ \bibinfo {author} {\bibfnamefont {O.~V.}\ \bibnamefont {Prezhdo}},\ }\textsl {\enquote {\bibinfo {title} {{CO Adsorbate Promotes Polaron Photoactivity on the Reduced Rutile TiO2(110) Surface}},}\ }\href {\doibase10.1021/jacsau.1c00508} {\bibfield  {journal} {\bibinfo  {journal} {JACS Au}\ }\bibinfo {volume} {2},\ \bibinfo {pages} {234} (\bibinfo {year} {2022})}\BibitemShut {NoStop}%
\bibitem [{\citenamefont {Yim}\ \emph {et~al.}(2018)\citenamefont {Yim}, \citenamefont {Chen}, \citenamefont {Zhang}, \citenamefont {Shaw}, \citenamefont {Pang}, \citenamefont {Grinter}, \citenamefont {Bluhm}, \citenamefont {Salmeron}, \citenamefont {Muryn}, \citenamefont {Michaelides},\ and\ \citenamefont {Thornton}}]{Yim2018}%
  \BibitemOpen
  \bibfield  {author} {\bibinfo {author} {\bibfnamefont {C.~M.}\ \bibnamefont {Yim}}, \bibinfo {author} {\bibfnamefont {J.}~\bibnamefont {Chen}}, \bibinfo {author} {\bibfnamefont {Y.}~\bibnamefont {Zhang}}, \bibinfo {author} {\bibfnamefont {B.~J.}\ \bibnamefont {Shaw}}, \bibinfo {author} {\bibfnamefont {C.~L.}\ \bibnamefont {Pang}}, \bibinfo {author} {\bibfnamefont {D.~C.}\ \bibnamefont {Grinter}}, \bibinfo {author} {\bibfnamefont {H.}~\bibnamefont {Bluhm}}, \bibinfo {author} {\bibfnamefont {M.}~\bibnamefont {Salmeron}}, \bibinfo {author} {\bibfnamefont {C.~A.}\ \bibnamefont {Muryn}}, \bibinfo {author} {\bibfnamefont {A.}~\bibnamefont {Michaelides}}, \ and\ \bibinfo {author} {\bibfnamefont {G.}~\bibnamefont {Thornton}},\ }\textsl {\enquote {\bibinfo {title} {{Visualization of Water-Induced Surface Segregation of Polarons on Rutile TiO2(110)}},}\ }\href {\doibase10.1021/acs.jpclett.8b01904} {\bibfield  {journal} {\bibinfo  {journal} {Journal of Physical Chemistry Letters}\ }\bibinfo {volume} {9},\ \bibinfo
  {pages} {4865} (\bibinfo {year} {2018})}\BibitemShut {NoStop}%
\bibitem [{\citenamefont {De~Lile}\ \emph {et~al.}(2022)\citenamefont {De~Lile}, \citenamefont {Bahadoran}, \citenamefont {Zhou},\ and\ \citenamefont {Zhang}}]{DeLile2021}%
  \BibitemOpen
  \bibfield  {author} {\bibinfo {author} {\bibfnamefont {J.~R.}\ \bibnamefont {De~Lile}}, \bibinfo {author} {\bibfnamefont {A.}~\bibnamefont {Bahadoran}}, \bibinfo {author} {\bibfnamefont {S.}~\bibnamefont {Zhou}}, \ and\ \bibinfo {author} {\bibfnamefont {J.}~\bibnamefont {Zhang}},\ }\textsl {\enquote {\bibinfo {title} {{Polaron in TiO2 from First-Principles: A Review}},}\ }\href {\doibase10.1002/adts.202100244} {\bibfield  {journal} {\bibinfo  {journal} {Advanced Theory and Simulations}\ }\bibinfo {volume} {5},\ \bibinfo {pages} {2100244} (\bibinfo {year} {2022})}\BibitemShut {NoStop}%
\bibitem [{\citenamefont {Sio}\ \emph {et~al.}(2019)\citenamefont {Sio}, \citenamefont {Verdi}, \citenamefont {Ponc{\'{e}}},\ and\ \citenamefont {Giustino}}]{Sio2019a}%
  \BibitemOpen
  \bibfield  {author} {\bibinfo {author} {\bibfnamefont {W.~H.}\ \bibnamefont {Sio}}, \bibinfo {author} {\bibfnamefont {C.}~\bibnamefont {Verdi}}, \bibinfo {author} {\bibfnamefont {S.}~\bibnamefont {Ponc{\'{e}}}}, \ and\ \bibinfo {author} {\bibfnamefont {F.}~\bibnamefont {Giustino}},\ }\textsl {\enquote {\bibinfo {title} {{Polarons from First Principles, without Supercells}},}\ }\href {\doibase10.1103/PhysRevLett.122.246403} {\bibfield  {journal} {\bibinfo  {journal} {Physical Review Letters}\ }\bibinfo {volume} {122},\ \bibinfo {pages} {246403} (\bibinfo {year} {2019})}\BibitemShut {NoStop}%
\bibitem [{\citenamefont {Pacchioni}(2015)}]{Pacchioni2015chapter}%
  \BibitemOpen
  \bibfield  {author} {\textsl {\bibinfo {author} {\bibfnamefont {G.}~\bibnamefont {Pacchioni}}},\ }in\ \href {\doibase10.1007/978-3-319-14367-5{\_}1} {\emph {\bibinfo {booktitle} {Defects at Oxide Surfaces}}},\ Vol.~\bibinfo {volume} {58},\ \bibinfo {editor} {edited by\ \bibinfo {editor} {\bibfnamefont {J.}~\bibnamefont {Jupille}}\ and\ \bibinfo {editor} {\bibfnamefont {G.}~\bibnamefont {Thornton}}}\ (\bibinfo  {publisher} {Springer},\ \bibinfo {year} {2015})\ pp.\ \bibinfo {pages} {1--28}\BibitemShut {NoStop}%
\bibitem [{\citenamefont {Ambrosio}\ and\ \citenamefont {Wiktor}(2025)}]{Ambrosio2025}%
  \BibitemOpen
  \bibfield  {author} {\bibinfo {author} {\bibfnamefont {F.}~\bibnamefont {Ambrosio}}\ and\ \bibinfo {author} {\bibfnamefont {J.}~\bibnamefont {Wiktor}},\ }\textsl {\enquote {\bibinfo {title} {{Charge localization in optoelectronic and photocatalytic applications: Computational perspective}},}\ }\href {\doibase10.1063/5.0257250} {\bibfield  {journal} {\bibinfo  {journal} {Applied Physics Letters}\ }\bibinfo {volume} {126} (\bibinfo {year} {2025}),\ 10.1063/5.0257250}\BibitemShut {NoStop}%
\bibitem [{\citenamefont {Gionco}\ \emph {et~al.}(2015)\citenamefont {Gionco}, \citenamefont {Livraghi}, \citenamefont {Maurelli}, \citenamefont {Giamello}, \citenamefont {Tosoni}, \citenamefont {Di~Valentin},\ and\ \citenamefont {Pacchioni}}]{Gionco2015}%
  \BibitemOpen
  \bibfield  {author} {\bibinfo {author} {\bibfnamefont {C.}~\bibnamefont {Gionco}}, \bibinfo {author} {\bibfnamefont {S.}~\bibnamefont {Livraghi}}, \bibinfo {author} {\bibfnamefont {S.}~\bibnamefont {Maurelli}}, \bibinfo {author} {\bibfnamefont {E.}~\bibnamefont {Giamello}}, \bibinfo {author} {\bibfnamefont {S.}~\bibnamefont {Tosoni}}, \bibinfo {author} {\bibfnamefont {C.}~\bibnamefont {Di~Valentin}}, \ and\ \bibinfo {author} {\bibfnamefont {G.}~\bibnamefont {Pacchioni}},\ }\textsl {\enquote {\bibinfo {title} {{Al- and Ga-Doped TiO2, ZrO2, and HfO2: The Nature of O 2p Trapped Holes from a Combined Electron Paramagnetic Resonance (EPR) and Density Functional Theory (DFT) Study}},}\ }\href {\doibase10.1021/acs.chemmater.5b00800} {\bibfield  {journal} {\bibinfo  {journal} {Chemistry of Materials}\ }\bibinfo {volume} {27},\ \bibinfo {pages} {3936} (\bibinfo {year} {2015})}\BibitemShut {NoStop}%
\bibitem [{\citenamefont {Reticcioli}\ \emph {et~al.}(2022)\citenamefont {Reticcioli}, \citenamefont {Diebold},\ and\ \citenamefont {Franchini}}]{Reticcioli2022}%
  \BibitemOpen
  \bibfield  {author} {\bibinfo {author} {\bibfnamefont {M.}~\bibnamefont {Reticcioli}}, \bibinfo {author} {\bibfnamefont {U.}~\bibnamefont {Diebold}}, \ and\ \bibinfo {author} {\bibfnamefont {C.}~\bibnamefont {Franchini}},\ }\textsl {\enquote {\bibinfo {title} {{Modeling polarons in density functional theory: Lessons learned from TiO2}},}\ }\href {\doibase10.1088/1361-648X/ac58d7} {\bibfield  {journal} {\bibinfo  {journal} {Journal of Physics Condensed Matter}\ }\bibinfo {volume} {34},\ \bibinfo {pages} {204006} (\bibinfo {year} {2022})}\BibitemShut {NoStop}%
\bibitem [{\citenamefont {Deskins}\ \emph {et~al.}(2009)\citenamefont {Deskins}, \citenamefont {Rousseau},\ and\ \citenamefont {Dupuis}}]{Deskins2009}%
  \BibitemOpen
  \bibfield  {author} {\bibinfo {author} {\bibfnamefont {N.~A.}\ \bibnamefont {Deskins}}, \bibinfo {author} {\bibfnamefont {R.}~\bibnamefont {Rousseau}}, \ and\ \bibinfo {author} {\bibfnamefont {M.}~\bibnamefont {Dupuis}},\ }\textsl {\enquote {\bibinfo {title} {{Localized electronic states from surface hydroxyls and polarons in TiO 2(110)}},}\ }\href {\doibase10.1021/jp9037655} {\bibfield  {journal} {\bibinfo  {journal} {Journal of Physical Chemistry C}\ }\bibinfo {volume} {113},\ \bibinfo {pages} {14583} (\bibinfo {year} {2009})}\BibitemShut {NoStop}%
\bibitem [{\citenamefont {Yang}\ \emph {et~al.}(2014)\citenamefont {Yang}, \citenamefont {Yin}, \citenamefont {Bebensee}, \citenamefont {Buchholz}, \citenamefont {Sezen}, \citenamefont {Heissler}, \citenamefont {Chen}, \citenamefont {Nefedov}, \citenamefont {Idriss}, \citenamefont {Gong},\ and\ \citenamefont {W{\"{o}}ll}}]{Yang2014}%
  \BibitemOpen
  \bibfield  {author} {\bibinfo {author} {\bibfnamefont {C.}~\bibnamefont {Yang}}, \bibinfo {author} {\bibfnamefont {L.~L.}\ \bibnamefont {Yin}}, \bibinfo {author} {\bibfnamefont {F.}~\bibnamefont {Bebensee}}, \bibinfo {author} {\bibfnamefont {M.}~\bibnamefont {Buchholz}}, \bibinfo {author} {\bibfnamefont {H.}~\bibnamefont {Sezen}}, \bibinfo {author} {\bibfnamefont {S.}~\bibnamefont {Heissler}}, \bibinfo {author} {\bibfnamefont {J.}~\bibnamefont {Chen}}, \bibinfo {author} {\bibfnamefont {A.}~\bibnamefont {Nefedov}}, \bibinfo {author} {\bibfnamefont {H.}~\bibnamefont {Idriss}}, \bibinfo {author} {\bibfnamefont {X.~Q.}\ \bibnamefont {Gong}}, \ and\ \bibinfo {author} {\bibfnamefont {C.}~\bibnamefont {W{\"{o}}ll}},\ }\textsl {\enquote {\bibinfo {title} {{Chemical activity of oxygen vacancies on ceria: A combined experimental and theoretical study on CeO2(111)}},}\ }\href {\doibase10.1039/c4cp02372b} {\bibfield  {journal} {\bibinfo  {journal} {Physical Chemistry Chemical Physics}\ }\bibinfo {volume} {16},\
  \bibinfo {pages} {24165} (\bibinfo {year} {2014})}\BibitemShut {NoStop}%
\bibitem [{\citenamefont {Tanner}\ \emph {et~al.}(2021)\citenamefont {Tanner}, \citenamefont {Wen}, \citenamefont {Ontaneda}, \citenamefont {Zhang}, \citenamefont {Grau-Crespo}, \citenamefont {Fielding}, \citenamefont {Selloni},\ and\ \citenamefont {Thornton}}]{Tanner2021}%
  \BibitemOpen
  \bibfield  {author} {\bibinfo {author} {\bibfnamefont {A.~J.}\ \bibnamefont {Tanner}}, \bibinfo {author} {\bibfnamefont {B.}~\bibnamefont {Wen}}, \bibinfo {author} {\bibfnamefont {J.}~\bibnamefont {Ontaneda}}, \bibinfo {author} {\bibfnamefont {Y.}~\bibnamefont {Zhang}}, \bibinfo {author} {\bibfnamefont {R.}~\bibnamefont {Grau-Crespo}}, \bibinfo {author} {\bibfnamefont {H.~H.}\ \bibnamefont {Fielding}}, \bibinfo {author} {\bibfnamefont {A.}~\bibnamefont {Selloni}}, \ and\ \bibinfo {author} {\bibfnamefont {G.}~\bibnamefont {Thornton}},\ }\textsl {\enquote {\bibinfo {title} {{Polaron-Adsorbate Coupling at the TiO2(110)-Carboxylate Interface}},}\ }\href {\doibase10.1021/acs.jpclett.1c00678} {\bibfield  {journal} {\bibinfo  {journal} {Journal of Physical Chemistry Letters}\ }\bibinfo {volume} {12},\ \bibinfo {pages} {3571} (\bibinfo {year} {2021})}\BibitemShut {NoStop}%
\bibitem [{\citenamefont {Preda}\ and\ \citenamefont {Pacchioni}(2011)}]{Preda2011}%
  \BibitemOpen
  \bibfield  {author} {\bibinfo {author} {\bibfnamefont {G.}~\bibnamefont {Preda}}\ and\ \bibinfo {author} {\bibfnamefont {G.}~\bibnamefont {Pacchioni}},\ }\textsl {\enquote {\bibinfo {title} {{Formation of oxygen active species in Ag-modified CeO2 catalyst for soot oxidation: A DFT study}},}\ }\href {\doibase10.1016/j.cattod.2011.04.036} {\bibfield  {journal} {\bibinfo  {journal} {Catalysis Today}\ }\bibinfo {volume} {177},\ \bibinfo {pages} {31} (\bibinfo {year} {2011})}\BibitemShut {NoStop}%
\bibitem [{\citenamefont {Sombut}\ \emph {et~al.}(2022)\citenamefont {Sombut}, \citenamefont {Puntscher}, \citenamefont {Atzmueller}, \citenamefont {Jakub}, \citenamefont {Reticcioli}, \citenamefont {Meier}, \citenamefont {Parkinson},\ and\ \citenamefont {Franchini}}]{Sombut2022}%
  \BibitemOpen
  \bibfield  {author} {\bibinfo {author} {\bibfnamefont {P.}~\bibnamefont {Sombut}}, \bibinfo {author} {\bibfnamefont {L.}~\bibnamefont {Puntscher}}, \bibinfo {author} {\bibfnamefont {M.}~\bibnamefont {Atzmueller}}, \bibinfo {author} {\bibfnamefont {Z.}~\bibnamefont {Jakub}}, \bibinfo {author} {\bibfnamefont {M.}~\bibnamefont {Reticcioli}}, \bibinfo {author} {\bibfnamefont {M.}~\bibnamefont {Meier}}, \bibinfo {author} {\bibfnamefont {G.~S.}\ \bibnamefont {Parkinson}}, \ and\ \bibinfo {author} {\bibfnamefont {C.}~\bibnamefont {Franchini}},\ }\textsl {\enquote {\bibinfo {title} {{Role of Polarons in Single-Atom Catalysts: Case Study of Me1 [Au1, Pt1, and Rh1] on TiO2(110)}},}\ }\href {\doibase10.1007/s11244-022-01651-0} {\bibfield  {journal} {\bibinfo  {journal} {Topics in Catalysis}\ }\bibinfo {volume} {65},\ \bibinfo {pages} {1620} (\bibinfo {year} {2022})}\BibitemShut {NoStop}%
\bibitem [{\citenamefont {L{\'{o}}pez-Caballero}\ \emph {et~al.}(2020{\natexlab{a}})\citenamefont {L{\'{o}}pez-Caballero}, \citenamefont {Ramallo-L{\'{o}}pez}, \citenamefont {Giovanetti}, \citenamefont {Buceta}, \citenamefont {Miret-Art{\'{e}}s}, \citenamefont {L{\'{o}}pez-Quintela}, \citenamefont {Requejo},\ and\ \citenamefont {De~Lara-Castells}}]{Lopez-Caballero2020}%
  \BibitemOpen
  \bibfield  {author} {\bibinfo {author} {\bibfnamefont {P.}~\bibnamefont {L{\'{o}}pez-Caballero}}, \bibinfo {author} {\bibfnamefont {J.~M.}\ \bibnamefont {Ramallo-L{\'{o}}pez}}, \bibinfo {author} {\bibfnamefont {L.~J.}\ \bibnamefont {Giovanetti}}, \bibinfo {author} {\bibfnamefont {D.}~\bibnamefont {Buceta}}, \bibinfo {author} {\bibfnamefont {S.}~\bibnamefont {Miret-Art{\'{e}}s}}, \bibinfo {author} {\bibfnamefont {M.~A.}\ \bibnamefont {L{\'{o}}pez-Quintela}}, \bibinfo {author} {\bibfnamefont {F.~G.}\ \bibnamefont {Requejo}}, \ and\ \bibinfo {author} {\bibfnamefont {M.~P.}\ \bibnamefont {De~Lara-Castells}},\ }\textsl {\enquote {\bibinfo {title} {{Exploring the properties of Ag5-TiO2 interfaces: Stable surface polaron formation, UV-Vis optical response, and CO2 photoactivation}},}\ }\href {\doibase10.1039/d0ta00062k} {\bibfield  {journal} {\bibinfo  {journal} {Journal of Materials Chemistry A}\ }\bibinfo {volume} {8},\ \bibinfo {pages} {6842} (\bibinfo {year} {2020}{\natexlab{a}})}\BibitemShut {NoStop}%
\bibitem [{\citenamefont {L{\'{o}}pez-Caballero}\ \emph {et~al.}(2020{\natexlab{b}})\citenamefont {L{\'{o}}pez-Caballero}, \citenamefont {Miret-Art{\'{e}}s}, \citenamefont {Mitrushchenkov},\ and\ \citenamefont {De~Lara-Castells}}]{Lopez-Caballero2020a}%
  \BibitemOpen
  \bibfield  {author} {\bibinfo {author} {\bibfnamefont {P.}~\bibnamefont {L{\'{o}}pez-Caballero}}, \bibinfo {author} {\bibfnamefont {S.}~\bibnamefont {Miret-Art{\'{e}}s}}, \bibinfo {author} {\bibfnamefont {A.~O.}\ \bibnamefont {Mitrushchenkov}}, \ and\ \bibinfo {author} {\bibfnamefont {M.~P.}\ \bibnamefont {De~Lara-Castells}},\ }\textsl {\enquote {\bibinfo {title} {{Ag5-induced stabilization of multiple surface polarons on perfect and reduced TiO2rutile (110)}},}\ }\href {\doibase10.1063/5.0029099} {\bibfield  {journal} {\bibinfo  {journal} {Journal of Chemical Physics}\ }\bibinfo {volume} {153},\ \bibinfo {pages} {164702} (\bibinfo {year} {2020}{\natexlab{b}})}\BibitemShut {NoStop}%
\bibitem [{\citenamefont {Reticcioli}\ \emph {et~al.}(2018)\citenamefont {Reticcioli}, \citenamefont {Setvin}, \citenamefont {Schmid}, \citenamefont {Diebold},\ and\ \citenamefont {Franchini}}]{Reticcioli2018a}%
  \BibitemOpen
  \bibfield  {author} {\bibinfo {author} {\bibfnamefont {M.}~\bibnamefont {Reticcioli}}, \bibinfo {author} {\bibfnamefont {M.}~\bibnamefont {Setvin}}, \bibinfo {author} {\bibfnamefont {M.}~\bibnamefont {Schmid}}, \bibinfo {author} {\bibfnamefont {U.}~\bibnamefont {Diebold}}, \ and\ \bibinfo {author} {\bibfnamefont {C.}~\bibnamefont {Franchini}},\ }\textsl {\enquote {\bibinfo {title} {{Formation and dynamics of small polarons on the rutile TiO2(110) surface}},}\ }\href {\doibase10.1103/PhysRevB.98.045306} {\bibfield  {journal} {\bibinfo  {journal} {Physical Review B}\ }\bibinfo {volume} {98},\ \bibinfo {pages} {045306} (\bibinfo {year} {2018})}\BibitemShut {NoStop}%
\bibitem [{\citenamefont {Pham}\ and\ \citenamefont {Deskins}(2020)}]{Pham2020}%
  \BibitemOpen
  \bibfield  {author} {\bibinfo {author} {\bibfnamefont {T.~D.}\ \bibnamefont {Pham}}\ and\ \bibinfo {author} {\bibfnamefont {N.~A.}\ \bibnamefont {Deskins}},\ }\textsl {\enquote {\bibinfo {title} {{Efficient Method for Modeling Polarons Using Electronic Structure Methods}},}\ }\href {\doibase10.1021/acs.jctc.0c00374} {\bibfield  {journal} {\bibinfo  {journal} {Journal of Chemical Theory and Computation}\ }\bibinfo {volume} {16},\ \bibinfo {pages} {5264} (\bibinfo {year} {2020})}\BibitemShut {NoStop}%
\bibitem [{\citenamefont {Allen}\ and\ \citenamefont {Watson}(2014)}]{Allen2014}%
  \BibitemOpen
  \bibfield  {author} {\bibinfo {author} {\bibfnamefont {J.~P.}\ \bibnamefont {Allen}}\ and\ \bibinfo {author} {\bibfnamefont {G.~W.}\ \bibnamefont {Watson}},\ }\textsl {\enquote {\bibinfo {title} {{Occupation matrix control of d- and f-electron localisations using DFT + U}},}\ }\href {\doibase10.1039/c4cp01083c} {\bibfield  {journal} {\bibinfo  {journal} {Physical Chemistry Chemical Physics}\ }\bibinfo {volume} {16},\ \bibinfo {pages} {21016} (\bibinfo {year} {2014})}\BibitemShut {NoStop}%
\bibitem [{\citenamefont {Reticcioli}\ \emph {et~al.}(2020)\citenamefont {Reticcioli}, \citenamefont {Diebold}, \citenamefont {Kresse},\ and\ \citenamefont {Franchini}}]{Reticcioli2019b}%
  \BibitemOpen
  \bibfield  {author} {\textsl {\bibinfo {author} {\bibfnamefont {M.}~\bibnamefont {Reticcioli}}, \bibinfo {author} {\bibfnamefont {U.}~\bibnamefont {Diebold}}, \bibinfo {author} {\bibfnamefont {G.}~\bibnamefont {Kresse}}, \ and\ \bibinfo {author} {\bibfnamefont {C.}~\bibnamefont {Franchini}}},\ }in\ \href {\doibase10.1007/978-3-319-50257-1} {\emph {\bibinfo {booktitle} {Handbook of Materials Modeling}}},\ \bibinfo {editor} {edited by\ \bibinfo {editor} {\bibfnamefont {W.}~\bibnamefont {Andreoni}}\ and\ \bibinfo {editor} {\bibfnamefont {S.}~\bibnamefont {Yip}}}\ (\bibinfo  {publisher} {Springer International Publishing},\ \bibinfo {address} {Cham},\ \bibinfo {year} {2020})\ pp.\ \bibinfo {pages} {1--39}\BibitemShut {NoStop}%
\bibitem [{\citenamefont {Falletta}\ and\ \citenamefont {Pasquarello}(2022)}]{Falletta2022a}%
  \BibitemOpen
  \bibfield  {author} {\bibinfo {author} {\bibfnamefont {S.}~\bibnamefont {Falletta}}\ and\ \bibinfo {author} {\bibfnamefont {A.}~\bibnamefont {Pasquarello}},\ }\textsl {\enquote {\bibinfo {title} {{Polarons free from many-body self-interaction in density functional theory}},}\ }\href {\doibase10.1103/PhysRevB.106.125119} {\bibfield  {journal} {\bibinfo  {journal} {Physical Review B}\ }\bibinfo {volume} {106},\ \bibinfo {pages} {125119} (\bibinfo {year} {2022})}\BibitemShut {NoStop}%
\bibitem [{\citenamefont {Chen}\ \emph {et~al.}(2022)\citenamefont {Chen}, \citenamefont {Grieder}, \citenamefont {Smart}, \citenamefont {Mayford}, \citenamefont {McNair}, \citenamefont {Pinongcos}, \citenamefont {Eisenberg}, \citenamefont {Bridges}, \citenamefont {Li},\ and\ \citenamefont {Ping}}]{Chen2023}%
  \BibitemOpen
  \bibfield  {author} {\bibinfo {author} {\bibfnamefont {M.}~\bibnamefont {Chen}}, \bibinfo {author} {\bibfnamefont {A.~C.}\ \bibnamefont {Grieder}}, \bibinfo {author} {\bibfnamefont {T.~J.}\ \bibnamefont {Smart}}, \bibinfo {author} {\bibfnamefont {K.}~\bibnamefont {Mayford}}, \bibinfo {author} {\bibfnamefont {S.}~\bibnamefont {McNair}}, \bibinfo {author} {\bibfnamefont {A.}~\bibnamefont {Pinongcos}}, \bibinfo {author} {\bibfnamefont {S.}~\bibnamefont {Eisenberg}}, \bibinfo {author} {\bibfnamefont {F.}~\bibnamefont {Bridges}}, \bibinfo {author} {\bibfnamefont {Y.}~\bibnamefont {Li}}, \ and\ \bibinfo {author} {\bibfnamefont {Y.}~\bibnamefont {Ping}},\ }\textsl {\enquote {\bibinfo {title} {{The impacts of dopants on the small polaron mobility and conductivity in hematite - the role of disorder}},}\ }\href {\doibase10.1039/d2nr04807h} {\bibfield  {journal} {\bibinfo  {journal} {Nanoscale}\ }\bibinfo {volume} {15},\ \bibinfo {pages} {1619} (\bibinfo {year} {2022})}\BibitemShut {NoStop}%
\bibitem [{\citenamefont {Ma}\ \emph {et~al.}(2025)\citenamefont {Ma}, \citenamefont {Yu}, \citenamefont {Zhong}, \citenamefont {Chen}, \citenamefont {Gong},\ and\ \citenamefont {Xiang}}]{Ma2025MLdefects}%
  \BibitemOpen
  \bibfield  {author} {\bibinfo {author} {\bibfnamefont {Y.}~\bibnamefont {Ma}}, \bibinfo {author} {\bibfnamefont {H.}~\bibnamefont {Yu}}, \bibinfo {author} {\bibfnamefont {Y.}~\bibnamefont {Zhong}}, \bibinfo {author} {\bibfnamefont {S.}~\bibnamefont {Chen}}, \bibinfo {author} {\bibfnamefont {X.}~\bibnamefont {Gong}}, \ and\ \bibinfo {author} {\bibfnamefont {H.}~\bibnamefont {Xiang}},\ }\textsl {\enquote {\bibinfo {title} {{Transferable machine learning approach for predicting electronic structures of charged defects}},}\ }\href {\doibase10.1063/5.0242683} {\bibfield  {journal} {\bibinfo  {journal} {Applied Physics Letters}\ }\bibinfo {volume} {126} (\bibinfo {year} {2025}),\ 10.1063/5.0242683}\BibitemShut {NoStop}%
\bibitem [{\citenamefont {Smart}\ and\ \citenamefont {Ping}(2017)}]{Smart2017}%
  \BibitemOpen
  \bibfield  {author} {\bibinfo {author} {\bibfnamefont {T.~J.}\ \bibnamefont {Smart}}\ and\ \bibinfo {author} {\bibfnamefont {Y.}~\bibnamefont {Ping}},\ }\textsl {\enquote {\bibinfo {title} {{Effect of defects on the small polaron formation and transport properties of hematite from first-principles calculations}},}\ }\href {\doibase10.1088/1361-648X/aa7e3d} {\bibfield  {journal} {\bibinfo  {journal} {Journal of Physics Condensed Matter}\ }\bibinfo {volume} {29},\ \bibinfo {pages} {394006} (\bibinfo {year} {2017})}\BibitemShut {NoStop}%
\bibitem [{\citenamefont {Yim}\ \emph {et~al.}(2010)\citenamefont {Yim}, \citenamefont {Pang},\ and\ \citenamefont {Thornton}}]{Yim2010}%
  \BibitemOpen
  \bibfield  {author} {\bibinfo {author} {\bibfnamefont {C.~M.}\ \bibnamefont {Yim}}, \bibinfo {author} {\bibfnamefont {C.~L.}\ \bibnamefont {Pang}}, \ and\ \bibinfo {author} {\bibfnamefont {G.}~\bibnamefont {Thornton}},\ }\textsl {\enquote {\bibinfo {title} {{Oxygen vacancy origin of the surface band-gap state of TiO2(110)}},}\ }\href {\doibase10.1103/PhysRevLett.104.036806} {\bibfield  {journal} {\bibinfo  {journal} {Physical Review Letters}\ }\bibinfo {volume} {104},\ \bibinfo {pages} {036806} (\bibinfo {year} {2010})}\BibitemShut {NoStop}%
\bibitem [{\citenamefont {Zhang}\ \emph {et~al.}(2019)\citenamefont {Zhang}, \citenamefont {Han}, \citenamefont {Murgida}, \citenamefont {Ganduglia-Pirovano},\ and\ \citenamefont {Gao}}]{Zhang2019}%
  \BibitemOpen
  \bibfield  {author} {\bibinfo {author} {\bibfnamefont {D.}~\bibnamefont {Zhang}}, \bibinfo {author} {\bibfnamefont {Z.~K.}\ \bibnamefont {Han}}, \bibinfo {author} {\bibfnamefont {G.~E.}\ \bibnamefont {Murgida}}, \bibinfo {author} {\bibfnamefont {M.~V.}\ \bibnamefont {Ganduglia-Pirovano}}, \ and\ \bibinfo {author} {\bibfnamefont {Y.}~\bibnamefont {Gao}},\ }\textsl {\enquote {\bibinfo {title} {{Oxygen-Vacancy Dynamics and Entanglement with Polaron Hopping at the Reduced CeO2 (111) Surface}},}\ }\href {\doibase10.1103/PhysRevLett.122.096101} {\bibfield  {journal} {\bibinfo  {journal} {Physical Review Letters}\ }\bibinfo {volume} {122},\ \bibinfo {pages} {096101} (\bibinfo {year} {2019})}\BibitemShut {NoStop}%
\bibitem [{\citenamefont {Reticcioli}\ \emph {et~al.}(2017)\citenamefont {Reticcioli}, \citenamefont {Setvin}, \citenamefont {Hao}, \citenamefont {Flauger}, \citenamefont {Kresse}, \citenamefont {Schmid}, \citenamefont {Diebold},\ and\ \citenamefont {Franchini}}]{Reticcioli2017d}%
  \BibitemOpen
  \bibfield  {author} {\bibinfo {author} {\bibfnamefont {M.}~\bibnamefont {Reticcioli}}, \bibinfo {author} {\bibfnamefont {M.}~\bibnamefont {Setvin}}, \bibinfo {author} {\bibfnamefont {X.}~\bibnamefont {Hao}}, \bibinfo {author} {\bibfnamefont {P.}~\bibnamefont {Flauger}}, \bibinfo {author} {\bibfnamefont {G.}~\bibnamefont {Kresse}}, \bibinfo {author} {\bibfnamefont {M.}~\bibnamefont {Schmid}}, \bibinfo {author} {\bibfnamefont {U.}~\bibnamefont {Diebold}}, \ and\ \bibinfo {author} {\bibfnamefont {C.}~\bibnamefont {Franchini}},\ }\textsl {\enquote {\bibinfo {title} {{Polaron-driven surface reconstructions}},}\ }\href {\doibase10.1103/PhysRevX.7.031053} {\bibfield  {journal} {\bibinfo  {journal} {Physical Review X}\ }\bibinfo {volume} {7},\ \bibinfo {pages} {031053} (\bibinfo {year} {2017})}\BibitemShut {NoStop}%
\bibitem [{\citenamefont {Yuan}\ \emph {et~al.}(2024)\citenamefont {Yuan}, \citenamefont {Chen}, \citenamefont {Han}, \citenamefont {You}, \citenamefont {Jiang}, \citenamefont {Qi}, \citenamefont {Li}, \citenamefont {Wu}, \citenamefont {Ganduglia-Pirovano},\ and\ \citenamefont {Wang}}]{Yuan2024}%
  \BibitemOpen
  \bibfield  {author} {\bibinfo {author} {\bibfnamefont {W.}~\bibnamefont {Yuan}}, \bibinfo {author} {\bibfnamefont {B.}~\bibnamefont {Chen}}, \bibinfo {author} {\bibfnamefont {Z.-K.}\ \bibnamefont {Han}}, \bibinfo {author} {\bibfnamefont {R.}~\bibnamefont {You}}, \bibinfo {author} {\bibfnamefont {Y.}~\bibnamefont {Jiang}}, \bibinfo {author} {\bibfnamefont {R.}~\bibnamefont {Qi}}, \bibinfo {author} {\bibfnamefont {G.}~\bibnamefont {Li}}, \bibinfo {author} {\bibfnamefont {H.}~\bibnamefont {Wu}}, \bibinfo {author} {\bibfnamefont {M.~V.}\ \bibnamefont {Ganduglia-Pirovano}}, \ and\ \bibinfo {author} {\bibfnamefont {Y.}~\bibnamefont {Wang}},\ }\textsl {\enquote {\bibinfo {title} {{Direct in-situ insights into the asymmetric surface reconstruction of rutile TiO2 (110)}},}\ }\href {\doibase10.1038/s41467-024-46011-6} {\bibfield  {journal} {\bibinfo  {journal} {Nature Communications}\ }\bibinfo {volume} {15},\ \bibinfo {pages} {1616} (\bibinfo {year} {2024})}\BibitemShut {NoStop}%
\bibitem [{\citenamefont {Geiger}\ and\ \citenamefont {L{\'{o}}pez}(2022)}]{Geiger2022}%
  \BibitemOpen
  \bibfield  {author} {\bibinfo {author} {\bibfnamefont {J.}~\bibnamefont {Geiger}}\ and\ \bibinfo {author} {\bibfnamefont {N.}~\bibnamefont {L{\'{o}}pez}},\ }\textsl {\enquote {\bibinfo {title} {{Coupling Metal and Support Redox Terms in Single-Atom Catalysts}},}\ }\href {\doibase10.1021/acs.jpcc.2c03710} {\bibfield  {journal} {\bibinfo  {journal} {Journal of Physical Chemistry C}\ }\bibinfo {volume} {126},\ \bibinfo {pages} {13698} (\bibinfo {year} {2022})}\BibitemShut {NoStop}%
\bibitem [{\citenamefont {{\"{O}}sterbacka}\ \emph {et~al.}(2022)\citenamefont {{\"{O}}sterbacka}, \citenamefont {Ambrosio},\ and\ \citenamefont {Wiktor}}]{oesterbacka2022bivo4}%
  \BibitemOpen
  \bibfield  {author} {\bibinfo {author} {\bibfnamefont {N.}~\bibnamefont {{\"{O}}sterbacka}}, \bibinfo {author} {\bibfnamefont {F.}~\bibnamefont {Ambrosio}}, \ and\ \bibinfo {author} {\bibfnamefont {J.}~\bibnamefont {Wiktor}},\ }\textsl {\enquote {\bibinfo {title} {{Charge Localization in Defective BiVO4}},}\ }\href {\doibase10.1021/acs.jpcc.1c09990} {\bibfield  {journal} {\bibinfo  {journal} {Journal of Physical Chemistry C}\ }\bibinfo {volume} {126},\ \bibinfo {pages} {2960} (\bibinfo {year} {2022})}\BibitemShut {NoStop}%
\bibitem [{\citenamefont {Kowalski}\ \emph {et~al.}(2010)\citenamefont {Kowalski}, \citenamefont {Camellone}, \citenamefont {Nair}, \citenamefont {Meyer},\ and\ \citenamefont {Marx}}]{Kowalski2010}%
  \BibitemOpen
  \bibfield  {author} {\bibinfo {author} {\bibfnamefont {P.~M.}\ \bibnamefont {Kowalski}}, \bibinfo {author} {\bibfnamefont {M.~F.}\ \bibnamefont {Camellone}}, \bibinfo {author} {\bibfnamefont {N.~N.}\ \bibnamefont {Nair}}, \bibinfo {author} {\bibfnamefont {B.}~\bibnamefont {Meyer}}, \ and\ \bibinfo {author} {\bibfnamefont {D.}~\bibnamefont {Marx}},\ }\textsl {\enquote {\bibinfo {title} {{Charge localization dynamics induced by oxygen vacancies on the TiO2(110) surface}},}\ }\href {\doibase10.1103/PhysRevLett.105.146405} {\bibfield  {journal} {\bibinfo  {journal} {Physical Review Letters}\ }\bibinfo {volume} {105},\ \bibinfo {pages} {146405} (\bibinfo {year} {2010})}\BibitemShut {NoStop}%
\bibitem [{\citenamefont {Birschitzky}\ \emph {et~al.}(2022)\citenamefont {Birschitzky}, \citenamefont {Ellinger}, \citenamefont {Diebold}, \citenamefont {Reticcioli},\ and\ \citenamefont {Franchini}}]{Birschitzky2022}%
  \BibitemOpen
  \bibfield  {author} {\bibinfo {author} {\bibfnamefont {V.~C.}\ \bibnamefont {Birschitzky}}, \bibinfo {author} {\bibfnamefont {F.}~\bibnamefont {Ellinger}}, \bibinfo {author} {\bibfnamefont {U.}~\bibnamefont {Diebold}}, \bibinfo {author} {\bibfnamefont {M.}~\bibnamefont {Reticcioli}}, \ and\ \bibinfo {author} {\bibfnamefont {C.}~\bibnamefont {Franchini}},\ }\textsl {\enquote {\bibinfo {title} {{Machine learning for exploring small polaron configurational space}},}\ }\href {\doibase10.1038/s41524-022-00805-8} {\bibfield  {journal} {\bibinfo  {journal} {npj Computational Materials}\ }\bibinfo {volume} {8},\ \bibinfo {pages} {125} (\bibinfo {year} {2022})}\BibitemShut {NoStop}%
\bibitem [{\citenamefont {Li}\ \emph {et~al.}(2024)\citenamefont {Li}, \citenamefont {Xu}, \citenamefont {Zhang}, \citenamefont {Liu}, \citenamefont {Wang}, \citenamefont {Ma}, \citenamefont {Fu}, \citenamefont {Hu}, \citenamefont {Xu},\ and\ \citenamefont {Han}}]{Li2024clusterex}%
  \BibitemOpen
  \bibfield  {author} {\bibinfo {author} {\bibfnamefont {Z.}~\bibnamefont {Li}}, \bibinfo {author} {\bibfnamefont {N.}~\bibnamefont {Xu}}, \bibinfo {author} {\bibfnamefont {Y.}~\bibnamefont {Zhang}}, \bibinfo {author} {\bibfnamefont {W.}~\bibnamefont {Liu}}, \bibinfo {author} {\bibfnamefont {J.}~\bibnamefont {Wang}}, \bibinfo {author} {\bibfnamefont {M.}~\bibnamefont {Ma}}, \bibinfo {author} {\bibfnamefont {X.}~\bibnamefont {Fu}}, \bibinfo {author} {\bibfnamefont {X.}~\bibnamefont {Hu}}, \bibinfo {author} {\bibfnamefont {W.}~\bibnamefont {Xu}}, \ and\ \bibinfo {author} {\bibfnamefont {Z.-K.}\ \bibnamefont {Han}},\ }\textsl {\enquote {\bibinfo {title} {{Unveiling the Structure of Oxygen Vacancies in Bulk Ceria and the Physical Mechanisms behind Their Formation}},}\ }\href {\doibase10.1021/acs.jpclett.4c00889} {\bibfield  {journal} {\bibinfo  {journal} {The Journal of Physical Chemistry Letters}\ ,\ \bibinfo {pages} {5868}} (\bibinfo {year} {2024})}\BibitemShut {NoStop}%
\bibitem [{\citenamefont {Setvin}\ \emph {et~al.}(2014)\citenamefont {Setvin}, \citenamefont {Franchini}, \citenamefont {Hao}, \citenamefont {Schmid}, \citenamefont {Janotti}, \citenamefont {Kaltak}, \citenamefont {Van De~Walle}, \citenamefont {Kresse},\ and\ \citenamefont {Diebold}}]{Setvin2014}%
  \BibitemOpen
  \bibfield  {author} {\bibinfo {author} {\bibfnamefont {M.}~\bibnamefont {Setvin}}, \bibinfo {author} {\bibfnamefont {C.}~\bibnamefont {Franchini}}, \bibinfo {author} {\bibfnamefont {X.}~\bibnamefont {Hao}}, \bibinfo {author} {\bibfnamefont {M.}~\bibnamefont {Schmid}}, \bibinfo {author} {\bibfnamefont {A.}~\bibnamefont {Janotti}}, \bibinfo {author} {\bibfnamefont {M.}~\bibnamefont {Kaltak}}, \bibinfo {author} {\bibfnamefont {C.~G.}\ \bibnamefont {Van De~Walle}}, \bibinfo {author} {\bibfnamefont {G.}~\bibnamefont {Kresse}}, \ and\ \bibinfo {author} {\bibfnamefont {U.}~\bibnamefont {Diebold}},\ }\textsl {\enquote {\bibinfo {title} {{Direct view at excess electrons in TiO2 rutile and anatase}},}\ }\href {\doibase10.1103/PhysRevLett.113.086402} {\bibfield  {journal} {\bibinfo  {journal} {Physical Review Letters}\ }\bibinfo {volume} {113},\ \bibinfo {pages} {086402} (\bibinfo {year} {2014})}\BibitemShut {NoStop}%
\bibitem [{\citenamefont {Diebold}(2002)}]{Diebold2002}%
  \BibitemOpen
  \bibfield  {author} {\bibinfo {author} {\bibfnamefont {U.}~\bibnamefont {Diebold}},\ }\textsl {\enquote {\bibinfo {title} {The surface science of titanium dioxide},}\ }\href@noop {} {\bibfield  {journal} {\bibinfo  {journal} {Surface Science Reports}\ }\bibinfo {volume} {48},\ \bibinfo {pages} {53} (\bibinfo {year} {2002})}\BibitemShut {NoStop}%
\bibitem [{\citenamefont {De~Lile}\ \emph {et~al.}(2019)\citenamefont {De~Lile}, \citenamefont {Kang}, \citenamefont {Son},\ and\ \citenamefont {Lee}}]{DeLile2019a}%
  \BibitemOpen
  \bibfield  {author} {\bibinfo {author} {\bibfnamefont {J.~R.}\ \bibnamefont {De~Lile}}, \bibinfo {author} {\bibfnamefont {S.~G.}\ \bibnamefont {Kang}}, \bibinfo {author} {\bibfnamefont {Y.~A.}\ \bibnamefont {Son}}, \ and\ \bibinfo {author} {\bibfnamefont {S.~G.}\ \bibnamefont {Lee}},\ }\textsl {\enquote {\bibinfo {title} {{Investigating Polaron Formation in Anatase and Brookite TiO 2 by Density Functional Theory with Hybrid-Functional and DFT + U Methods}},}\ }\href {\doibase10.1021/acsomega.9b00443} {\bibfield  {journal} {\bibinfo  {journal} {ACS Omega}\ }\bibinfo {volume} {4},\ \bibinfo {pages} {8056} (\bibinfo {year} {2019})}\BibitemShut {NoStop}%
\bibitem [{\citenamefont {Jinnouchi}\ \emph {et~al.}(2019{\natexlab{a}})\citenamefont {Jinnouchi}, \citenamefont {Karsai},\ and\ \citenamefont {Kresse}}]{Jinnouchi2019a}%
  \BibitemOpen
  \bibfield  {author} {\bibinfo {author} {\bibfnamefont {R.}~\bibnamefont {Jinnouchi}}, \bibinfo {author} {\bibfnamefont {F.}~\bibnamefont {Karsai}}, \ and\ \bibinfo {author} {\bibfnamefont {G.}~\bibnamefont {Kresse}},\ }\textsl {\enquote {\bibinfo {title} {{On-the-fly machine learning force field generation: Application to melting points}},}\ }\href {\doibase10.1103/PhysRevB.100.014105} {\bibfield  {journal} {\bibinfo  {journal} {Physical Review B}\ }\bibinfo {volume} {100},\ \bibinfo {pages} {014105} (\bibinfo {year} {2019}{\natexlab{a}})}\BibitemShut {NoStop}%
\bibitem [{\citenamefont {Jinnouchi}\ \emph {et~al.}(2019{\natexlab{b}})\citenamefont {Jinnouchi}, \citenamefont {Lahnsteiner}, \citenamefont {Karsai}, \citenamefont {Kresse},\ and\ \citenamefont {Bokdam}}]{Jinnouchi2019}%
  \BibitemOpen
  \bibfield  {author} {\bibinfo {author} {\bibfnamefont {R.}~\bibnamefont {Jinnouchi}}, \bibinfo {author} {\bibfnamefont {J.}~\bibnamefont {Lahnsteiner}}, \bibinfo {author} {\bibfnamefont {F.}~\bibnamefont {Karsai}}, \bibinfo {author} {\bibfnamefont {G.}~\bibnamefont {Kresse}}, \ and\ \bibinfo {author} {\bibfnamefont {M.}~\bibnamefont {Bokdam}},\ }\textsl {\enquote {\bibinfo {title} {{Phase Transitions of Hybrid Perovskites Simulated by Machine-Learning Force Fields Trained on the Fly with Bayesian Inference}},}\ }\href {\doibase10.1103/PhysRevLett.122.225701} {\bibfield  {journal} {\bibinfo  {journal} {Physical Review Letters}\ }\bibinfo {volume} {122},\ \bibinfo {pages} {225701} (\bibinfo {year} {2019}{\natexlab{b}})}\BibitemShut {NoStop}%
\bibitem [{\citenamefont {Birschitzky}\ \emph {et~al.}(2025)\citenamefont {Birschitzky}, \citenamefont {Leoni}, \citenamefont {Reticcioli},\ and\ \citenamefont {Franchini}}]{Birschitzky2025MLMD}%
  \BibitemOpen
  \bibfield  {author} {\bibinfo {author} {\bibfnamefont {V.~C.}\ \bibnamefont {Birschitzky}}, \bibinfo {author} {\bibfnamefont {L.}~\bibnamefont {Leoni}}, \bibinfo {author} {\bibfnamefont {M.}~\bibnamefont {Reticcioli}}, \ and\ \bibinfo {author} {\bibfnamefont {C.}~\bibnamefont {Franchini}},\ }\textsl {\enquote {\bibinfo {title} {{Machine Learning Small Polaron Dynamics}},}\ }\href {\doibase10.1103/PhysRevLett.134.216301} {\bibfield  {journal} {\bibinfo  {journal} {Physical Review Letters}\ }\bibinfo {volume} {134},\ \bibinfo {pages} {216301} (\bibinfo {year} {2025})}\BibitemShut {NoStop}%
\bibitem [{\citenamefont {Trivisonne}\ and\ \citenamefont {Reticcioli}(2019)}]{Trivisonne2019BSc}%
  \BibitemOpen
  \bibfield  {author} {\bibinfo {author} {\bibfnamefont {S.}~\bibnamefont {Trivisonne}}\ and\ \bibinfo {author} {\bibfnamefont {M.}~\bibnamefont {Reticcioli}},\ }\textsl {\enquote {\bibinfo {title} {{Follow the Polarons – Molecular Dynamics and Machine Learning}},}\ }\href {https://homepage.univie.ac.at/michele.reticcioli/linked/BSc_Thesis_Trivisonne_v7_FINAL.pdf} {\bibfield  {journal} {\bibinfo  {journal} {University of Vienna, Bachelor Thesis}\ }\bibinfo {volume} {1},\ \bibinfo {pages} {16} (\bibinfo {year} {2019})}\BibitemShut {NoStop}%
\bibitem [{\citenamefont {Trivisonne}\ \emph {et~al.}(2024)\citenamefont {Trivisonne}, \citenamefont {Franchini},\ and\ \citenamefont {Reticcioli}}]{Trivisonne2024MSc}%
  \BibitemOpen
  \bibfield  {author} {\bibinfo {author} {\bibfnamefont {S.}~\bibnamefont {Trivisonne}}, \bibinfo {author} {\bibfnamefont {C.}~\bibnamefont {Franchini}}, \ and\ \bibinfo {author} {\bibfnamefont {M.}~\bibnamefont {Reticcioli}},\ }\textsl {\enquote {\bibinfo {title} {{Polaron Dynamics in Force Field Molecular Dynamics via Machine Learning}},}\ }\href {\doibase10.25365/thesis.76711} {\bibfield  {journal} {\bibinfo  {journal} {University of Vienna, Master Thesis}\ } (\bibinfo {year} {2024}),\ 10.25365/thesis.76711}\BibitemShut {NoStop}%
\bibitem [{\citenamefont {Jain}\ \emph {et~al.}(2015)\citenamefont {Jain}, \citenamefont {Ong}, \citenamefont {Chen}, \citenamefont {Medasani}, \citenamefont {Qu}, \citenamefont {Kocher}, \citenamefont {Brafman}, \citenamefont {Petretto}, \citenamefont {Rignanese}, \citenamefont {Hautier}, \citenamefont {Gunter},\ and\ \citenamefont {Persson}}]{CPE:CPE3505}%
  \BibitemOpen
  \bibfield  {author} {\bibinfo {author} {\bibfnamefont {A.}~\bibnamefont {Jain}}, \bibinfo {author} {\bibfnamefont {S.~P.}\ \bibnamefont {Ong}}, \bibinfo {author} {\bibfnamefont {W.}~\bibnamefont {Chen}}, \bibinfo {author} {\bibfnamefont {B.}~\bibnamefont {Medasani}}, \bibinfo {author} {\bibfnamefont {X.}~\bibnamefont {Qu}}, \bibinfo {author} {\bibfnamefont {M.}~\bibnamefont {Kocher}}, \bibinfo {author} {\bibfnamefont {M.}~\bibnamefont {Brafman}}, \bibinfo {author} {\bibfnamefont {G.}~\bibnamefont {Petretto}}, \bibinfo {author} {\bibfnamefont {G.-M.}\ \bibnamefont {Rignanese}}, \bibinfo {author} {\bibfnamefont {G.}~\bibnamefont {Hautier}}, \bibinfo {author} {\bibfnamefont {D.}~\bibnamefont {Gunter}}, \ and\ \bibinfo {author} {\bibfnamefont {K.~A.}\ \bibnamefont {Persson}},\ }\textsl {\enquote {\bibinfo {title} {Fireworks: a dynamic workflow system designed for high-throughput applications},}\ }\href {\doibase10.1002/cpe.3505} {\bibfield  {journal} {\bibinfo  {journal} {Concurrency and Computation: Practice
  and Experience}\ }\bibinfo {volume} {27},\ \bibinfo {pages} {5037} (\bibinfo {year} {2015})},\ \bibinfo {note} {cPE-14-0307.R2}\BibitemShut {NoStop}%
\bibitem [{\citenamefont {Mathew}\ \emph {et~al.}(2017)\citenamefont {Mathew}, \citenamefont {Montoya}, \citenamefont {Faghaninia}, \citenamefont {Dwarakanath}, \citenamefont {Aykol}, \citenamefont {Tang}, \citenamefont {heng Chu}, \citenamefont {Smidt}, \citenamefont {Bocklund}, \citenamefont {Horton}, \citenamefont {Dagdelen}, \citenamefont {Wood}, \citenamefont {Liu}, \citenamefont {Neaton}, \citenamefont {Ong}, \citenamefont {Persson},\ and\ \citenamefont {Jain}}]{MATHEW2017140}%
  \BibitemOpen
  \bibfield  {author} {\bibinfo {author} {\bibfnamefont {K.}~\bibnamefont {Mathew}}, \bibinfo {author} {\bibfnamefont {J.~H.}\ \bibnamefont {Montoya}}, \bibinfo {author} {\bibfnamefont {A.}~\bibnamefont {Faghaninia}}, \bibinfo {author} {\bibfnamefont {S.}~\bibnamefont {Dwarakanath}}, \bibinfo {author} {\bibfnamefont {M.}~\bibnamefont {Aykol}}, \bibinfo {author} {\bibfnamefont {H.}~\bibnamefont {Tang}}, \bibinfo {author} {\bibfnamefont {I.}~\bibnamefont {heng Chu}}, \bibinfo {author} {\bibfnamefont {T.}~\bibnamefont {Smidt}}, \bibinfo {author} {\bibfnamefont {B.}~\bibnamefont {Bocklund}}, \bibinfo {author} {\bibfnamefont {M.}~\bibnamefont {Horton}}, \bibinfo {author} {\bibfnamefont {J.}~\bibnamefont {Dagdelen}}, \bibinfo {author} {\bibfnamefont {B.}~\bibnamefont {Wood}}, \bibinfo {author} {\bibfnamefont {Z.~K.}\ \bibnamefont {Liu}}, \bibinfo {author} {\bibfnamefont {J.}~\bibnamefont {Neaton}}, \bibinfo {author} {\bibfnamefont {S.~P.}\ \bibnamefont {Ong}}, \bibinfo {author} {\bibfnamefont {K.}~\bibnamefont
  {Persson}}, \ and\ \bibinfo {author} {\bibfnamefont {A.}~\bibnamefont {Jain}},\ }\textsl {\enquote {\bibinfo {title} {Atomate: {{A}} high-level interface to generate, execute, and analyze computational materials science workflows},}\ }\href {\doibase10.1016/j.commatsci.2017.07.030} {\bibfield  {journal} {\bibinfo  {journal} {Computational Materials Science}\ }\bibinfo {volume} {139},\ \bibinfo {pages} {140} (\bibinfo {year} {2017})}\BibitemShut {NoStop}%
\bibitem [{\citenamefont {Ong}\ \emph {et~al.}(2013)\citenamefont {Ong}, \citenamefont {Richards}, \citenamefont {Jain}, \citenamefont {Hautier}, \citenamefont {Kocher}, \citenamefont {Cholia}, \citenamefont {Gunter}, \citenamefont {Chevrier}, \citenamefont {Persson},\ and\ \citenamefont {Ceder}}]{ONG2013314}%
  \BibitemOpen
  \bibfield  {author} {\bibinfo {author} {\bibfnamefont {S.~P.}\ \bibnamefont {Ong}}, \bibinfo {author} {\bibfnamefont {W.~D.}\ \bibnamefont {Richards}}, \bibinfo {author} {\bibfnamefont {A.}~\bibnamefont {Jain}}, \bibinfo {author} {\bibfnamefont {G.}~\bibnamefont {Hautier}}, \bibinfo {author} {\bibfnamefont {M.}~\bibnamefont {Kocher}}, \bibinfo {author} {\bibfnamefont {S.}~\bibnamefont {Cholia}}, \bibinfo {author} {\bibfnamefont {D.}~\bibnamefont {Gunter}}, \bibinfo {author} {\bibfnamefont {V.~L.}\ \bibnamefont {Chevrier}}, \bibinfo {author} {\bibfnamefont {K.~A.}\ \bibnamefont {Persson}}, \ and\ \bibinfo {author} {\bibfnamefont {G.}~\bibnamefont {Ceder}},\ }\textsl {\enquote {\bibinfo {title} {Python {{Materials Genomics}} (pymatgen): {{A}} robust, open-source python library for materials analysis},}\ }\href {\doibase10.1016/j.commatsci.2012.10.028} {\bibfield  {journal} {\bibinfo  {journal} {Computational Materials Science}\ }\bibinfo {volume} {68},\ \bibinfo {pages} {314} (\bibinfo {year}
  {2013})}\BibitemShut {NoStop}%
\bibitem [{\citenamefont {Jain}\ \emph {et~al.}(2013)\citenamefont {Jain}, \citenamefont {Ong}, \citenamefont {Hautier}, \citenamefont {Chen}, \citenamefont {Richards}, \citenamefont {Dacek}, \citenamefont {Cholia}, \citenamefont {Gunter}, \citenamefont {Skinner}, \citenamefont {Ceder},\ and\ \citenamefont {Persson}}]{Jain2013}%
  \BibitemOpen
  \bibfield  {author} {\bibinfo {author} {\bibfnamefont {A.}~\bibnamefont {Jain}}, \bibinfo {author} {\bibfnamefont {S.~P.}\ \bibnamefont {Ong}}, \bibinfo {author} {\bibfnamefont {G.}~\bibnamefont {Hautier}}, \bibinfo {author} {\bibfnamefont {W.}~\bibnamefont {Chen}}, \bibinfo {author} {\bibfnamefont {W.~D.}\ \bibnamefont {Richards}}, \bibinfo {author} {\bibfnamefont {S.}~\bibnamefont {Dacek}}, \bibinfo {author} {\bibfnamefont {S.}~\bibnamefont {Cholia}}, \bibinfo {author} {\bibfnamefont {D.}~\bibnamefont {Gunter}}, \bibinfo {author} {\bibfnamefont {D.}~\bibnamefont {Skinner}}, \bibinfo {author} {\bibfnamefont {G.}~\bibnamefont {Ceder}}, \ and\ \bibinfo {author} {\bibfnamefont {K.~A.}\ \bibnamefont {Persson}},\ }\textsl {\enquote {\bibinfo {title} {Commentary: {{The}} materials project: {{A}} materials genome approach to accelerating materials innovation},}\ }\href {\doibase10.1063/1.4812323} {\bibfield  {journal} {\bibinfo  {journal} {APL Materials}\ }\bibinfo {volume} {1} (\bibinfo {year} {2013}),\
  10.1063/1.4812323}\BibitemShut {NoStop}%
\bibitem [{\citenamefont {Wang}\ \emph {et~al.}(2023)\citenamefont {Wang}, \citenamefont {Cheng},\ and\ \citenamefont {Zhou}}]{Wang2023a}%
  \BibitemOpen
  \bibfield  {author} {\bibinfo {author} {\bibfnamefont {K.}~\bibnamefont {Wang}}, \bibinfo {author} {\bibfnamefont {D.}~\bibnamefont {Cheng}}, \ and\ \bibinfo {author} {\bibfnamefont {B.-C.}\ \bibnamefont {Zhou}},\ }\textsl {\enquote {\bibinfo {title} {{Generalization of the mixed-space cluster expansion method for arbitrary lattices}},}\ }\href {\doibase10.1038/s41524-023-01029-0} {\bibfield  {journal} {\bibinfo  {journal} {npj Computational Materials}\ }\bibinfo {volume} {9},\ \bibinfo {pages} {75} (\bibinfo {year} {2023})}\BibitemShut {NoStop}%
\bibitem [{\citenamefont {Bradbury}\ \emph {et~al.}(2018)\citenamefont {Bradbury}, \citenamefont {Frostig}, \citenamefont {Hawkins}, \citenamefont {Johnson}, \citenamefont {Leary}, \citenamefont {Maclaurin}, \citenamefont {Necula}, \citenamefont {Paszke}, \citenamefont {Vander{P}las}, \citenamefont {Wanderman-{M}ilne},\ and\ \citenamefont {Zhang}}]{jax2018github}%
  \BibitemOpen
  \bibfield  {author} {\bibinfo {author} {\bibfnamefont {J.}~\bibnamefont {Bradbury}}, \bibinfo {author} {\bibfnamefont {R.}~\bibnamefont {Frostig}}, \bibinfo {author} {\bibfnamefont {P.}~\bibnamefont {Hawkins}}, \bibinfo {author} {\bibfnamefont {M.~J.}\ \bibnamefont {Johnson}}, \bibinfo {author} {\bibfnamefont {C.}~\bibnamefont {Leary}}, \bibinfo {author} {\bibfnamefont {D.}~\bibnamefont {Maclaurin}}, \bibinfo {author} {\bibfnamefont {G.}~\bibnamefont {Necula}}, \bibinfo {author} {\bibfnamefont {A.}~\bibnamefont {Paszke}}, \bibinfo {author} {\bibfnamefont {J.}~\bibnamefont {Vander{P}las}}, \bibinfo {author} {\bibfnamefont {S.}~\bibnamefont {Wanderman-{M}ilne}}, \ and\ \bibinfo {author} {\bibfnamefont {Q.}~\bibnamefont {Zhang}},\ }\href {http://github.com/jax-ml/jax} {\enquote {\bibinfo {title} {{JAX}: composable transformations of {P}ython+{N}um{P}y programs},}\ }\bibinfo {howpublished} {GitHub} (\bibinfo {year} {2018})\BibitemShut {NoStop}%
\bibitem [{\citenamefont {Kresse}\ and\ \citenamefont {Furthm{\"{u}}ller}(1996)}]{VASP1}%
  \BibitemOpen
  \bibfield  {author} {\bibinfo {author} {\bibfnamefont {G.}~\bibnamefont {Kresse}}\ and\ \bibinfo {author} {\bibfnamefont {J.}~\bibnamefont {Furthm{\"{u}}ller}},\ }\textsl {\enquote {\bibinfo {title} {{Efficient iterative schemes for ab initio total-energy calculations using a plane-wave basis set}},}\ }\href {\doibase10.1103/PhysRevB.54.11169} {\bibfield  {journal} {\bibinfo  {journal} {Physical Review B}\ }\bibinfo {volume} {54},\ \bibinfo {pages} {11169} (\bibinfo {year} {1996})}\BibitemShut {NoStop}%
\bibitem [{\citenamefont {Kresse}\ and\ \citenamefont {Joubert}(1999)}]{VASP2}%
  \BibitemOpen
  \bibfield  {author} {\bibinfo {author} {\bibfnamefont {G.}~\bibnamefont {Kresse}}\ and\ \bibinfo {author} {\bibfnamefont {D.}~\bibnamefont {Joubert}},\ }\textsl {\enquote {\bibinfo {title} {{From ultrasoft pseudopotentials to the projector augmented-wave method}},}\ }\href {\doibase10.1103/PhysRevB.59.1758} {\bibfield  {journal} {\bibinfo  {journal} {Physical Review B}\ }\bibinfo {volume} {59},\ \bibinfo {pages} {1758} (\bibinfo {year} {1999})}\BibitemShut {NoStop}%
\bibitem [{\citenamefont {Perdew}\ \emph {et~al.}(1996)\citenamefont {Perdew}, \citenamefont {Burke},\ and\ \citenamefont {Ernzerhof}}]{Perdew1996a}%
  \BibitemOpen
  \bibfield  {author} {\bibinfo {author} {\bibfnamefont {J.~P.}\ \bibnamefont {Perdew}}, \bibinfo {author} {\bibfnamefont {K.}~\bibnamefont {Burke}}, \ and\ \bibinfo {author} {\bibfnamefont {M.}~\bibnamefont {Ernzerhof}},\ }\textsl {\enquote {\bibinfo {title} {{Generalized gradient approximation made simple}},}\ }\href {\doibase10.1103/PhysRevLett.77.3865} {\bibfield  {journal} {\bibinfo  {journal} {Physical Review Letters}\ }\bibinfo {volume} {77},\ \bibinfo {pages} {3865} (\bibinfo {year} {1996})}\BibitemShut {NoStop}%
\bibitem [{\citenamefont {Dudarev}\ and\ \citenamefont {Botton}(1998)}]{Dudarev1998}%
  \BibitemOpen
  \bibfield  {author} {\bibinfo {author} {\bibfnamefont {S.}~\bibnamefont {Dudarev}}\ and\ \bibinfo {author} {\bibfnamefont {G.}~\bibnamefont {Botton}},\ }\textsl {\enquote {\bibinfo {title} {{Electron-energy-loss spectra and the structural stability of nickel oxide: An LSDA+U study}},}\ }\href {\doibase10.1103/PhysRevB.57.1505} {\bibfield  {journal} {\bibinfo  {journal} {Physical Review B - Condensed Matter and Materials Physics}\ }\bibinfo {volume} {57},\ \bibinfo {pages} {1505} (\bibinfo {year} {1998})}\BibitemShut {NoStop}%
\bibitem [{\citenamefont {Jinnouchi}\ \emph {et~al.}(2020)\citenamefont {Jinnouchi}, \citenamefont {Karsai}, \citenamefont {Verdi}, \citenamefont {Asahi},\ and\ \citenamefont {Kresse}}]{Jinnouchi2020}%
  \BibitemOpen
  \bibfield  {author} {\bibinfo {author} {\bibfnamefont {R.}~\bibnamefont {Jinnouchi}}, \bibinfo {author} {\bibfnamefont {F.}~\bibnamefont {Karsai}}, \bibinfo {author} {\bibfnamefont {C.}~\bibnamefont {Verdi}}, \bibinfo {author} {\bibfnamefont {R.}~\bibnamefont {Asahi}}, \ and\ \bibinfo {author} {\bibfnamefont {G.}~\bibnamefont {Kresse}},\ }\textsl {\enquote {\bibinfo {title} {{Descriptors representing two-and three-body atomic distributions and their effects on the accuracy of machine-learned inter-atomic potentials}},}\ }\href {\doibase10.1063/5.0009491} {\bibfield  {journal} {\bibinfo  {journal} {Journal of Chemical Physics}\ }\bibinfo {volume} {152} (\bibinfo {year} {2020}),\ 10.1063/5.0009491}\BibitemShut {NoStop}%
\bibitem [{\citenamefont {Peng}\ \emph {et~al.}(2016)\citenamefont {Peng}, \citenamefont {Yang}, \citenamefont {Perdew},\ and\ \citenamefont {Sun}}]{Peng2016}%
  \BibitemOpen
  \bibfield  {author} {\bibinfo {author} {\bibfnamefont {H.}~\bibnamefont {Peng}}, \bibinfo {author} {\bibfnamefont {Z.~H.}\ \bibnamefont {Yang}}, \bibinfo {author} {\bibfnamefont {J.~P.}\ \bibnamefont {Perdew}}, \ and\ \bibinfo {author} {\bibfnamefont {J.}~\bibnamefont {Sun}},\ }\textsl {\enquote {\bibinfo {title} {{Versatile van der Waals density functional based on a meta-generalized gradient approximation}},}\ }\href {\doibase10.1103/PhysRevX.6.041005} {\bibfield  {journal} {\bibinfo  {journal} {Physical Review X}\ }\bibinfo {volume} {6},\ \bibinfo {pages} {041005} (\bibinfo {year} {2016})}\BibitemShut {NoStop}%
\bibitem [{\citenamefont {Sun}\ \emph {et~al.}(2016)\citenamefont {Sun}, \citenamefont {Remsing}, \citenamefont {Zhang}, \citenamefont {Sun}, \citenamefont {Ruzsinszky}, \citenamefont {Peng}, \citenamefont {Yang}, \citenamefont {Paul}, \citenamefont {Waghmare}, \citenamefont {Wu}, \citenamefont {Klein},\ and\ \citenamefont {Perdew}}]{Sun2016}%
  \BibitemOpen
  \bibfield  {author} {\bibinfo {author} {\bibfnamefont {J.}~\bibnamefont {Sun}}, \bibinfo {author} {\bibfnamefont {R.~C.}\ \bibnamefont {Remsing}}, \bibinfo {author} {\bibfnamefont {Y.}~\bibnamefont {Zhang}}, \bibinfo {author} {\bibfnamefont {Z.}~\bibnamefont {Sun}}, \bibinfo {author} {\bibfnamefont {A.}~\bibnamefont {Ruzsinszky}}, \bibinfo {author} {\bibfnamefont {H.}~\bibnamefont {Peng}}, \bibinfo {author} {\bibfnamefont {Z.}~\bibnamefont {Yang}}, \bibinfo {author} {\bibfnamefont {A.}~\bibnamefont {Paul}}, \bibinfo {author} {\bibfnamefont {U.}~\bibnamefont {Waghmare}}, \bibinfo {author} {\bibfnamefont {X.}~\bibnamefont {Wu}}, \bibinfo {author} {\bibfnamefont {M.~L.}\ \bibnamefont {Klein}}, \ and\ \bibinfo {author} {\bibfnamefont {J.~P.}\ \bibnamefont {Perdew}},\ }\textsl {\enquote {\bibinfo {title} {{Accurate first-principles structures and energies of diversely bonded systems from an efficient density functional}},}\ }\href {\doibase10.1038/nchem.2535} {\bibfield  {journal} {\bibinfo  {journal} {Nature
  Chemistry}\ }\bibinfo {volume} {8},\ \bibinfo {pages} {831} (\bibinfo {year} {2016})}\BibitemShut {NoStop}%
\bibitem [{\citenamefont {Sun}\ \emph {et~al.}(2015)\citenamefont {Sun}, \citenamefont {Ruzsinszky},\ and\ \citenamefont {Perdew}}]{Sun2015SCAN}%
  \BibitemOpen
  \bibfield  {author} {\bibinfo {author} {\bibfnamefont {J.}~\bibnamefont {Sun}}, \bibinfo {author} {\bibfnamefont {A.}~\bibnamefont {Ruzsinszky}}, \ and\ \bibinfo {author} {\bibfnamefont {J.}~\bibnamefont {Perdew}},\ }\textsl {\enquote {\bibinfo {title} {{Strongly Constrained and Appropriately Normed Semilocal Density Functional}},}\ }\href {\doibase10.1103/PhysRevLett.115.036402} {\bibfield  {journal} {\bibinfo  {journal} {Physical Review Letters}\ }\bibinfo {volume} {115},\ \bibinfo {pages} {036402} (\bibinfo {year} {2015})}\BibitemShut {NoStop}%
\bibitem [{\citenamefont {Tersoff}\ and\ \citenamefont {Hamann}(1985)}]{Tersoff1985}%
  \BibitemOpen
  \bibfield  {author} {\bibinfo {author} {\bibfnamefont {J.}~\bibnamefont {Tersoff}}\ and\ \bibinfo {author} {\bibfnamefont {D.~R.}\ \bibnamefont {Hamann}},\ }\textsl {\enquote {\bibinfo {title} {{Theory of the scanning tunneling microscope}},}\ }\href {\doibase10.1103/PhysRevB.31.805} {\bibfield  {journal} {\bibinfo  {journal} {Physical Review B}\ }\bibinfo {volume} {31},\ \bibinfo {pages} {805} (\bibinfo {year} {1985})}\BibitemShut {NoStop}%
\bibitem [{\citenamefont {Kirby}(1967)}]{Kirby1967rutileT}%
  \BibitemOpen
  \bibfield  {author} {\bibinfo {author} {\bibfnamefont {R.~K.}\ \bibnamefont {Kirby}},\ }\textsl {\enquote {\bibinfo {title} {{Thermal expansion of rutile from 100 to 700 K}},}\ }\href {\doibase10.6028/jres.071A.041} {\bibfield  {journal} {\bibinfo  {journal} {Journal of Research of the National Bureau of Standards Section A: Physics and Chemistry}\ }\bibinfo {volume} {71A},\ \bibinfo {pages} {363} (\bibinfo {year} {1967})}\BibitemShut {NoStop}%
\bibitem [{\citenamefont {Hummer}\ \emph {et~al.}(2007)\citenamefont {Hummer}, \citenamefont {Heaney},\ and\ \citenamefont {Post}}]{Hummer2007rutileT}%
  \BibitemOpen
  \bibfield  {author} {\bibinfo {author} {\bibfnamefont {D.~R.}\ \bibnamefont {Hummer}}, \bibinfo {author} {\bibfnamefont {P.~J.}\ \bibnamefont {Heaney}}, \ and\ \bibinfo {author} {\bibfnamefont {J.~E.}\ \bibnamefont {Post}},\ }\textsl {\enquote {\bibinfo {title} {{Thermal expansion of anatase and rutile between 300 and 575 K using synchrotron powder X-ray diffraction}},}\ }\href {\doibase10.1154/1.2790965} {\bibfield  {journal} {\bibinfo  {journal} {Powder Diffraction}\ }\bibinfo {volume} {22},\ \bibinfo {pages} {352} (\bibinfo {year} {2007})}\BibitemShut {NoStop}%
\bibitem [{\citenamefont {Sasahara}\ and\ \citenamefont {Tomitori}(2013)}]{Sasahara2013}%
  \BibitemOpen
  \bibfield  {author} {\bibinfo {author} {\bibfnamefont {A.}~\bibnamefont {Sasahara}}\ and\ \bibinfo {author} {\bibfnamefont {M.}~\bibnamefont {Tomitori}},\ }\textsl {\enquote {\bibinfo {title} {{XPS and STM Study of Nb-Doped TiO 2 (110)-(1 × 1) Surfaces}},}\ }\href {\doibase10.1021/jp4057576} {\bibfield  {journal} {\bibinfo  {journal} {The Journal of Physical Chemistry C}\ }\bibinfo {volume} {117},\ \bibinfo {pages} {17680} (\bibinfo {year} {2013})}\BibitemShut {NoStop}%
\bibitem [{\citenamefont {Dohn{\'{a}}lek}\ \emph {et~al.}(2010)\citenamefont {Dohn{\'{a}}lek}, \citenamefont {Lyubinetsky},\ and\ \citenamefont {Rousseau}}]{Dohnalek2010}%
  \BibitemOpen
  \bibfield  {author} {\bibinfo {author} {\bibfnamefont {Z.}~\bibnamefont {Dohn{\'{a}}lek}}, \bibinfo {author} {\bibfnamefont {I.}~\bibnamefont {Lyubinetsky}}, \ and\ \bibinfo {author} {\bibfnamefont {R.}~\bibnamefont {Rousseau}},\ }\textsl {\enquote {\bibinfo {title} {{Thermally-driven processes on rutile TiO2(1 1 0)-(1 × 1): A direct view at the atomic scale}},}\ }\href {\doibase10.1016/j.progsurf.2010.03.001} {\bibfield  {journal} {\bibinfo  {journal} {Progress in Surface Science}\ }\bibinfo {volume} {85},\ \bibinfo {pages} {161} (\bibinfo {year} {2010})}\BibitemShut {NoStop}%
\bibitem [{\citenamefont {Onishi}\ and\ \citenamefont {Iwasawa}(1994)}]{Onishi1994}%
  \BibitemOpen
  \bibfield  {author} {\bibinfo {author} {\bibfnamefont {H.}~\bibnamefont {Onishi}}\ and\ \bibinfo {author} {\bibfnamefont {Y.}~\bibnamefont {Iwasawa}},\ }\textsl {\enquote {\bibinfo {title} {{Reconstruction of TiO2(110) surface: STM study with atomic-scale resolution}},}\ }\href {\doibase10.1016/0039-6028(94)91146-0} {\bibfield  {journal} {\bibinfo  {journal} {Surface Science}\ }\bibinfo {volume} {313},\ \bibinfo {pages} {L783} (\bibinfo {year} {1994})}\BibitemShut {NoStop}%
\bibitem [{\citenamefont {Li}\ \emph {et~al.}(1999)\citenamefont {Li}, \citenamefont {Hebenstreit}, \citenamefont {Gross}, \citenamefont {Diebold}, \citenamefont {Henderson}, \citenamefont {Jennison}, \citenamefont {Schultz},\ and\ \citenamefont {Sears}}]{Li1999a}%
  \BibitemOpen
  \bibfield  {author} {\bibinfo {author} {\bibfnamefont {M.}~\bibnamefont {Li}}, \bibinfo {author} {\bibfnamefont {W.}~\bibnamefont {Hebenstreit}}, \bibinfo {author} {\bibfnamefont {L.}~\bibnamefont {Gross}}, \bibinfo {author} {\bibfnamefont {U.}~\bibnamefont {Diebold}}, \bibinfo {author} {\bibfnamefont {M.~A.}\ \bibnamefont {Henderson}}, \bibinfo {author} {\bibfnamefont {D.~R.}\ \bibnamefont {Jennison}}, \bibinfo {author} {\bibfnamefont {P.~A.}\ \bibnamefont {Schultz}}, \ and\ \bibinfo {author} {\bibfnamefont {M.~P.}\ \bibnamefont {Sears}},\ }\textsl {\enquote {\bibinfo {title} {{Oxygen-induced restructuring of the TiO2(110) surface: a comprehensive study}},}\ }\href {\doibase10.1016/S0039-6028(99)00720-7} {\bibfield  {journal} {\bibinfo  {journal} {Surface Science}\ }\bibinfo {volume} {437},\ \bibinfo {pages} {173} (\bibinfo {year} {1999})}\BibitemShut {NoStop}%
\bibitem [{\citenamefont {McCarty}\ and\ \citenamefont {Bartelt}(2003)}]{McCarty2003b}%
  \BibitemOpen
  \bibfield  {author} {\bibinfo {author} {\bibfnamefont {K.~F.}\ \bibnamefont {McCarty}}\ and\ \bibinfo {author} {\bibfnamefont {N.~C.}\ \bibnamefont {Bartelt}},\ }\textsl {\enquote {\bibinfo {title} {{The 1 × 1/1 × 2 phase transition of the TiO2(1 1 0) surface-variation of transition temperature with crystal composition}},}\ }\href {\doibase10.1016/S0039-6028(03)00003-7} {\bibfield  {journal} {\bibinfo  {journal} {Surface Science}\ }\bibinfo {volume} {527},\ \bibinfo {pages} {L203} (\bibinfo {year} {2003})}\BibitemShut {NoStop}%
\bibitem [{\citenamefont {Wang}\ \emph {et~al.}(2014)\citenamefont {Wang}, \citenamefont {Oganov}, \citenamefont {Zhu},\ and\ \citenamefont {Zhou}}]{Wang2014a}%
  \BibitemOpen
  \bibfield  {author} {\bibinfo {author} {\bibfnamefont {Q.}~\bibnamefont {Wang}}, \bibinfo {author} {\bibfnamefont {A.~R.}\ \bibnamefont {Oganov}}, \bibinfo {author} {\bibfnamefont {Q.}~\bibnamefont {Zhu}}, \ and\ \bibinfo {author} {\bibfnamefont {X.~F.}\ \bibnamefont {Zhou}},\ }\textsl {\enquote {\bibinfo {title} {{New reconstructions of the (110) surface of rutile TiO2 predicted by an evolutionary method}},}\ }\href {\doibase10.1103/PhysRevLett.113.266101} {\bibfield  {journal} {\bibinfo  {journal} {Physical Review Letters}\ }\bibinfo {volume} {113},\ \bibinfo {pages} {266101} (\bibinfo {year} {2014})}\BibitemShut {NoStop}%
\bibitem [{\citenamefont {Mochizuki}\ \emph {et~al.}(2016)\citenamefont {Mochizuki}, \citenamefont {Ariga}, \citenamefont {Fukaya}, \citenamefont {Wada}, \citenamefont {Maekawa}, \citenamefont {Kawasuso}, \citenamefont {Shidara}, \citenamefont {Asakura},\ and\ \citenamefont {Hyodo}}]{Mochizuki2016}%
  \BibitemOpen
  \bibfield  {author} {\bibinfo {author} {\bibfnamefont {I.}~\bibnamefont {Mochizuki}}, \bibinfo {author} {\bibfnamefont {H.}~\bibnamefont {Ariga}}, \bibinfo {author} {\bibfnamefont {Y.}~\bibnamefont {Fukaya}}, \bibinfo {author} {\bibfnamefont {K.}~\bibnamefont {Wada}}, \bibinfo {author} {\bibfnamefont {M.}~\bibnamefont {Maekawa}}, \bibinfo {author} {\bibfnamefont {A.}~\bibnamefont {Kawasuso}}, \bibinfo {author} {\bibfnamefont {T.}~\bibnamefont {Shidara}}, \bibinfo {author} {\bibfnamefont {K.}~\bibnamefont {Asakura}}, \ and\ \bibinfo {author} {\bibfnamefont {T.}~\bibnamefont {Hyodo}},\ }\textsl {\enquote {\bibinfo {title} {{Structure determination of the rutile-TiO2(110)-(1 × 2) surface using total-reflection high-energy positron diffraction (TRHEPD)}},}\ }\href {\doibase10.1039/c5cp07892j} {\bibfield  {journal} {\bibinfo  {journal} {Physical Chemistry Chemical Physics}\ }\bibinfo {volume} {18},\ \bibinfo {pages} {7085} (\bibinfo {year} {2016})}\BibitemShut {NoStop}%
\bibitem [{\citenamefont {Henderson}(2011)}]{Henderson2011}%
  \BibitemOpen
  \bibfield  {author} {\bibinfo {author} {\bibfnamefont {M.~A.}\ \bibnamefont {Henderson}},\ }\textsl {\enquote {\bibinfo {title} {{A surface science perspective on TiO2 photocatalysis}},}\ }\href {\doibase10.1016/j.surfrep.2011.01.001} {\bibfield  {journal} {\bibinfo  {journal} {Surface Science Reports}\ }\bibinfo {volume} {66},\ \bibinfo {pages} {185} (\bibinfo {year} {2011})}\BibitemShut {NoStop}%
\bibitem [{\citenamefont {Adachi}\ \emph {et~al.}(2022)\citenamefont {Adachi}, \citenamefont {Sugawara},\ and\ \citenamefont {Li}}]{Adachi2021}%
  \BibitemOpen
  \bibfield  {author} {\bibinfo {author} {\bibfnamefont {Y.}~\bibnamefont {Adachi}}, \bibinfo {author} {\bibfnamefont {Y.}~\bibnamefont {Sugawara}}, \ and\ \bibinfo {author} {\bibfnamefont {Y.~J.}\ \bibnamefont {Li}},\ }\textsl {\enquote {\bibinfo {title} {{Probing CO on a rutile TiO2(110) surface using atomic force microscopy and Kelvin probe force microscopy}},}\ }\href {\doibase10.1007/s12274-021-3809-x} {\bibfield  {journal} {\bibinfo  {journal} {Nano Research}\ }\bibinfo {volume} {15},\ \bibinfo {pages} {1909} (\bibinfo {year} {2022})}\BibitemShut {NoStop}%
\bibitem [{\citenamefont {Wu}\ \emph {et~al.}(2020)\citenamefont {Wu}, \citenamefont {Wang}, \citenamefont {Xiong}, \citenamefont {Sun}, \citenamefont {Chai}, \citenamefont {Zhang}, \citenamefont {Xu}, \citenamefont {Fu},\ and\ \citenamefont {Huang}}]{Wu2020}%
  \BibitemOpen
  \bibfield  {author} {\bibinfo {author} {\bibfnamefont {L.}~\bibnamefont {Wu}}, \bibinfo {author} {\bibfnamefont {Z.}~\bibnamefont {Wang}}, \bibinfo {author} {\bibfnamefont {F.}~\bibnamefont {Xiong}}, \bibinfo {author} {\bibfnamefont {G.}~\bibnamefont {Sun}}, \bibinfo {author} {\bibfnamefont {P.}~\bibnamefont {Chai}}, \bibinfo {author} {\bibfnamefont {Z.}~\bibnamefont {Zhang}}, \bibinfo {author} {\bibfnamefont {H.}~\bibnamefont {Xu}}, \bibinfo {author} {\bibfnamefont {C.}~\bibnamefont {Fu}}, \ and\ \bibinfo {author} {\bibfnamefont {W.}~\bibnamefont {Huang}},\ }\textsl {\enquote {\bibinfo {title} {{Surface chemistry and photochemistry of small molecules on rutile TiO2(001) and TiO2(011)-(2 × 1) surfaces: The crucial roles of defects}},}\ }\href {\doibase10.1063/1.5135945} {\bibfield  {journal} {\bibinfo  {journal} {Journal of Chemical Physics}\ }\bibinfo {volume} {152},\ \bibinfo {pages} {044702} (\bibinfo {year} {2020})}\BibitemShut {NoStop}%
\bibitem [{\citenamefont {Petrik}\ and\ \citenamefont {Kimmel}(2012)}]{Petrik2012a}%
  \BibitemOpen
  \bibfield  {author} {\bibinfo {author} {\bibfnamefont {N.~G.}\ \bibnamefont {Petrik}}\ and\ \bibinfo {author} {\bibfnamefont {G.~A.}\ \bibnamefont {Kimmel}},\ }\textsl {\enquote {\bibinfo {title} {{Adsorption geometry of CO versus coverage on TiO2(110) from s- and p-polarized infrared spectroscopy}},}\ }\href {\doibase10.1021/jz301413v} {\bibfield  {journal} {\bibinfo  {journal} {Journal of Physical Chemistry Letters}\ }\bibinfo {volume} {3},\ \bibinfo {pages} {3425} (\bibinfo {year} {2012})}\BibitemShut {NoStop}%
\bibitem [{\citenamefont {Petrik}\ \emph {et~al.}(2018)\citenamefont {Petrik}, \citenamefont {Mu}, \citenamefont {Dahal}, \citenamefont {Wang}, \citenamefont {Lyubinetsky},\ and\ \citenamefont {Kimmel}}]{Petrik2018}%
  \BibitemOpen
  \bibfield  {author} {\bibinfo {author} {\bibfnamefont {N.~G.}\ \bibnamefont {Petrik}}, \bibinfo {author} {\bibfnamefont {R.}~\bibnamefont {Mu}}, \bibinfo {author} {\bibfnamefont {A.}~\bibnamefont {Dahal}}, \bibinfo {author} {\bibfnamefont {Z.}~\bibnamefont {Wang}}, \bibinfo {author} {\bibfnamefont {I.}~\bibnamefont {Lyubinetsky}}, \ and\ \bibinfo {author} {\bibfnamefont {G.~A.}\ \bibnamefont {Kimmel}},\ }\textsl {\enquote {\bibinfo {title} {{Diffusion and Photon-Stimulated Desorption of CO on TiO2(110)}},}\ }\href {\doibase10.1021/acs.jpcc.8b03418} {\bibfield  {journal} {\bibinfo  {journal} {Journal of Physical Chemistry C}\ }\bibinfo {volume} {122},\ \bibinfo {pages} {15382} (\bibinfo {year} {2018})}\BibitemShut {NoStop}%
\bibitem [{\citenamefont {Farnesi~Camellone}\ \emph {et~al.}(2011)\citenamefont {Farnesi~Camellone}, \citenamefont {Kowalski},\ and\ \citenamefont {Marx}}]{FarnesiCamellone2011a}%
  \BibitemOpen
  \bibfield  {author} {\bibinfo {author} {\bibfnamefont {M.}~\bibnamefont {Farnesi~Camellone}}, \bibinfo {author} {\bibfnamefont {P.~M.}\ \bibnamefont {Kowalski}}, \ and\ \bibinfo {author} {\bibfnamefont {D.}~\bibnamefont {Marx}},\ }\textsl {\enquote {\bibinfo {title} {{Ideal, defective, and gold-promoted rutile TiO2(110) surfaces interacting with CO, H2, and H2O: Structures, energies, thermodynamics, and dynamics from PBE+U}},}\ }\href {\doibase10.1103/PhysRevB.84.035413} {\bibfield  {journal} {\bibinfo  {journal} {Physical Review B - Condensed Matter and Materials Physics}\ }\bibinfo {volume} {84},\ \bibinfo {pages} {035413} (\bibinfo {year} {2011})}\BibitemShut {NoStop}%
\bibitem [{\citenamefont {Yu}\ and\ \citenamefont {Gong}(2015)}]{Yu2015a}%
  \BibitemOpen
  \bibfield  {author} {\bibinfo {author} {\bibfnamefont {Y.~Y.}\ \bibnamefont {Yu}}\ and\ \bibinfo {author} {\bibfnamefont {X.~Q.}\ \bibnamefont {Gong}},\ }\textsl {\enquote {\bibinfo {title} {{CO Oxidation at Rutile TiO2(110): Role of Oxygen Vacancies and Titanium Interstitials}},}\ }\href {\doibase10.1021/cs501900q} {\bibfield  {journal} {\bibinfo  {journal} {ACS Catalysis}\ }\bibinfo {volume} {5},\ \bibinfo {pages} {2042} (\bibinfo {year} {2015})}\BibitemShut {NoStop}%
\bibitem [{\citenamefont {Kunat}\ \emph {et~al.}(2009)\citenamefont {Kunat}, \citenamefont {Traeger}, \citenamefont {Silber}, \citenamefont {Qiu}, \citenamefont {Wang}, \citenamefont {Van~Veen}, \citenamefont {W{\"{o}}ll}, \citenamefont {Kowalski}, \citenamefont {Meyer}, \citenamefont {H{\"{a}}ttig},\ and\ \citenamefont {Marx}}]{Kunat2009}%
  \BibitemOpen
  \bibfield  {author} {\bibinfo {author} {\bibfnamefont {M.}~\bibnamefont {Kunat}}, \bibinfo {author} {\bibfnamefont {F.}~\bibnamefont {Traeger}}, \bibinfo {author} {\bibfnamefont {D.}~\bibnamefont {Silber}}, \bibinfo {author} {\bibfnamefont {H.}~\bibnamefont {Qiu}}, \bibinfo {author} {\bibfnamefont {Y.}~\bibnamefont {Wang}}, \bibinfo {author} {\bibfnamefont {A.~C.}\ \bibnamefont {Van~Veen}}, \bibinfo {author} {\bibfnamefont {C.}~\bibnamefont {W{\"{o}}ll}}, \bibinfo {author} {\bibfnamefont {P.~M.}\ \bibnamefont {Kowalski}}, \bibinfo {author} {\bibfnamefont {B.}~\bibnamefont {Meyer}}, \bibinfo {author} {\bibfnamefont {C.}~\bibnamefont {H{\"{a}}ttig}}, \ and\ \bibinfo {author} {\bibfnamefont {D.}~\bibnamefont {Marx}},\ }\textsl {\enquote {\bibinfo {title} {{Formation of weakly bound, ordered adlayers of CO on rutile TiO2 (110): A combined experimental and theoretical study}},}\ }\href {\doibase10.1063/1.3098318} {\bibfield  {journal} {\bibinfo  {journal} {Journal of Chemical Physics}\ }\bibinfo {volume} {130},\
  \bibinfo {pages} {144703} (\bibinfo {year} {2009})}\BibitemShut {NoStop}%
\bibitem [{\citenamefont {Dohn{\'{a}}lek}\ \emph {et~al.}(2006)\citenamefont {Dohn{\'{a}}lek}, \citenamefont {Kim}, \citenamefont {Bondarchuk}, \citenamefont {Mike~White},\ and\ \citenamefont {Kay}}]{Dohnalek2006}%
  \BibitemOpen
  \bibfield  {author} {\bibinfo {author} {\bibfnamefont {Z.}~\bibnamefont {Dohn{\'{a}}lek}}, \bibinfo {author} {\bibfnamefont {J.}~\bibnamefont {Kim}}, \bibinfo {author} {\bibfnamefont {O.}~\bibnamefont {Bondarchuk}}, \bibinfo {author} {\bibfnamefont {J.}~\bibnamefont {Mike~White}}, \ and\ \bibinfo {author} {\bibfnamefont {B.~D.}\ \bibnamefont {Kay}},\ }\textsl {\enquote {\bibinfo {title} {{Physisorption of N2, O2, and CO on fully oxidized TiO2(110)}},}\ }\href {\doibase10.1021/jp0564905} {\bibfield  {journal} {\bibinfo  {journal} {Journal of Physical Chemistry B}\ }\bibinfo {volume} {110},\ \bibinfo {pages} {6229} (\bibinfo {year} {2006})}\BibitemShut {NoStop}%
\bibitem [{\citenamefont {Mu}\ \emph {et~al.}(2017)\citenamefont {Mu}, \citenamefont {Dahal}, \citenamefont {Wang}, \citenamefont {Dohn{\'{a}}lek}, \citenamefont {Kimmel}, \citenamefont {Petrik},\ and\ \citenamefont {Lyubinetsky}}]{Mu2017}%
  \BibitemOpen
  \bibfield  {author} {\bibinfo {author} {\bibfnamefont {R.}~\bibnamefont {Mu}}, \bibinfo {author} {\bibfnamefont {A.}~\bibnamefont {Dahal}}, \bibinfo {author} {\bibfnamefont {Z.~T.}\ \bibnamefont {Wang}}, \bibinfo {author} {\bibfnamefont {Z.}~\bibnamefont {Dohn{\'{a}}lek}}, \bibinfo {author} {\bibfnamefont {G.~A.}\ \bibnamefont {Kimmel}}, \bibinfo {author} {\bibfnamefont {N.~G.}\ \bibnamefont {Petrik}}, \ and\ \bibinfo {author} {\bibfnamefont {I.}~\bibnamefont {Lyubinetsky}},\ }\textsl {\enquote {\bibinfo {title} {{Adsorption and Photodesorption of CO from Charged Point Defects on TiO2(110)}},}\ }\href {\doibase10.1021/acs.jpclett.7b02052} {\bibfield  {journal} {\bibinfo  {journal} {Journal of Physical Chemistry Letters}\ }\bibinfo {volume} {8},\ \bibinfo {pages} {4565} (\bibinfo {year} {2017})}\BibitemShut {NoStop}%
\bibitem [{\citenamefont {Zhao}\ \emph {et~al.}(2009)\citenamefont {Zhao}, \citenamefont {Wang}, \citenamefont {Cui}, \citenamefont {Huang}, \citenamefont {Wang}, \citenamefont {Luo}, \citenamefont {Yang},\ and\ \citenamefont {Hou}}]{Zhao2009b}%
  \BibitemOpen
  \bibfield  {author} {\bibinfo {author} {\bibfnamefont {Y.}~\bibnamefont {Zhao}}, \bibinfo {author} {\bibfnamefont {Z.}~\bibnamefont {Wang}}, \bibinfo {author} {\bibfnamefont {X.}~\bibnamefont {Cui}}, \bibinfo {author} {\bibfnamefont {T.}~\bibnamefont {Huang}}, \bibinfo {author} {\bibfnamefont {B.}~\bibnamefont {Wang}}, \bibinfo {author} {\bibfnamefont {Y.}~\bibnamefont {Luo}}, \bibinfo {author} {\bibfnamefont {J.}~\bibnamefont {Yang}}, \ and\ \bibinfo {author} {\bibfnamefont {J.}~\bibnamefont {Hou}},\ }\textsl {\enquote {\bibinfo {title} {{What are the adsorption sites for CO on the reduced TiO2(110)-1 x 1 surface?}}}\ }\href {\doibase10.1021/ja902259k} {\bibfield  {journal} {\bibinfo  {journal} {Journal of the American Chemical Society}\ }\bibinfo {volume} {131},\ \bibinfo {pages} {7958} (\bibinfo {year} {2009})}\BibitemShut {NoStop}%
\bibitem [{\citenamefont {Prates~Ramalho}\ \emph {et~al.}(2017)\citenamefont {Prates~Ramalho}, \citenamefont {Illas},\ and\ \citenamefont {Gomes}}]{PratesRamalho2017a}%
  \BibitemOpen
  \bibfield  {author} {\bibinfo {author} {\bibfnamefont {J.~P.}\ \bibnamefont {Prates~Ramalho}}, \bibinfo {author} {\bibfnamefont {F.}~\bibnamefont {Illas}}, \ and\ \bibinfo {author} {\bibfnamefont {J.~R.}\ \bibnamefont {Gomes}},\ }\textsl {\enquote {\bibinfo {title} {{Adsorption of CO on the rutile TiO2(110) surface: A dispersion-corrected density functional theory study}},}\ }\href {\doibase10.1039/c6cp06971a} {\bibfield  {journal} {\bibinfo  {journal} {Physical Chemistry Chemical Physics}\ }\bibinfo {volume} {19},\ \bibinfo {pages} {2487} (\bibinfo {year} {2017})}\BibitemShut {NoStop}%
\bibitem [{\citenamefont {Yalcin}(2025)}]{polflow_repo}%
  \BibitemOpen
  \bibfield  {author} {\bibinfo {author} {\bibfnamefont {F.}~\bibnamefont {Yalcin}},\ }\href {\doibase10.5281/zenodo.16090819} {\enquote {\bibinfo {title} {{polflow: v1.0.0 Initial Public Release}},}\ } (\bibinfo {year} {2025})\BibitemShut {NoStop}%
\end{thebibliography}%


\begin{thebibliography}{2}%
\makeatletter
\providecommand \@ifxundefined [1]{%
 \@ifx{#1\undefined}
}%
\providecommand \@ifnum [1]{%
 \ifnum #1\expandafter \@firstoftwo
 \else \expandafter \@secondoftwo
 \fi
}%
\providecommand \@ifx [1]{%
 \ifx #1\expandafter \@firstoftwo
 \else \expandafter \@secondoftwo
 \fi
}%
\providecommand \natexlab [1]{#1}%
\providecommand \enquote  [1]{``#1''}%
\providecommand \bibnamefont  [1]{#1}%
\providecommand \bibfnamefont [1]{#1}%
\providecommand \citenamefont [1]{#1}%
\providecommand \href@noop [0]{\@secondoftwo}%
\providecommand \href [0]{\begingroup \@sanitize@url \@href}%
\providecommand \@href[1]{\@@startlink{#1}\@@href}%
\providecommand \@@href[1]{\endgroup#1\@@endlink}%
\providecommand \@sanitize@url [0]{\catcode `\\12\catcode `\$12\catcode `\&12\catcode `\#12\catcode `\^12\catcode `\_12\catcode `\%12\relax}%
\providecommand \@@startlink[1]{}%
\providecommand \@@endlink[0]{}%
\providecommand \url  [0]{\begingroup\@sanitize@url \@url }%
\providecommand \@url [1]{\endgroup\@href {#1}{\urlprefix }}%
\providecommand \urlprefix  [0]{URL }%
\providecommand \Eprint [0]{\href }%
\providecommand \doibase [0]{http://dx.doi.org/}%
\providecommand \selectlanguage [0]{\@gobble}%
\providecommand \bibinfo  [0]{\@secondoftwo}%
\providecommand \bibfield  [0]{\@secondoftwo}%
\providecommand \translation [1]{[#1]}%
\providecommand \BibitemOpen [0]{}%
\providecommand \bibitemStop [0]{}%
\providecommand \bibitemNoStop [0]{.\EOS\space}%
\providecommand \EOS [0]{\spacefactor3000\relax}%
\providecommand \BibitemShut  [1]{\csname bibitem#1\endcsname}%
\let\auto@bib@innerbib\@empty
\bibitem [{\citenamefont {Yalcin}(2025)}]{polflow_repo}%
  \BibitemOpen
  \bibfield  {author} {\bibinfo {author} {\bibfnamefont {F.}~\bibnamefont {Yalcin}},\ }\href {\doibase 10.5281/zenodo.16090819} {\enquote {\bibinfo {title} {{polflow: v1.0.0 Initial Public Release}},}\ } (\bibinfo {year} {2025})\BibitemShut {NoStop}%
\bibitem [{\citenamefont {Birschitzky}\ \emph {et~al.}(2024)\citenamefont {Birschitzky}, \citenamefont {Sokolovi{\'{c}}}, \citenamefont {Prezzi}, \citenamefont {Palot{\'{a}}s}, \citenamefont {Setv{\'{i}}n}, \citenamefont {Diebold}, \citenamefont {Reticcioli},\ and\ \citenamefont {Franchini}}]{Birschitzky2024}%
  \BibitemOpen
  \bibfield  {author} {\bibinfo {author} {\bibfnamefont {V.~C.}\ \bibnamefont {Birschitzky}}, \bibinfo {author} {\bibfnamefont {I.}~\bibnamefont {Sokolovi{\'{c}}}}, \bibinfo {author} {\bibfnamefont {M.}~\bibnamefont {Prezzi}}, \bibinfo {author} {\bibfnamefont {K.}~\bibnamefont {Palot{\'{a}}s}}, \bibinfo {author} {\bibfnamefont {M.}~\bibnamefont {Setv{\'{i}}n}}, \bibinfo {author} {\bibfnamefont {U.}~\bibnamefont {Diebold}}, \bibinfo {author} {\bibfnamefont {M.}~\bibnamefont {Reticcioli}}, \ and\ \bibinfo {author} {\bibfnamefont {C.}~\bibnamefont {Franchini}},\ }\href {\doibase 10.1038/s41524-024-01289-4} {\bibfield  {journal} {\bibinfo  {journal} {npj Computational Materials}\ }\textbf {\bibinfo {volume} {10}},\ \bibinfo {pages} {89} (\bibinfo {year} {2024})}\BibitemShut {NoStop}%
\end{thebibliography}%

\end{document}


\author{Firat Yalcin}
\email[]{firat.yalcin@univie.ac.at}
\affiliation{University of Vienna, Faculty of Physics and Center for Computational Materials Science, Kolingasse 14-16, 1090 Vienna, Austria}

\author{Carla Verdi}
\email{c.verdi@uq.edu.au}
\affiliation{School of Mathematics and Physics, The University of Queensland, Brisbane 4072 QLD, Australia}

\author{Viktor C. Birschitzky}
\affiliation{University of Vienna, Faculty of Physics and Center for Computational Materials Science, Kolingasse 14-16, 1090 Vienna, Austria}

\author{Matthias Meier}
\affiliation{University of Vienna, Faculty of Physics and Center for Computational Materials Science, Kolingasse 14-16, 1090 Vienna, Austria}

\author{Michael Wolloch}
\affiliation{University of Vienna, Faculty of Physics and Center for Computational Materials Science, Kolingasse 14-16, 1090 Vienna, Austria}
\affiliation{VASP Software GmbH, Berggasse, 1090 Vienna, Austria}

\author{Michele Reticcioli}
\email[]{michele.reticcioli@univie.ac.at}
\affiliation{University of Vienna, Faculty of Physics and Center for Computational Materials Science, Kolingasse 14-16, 1090 Vienna, Austria}
\affiliation{National Research Council, CNR-SPIN, via Vetoio 42, 67100 L'Aquila, Italy}
\affiliation{University of L'Aquila, via Vetoio 10, 67100 L'Aquila, Italy}

\title{Supplementary Information:\\
Automated Modeling of Polarons: Defects and Reactivity on \tios\ Surfaces}

\maketitle

\tableofcontents

\clearpage

\section{Machine Learning Model Performance}
The active-learning approach implemented in the \wf\ package~\cite{polflow_repo} enabled efficient training of the \vbml\ model~\cite{Birschitzky2024} on a database including 1924 symmetrically distinct configurations.
Here we present a detailed analysis of the model's prediction accuracy across both the entire energy range and the low-energy region.
\begin{figure}[ht]
\centering
\includegraphics[width=\textwidth]{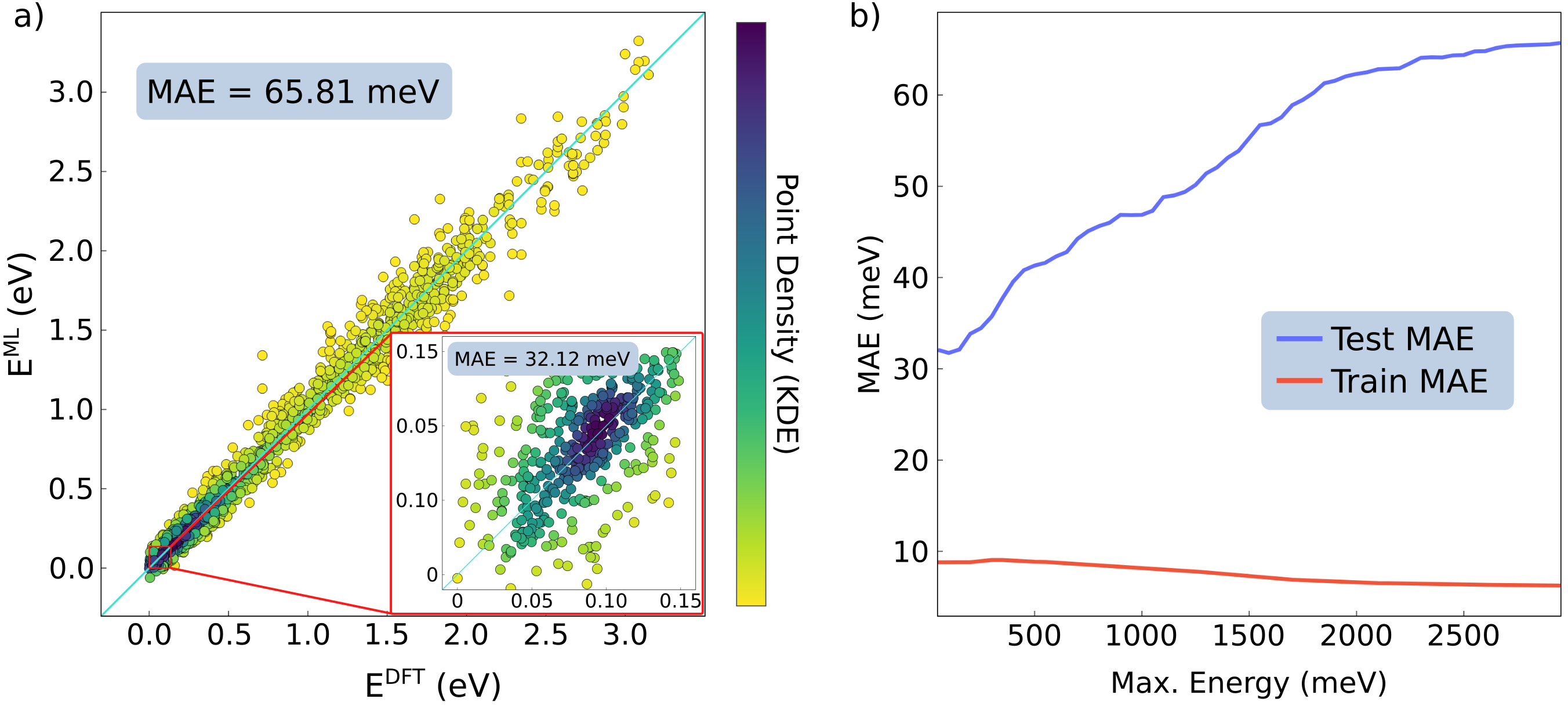}
\caption{a) Scatter plot comparing predicted energy values from the \vbml\ model to DFT energy values for training and test datasets. The plot shows an overall MAE of 65.81 meV for the full dataset, corresponding to approximately 13 meV per polaron. The red rectangle highlights a smaller energy range (up to 0.15 eV), where the model achieves improved accuracy with a MAE of 32.12 meV (6 meV per polaron). b) Mean absolute error for the test and train datasets plotted over maximum energy threshold.}
\label{fig_ConfMLbenchmark}
\end{figure}
As demonstrated in Figure~\ref{fig_ConfMLbenchmark}, our machine learning model can predict the energy of any defect-polaron configuration on \tios\ with a mean absolute error (MAE) of 65.81 meV for our 6$\times$3 slab, corresponding to approximately 13 meV per polaron. Notably, the model's accuracy is significantly enhanced in the low-energy region that is most relevant for low-temperature physical properties. By considering only configurations with energies up to 0.15 eV above the ground state (highlighted in the red rectangle), the prediction error is reduced to 32.12 meV for the entire slab, or just 6 meV per polaron.
As shown in panel b, the machine-learning test phase is highly sensitive to the energy range selected for the MAE analysis.
This improved accuracy at low energies is a direct result of our multi-phase active learning approach. The annealing-based training in Phase 3, as shown in Figure 3 of the main text, specifically prioritized the most stable defect-polaron configurations, allowing the model to learn the subtle energy differences between closely competing arrangements. This targeted training strategy ensures that the model performs best precisely where accuracy matters most: near the ground state configuration.

\begin{figure}[ht]
\centering
\includegraphics[width=0.65\textwidth]{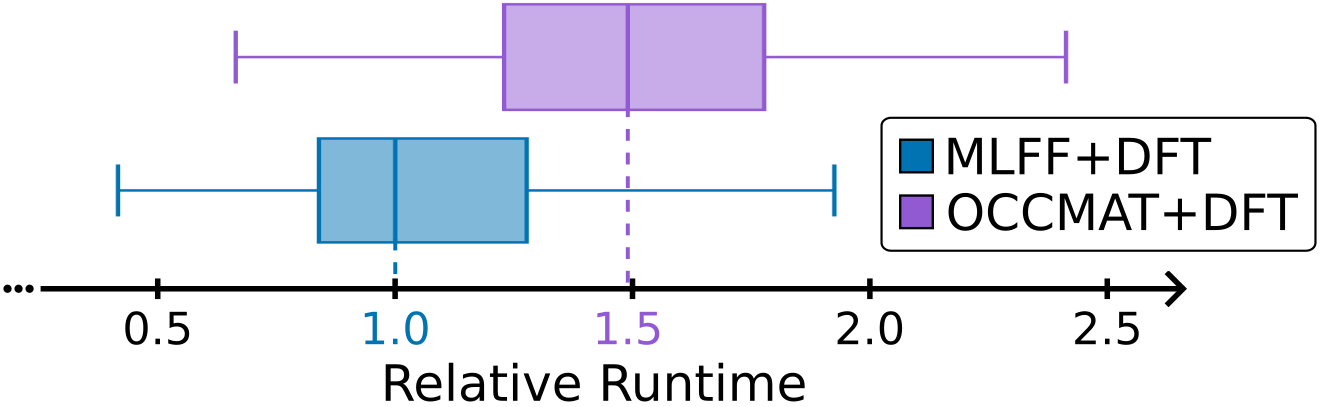}
\caption{Runtime of the MLFF+DFT and OCCMAT approaches in the full (low- and high-accuracy) runs (boxplots show the distribution of computational times normalized to the MLFF+DFT median).
The corresponding analysis for the low-accuracy set of calculations is shown in Fig.~3c in the main text.}
\label{figruntime}
\end{figure}

Figure~\ref{figruntime} complements Figure~3(c) in the main text.
Here, we show the overall runtime comparison between MLFF+DFT and OCCMAT+DFT.
In our system, the overall polaron localization process via the OCCMAT+DFT approach required on average 1.5~times the computing time compared to the MLFF+DFT strategy per defect-polaron configuration (corresponding to $7$ and $4.7$ hours, respectively).
Obviously, the gain of the MLFF-powered strategy is more evident in the comparison of the low-accuracy calculations (3~times faster, as described in the main text):
This is due to having trained the MLFF routine on a molecular dynamics simulation using the low-accuracy setup.

\clearpage

\section{Effect of Polaron-Polaron Interactions}
Our extensive sampling of the defect-polaron configuration space allows us to investigate not only defect-polaron interactions but also the effects of polaron-polaron interactions on the electronic structure. As mentioned in the main text, the repulsive interaction between polarons can significantly impact their stability and, under highly reducing conditions, even trigger surface reconstructions.

\begin{figure}[ht]
\centering
\includegraphics[width=0.7\textwidth]{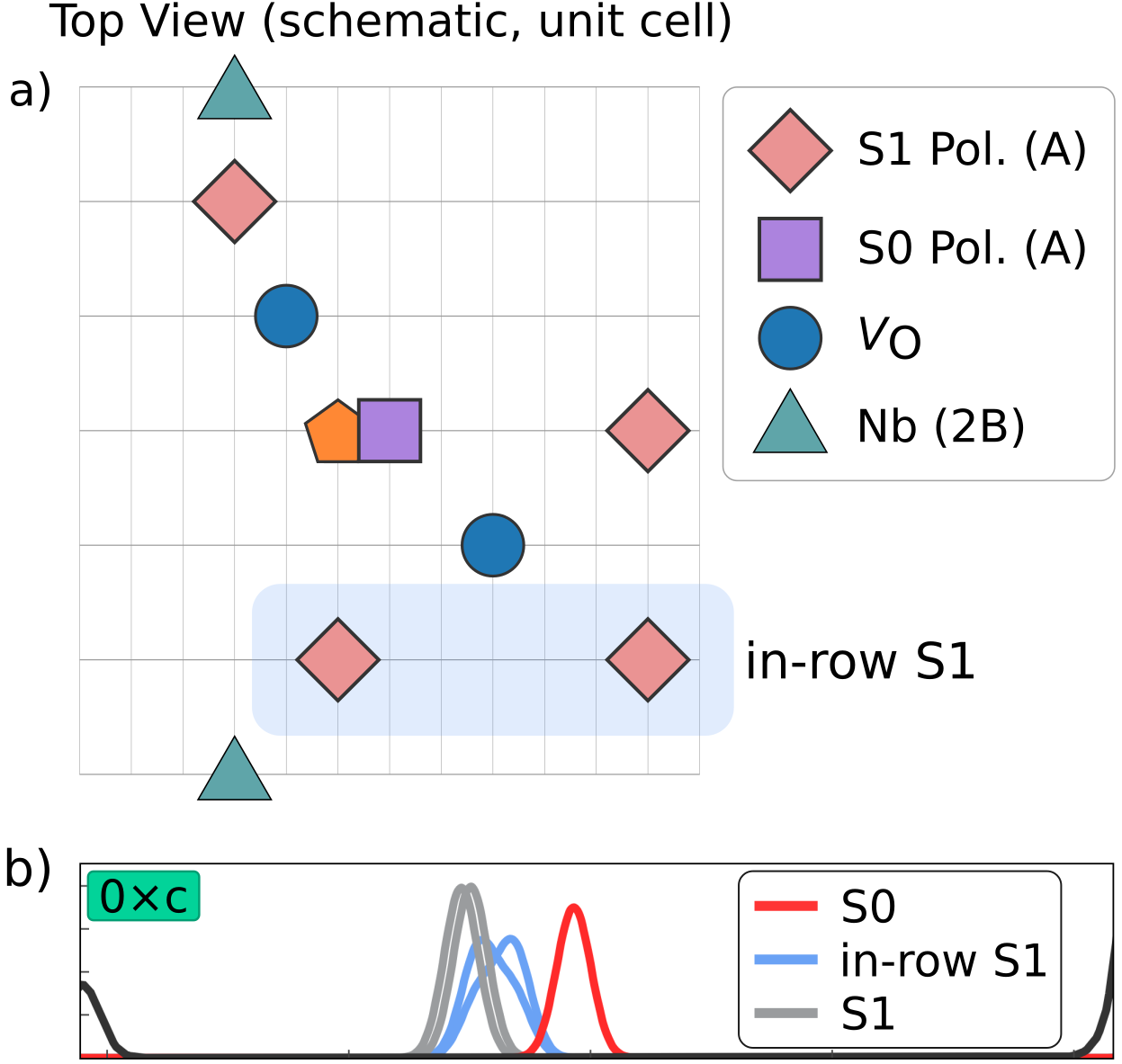}
\caption{Analysis of polaron-polaron interactions in the ground-state $2\times$c configuration. (a) Top and side views of the atomic structure, highlighting the oxygen vacancies (\VO) and the polaronic sites (yellow isosurfaces). Two polarons on the S1 layer are aligned along the same [001] Ti row ('in-row'). (b) Projected density of states (DOS) showing the sharp eigenstate of an isolated S1 polaron (gray) compared to the broadened eigenstate of an in-row S1 polaron (blue), indicating repulsive polaron-polaron interactions. The surface S0 polaron state is shown in red for reference.}
\label{fig_polaron_interaction}
\end{figure}

Figure~\ref{fig_polaron_interaction} illustrates this effect for our ground-state $2\times$c configuration. In this arrangement, multiple polarons are localized on the S1 subsurface layer. As shown in the structural model in Fig.~\ref{fig_polaron_interaction}(a), two of these S1 polarons are positioned along the same [001] Ti row, separated by three lattice sites. The projected density of states (DOS) in Fig.~\ref{fig_polaron_interaction}(b), which is also presented in the main text, reveals the electronic consequence of this proximity. The DOS peak for the 'in-row' S1 polaron (blue curve) is visibly broadened compared to the sharper peak of a more isolated S1 polaron within the same supercell (gray curve). This broadening is a characteristic signature of the repulsive interaction between the two nearby polarons, which destabilizes their energy levels. This finding corroborates previous studies and highlights the ability of our automated workflow to capture subtle but crucial electronic effects arising from the complex interplay of defects and charge carriers.

\clearpage

\section{Analysis of an S1 Nb Dopant Configuration}
\begin{figure}[ht]
\centering
\includegraphics[width=\textwidth]{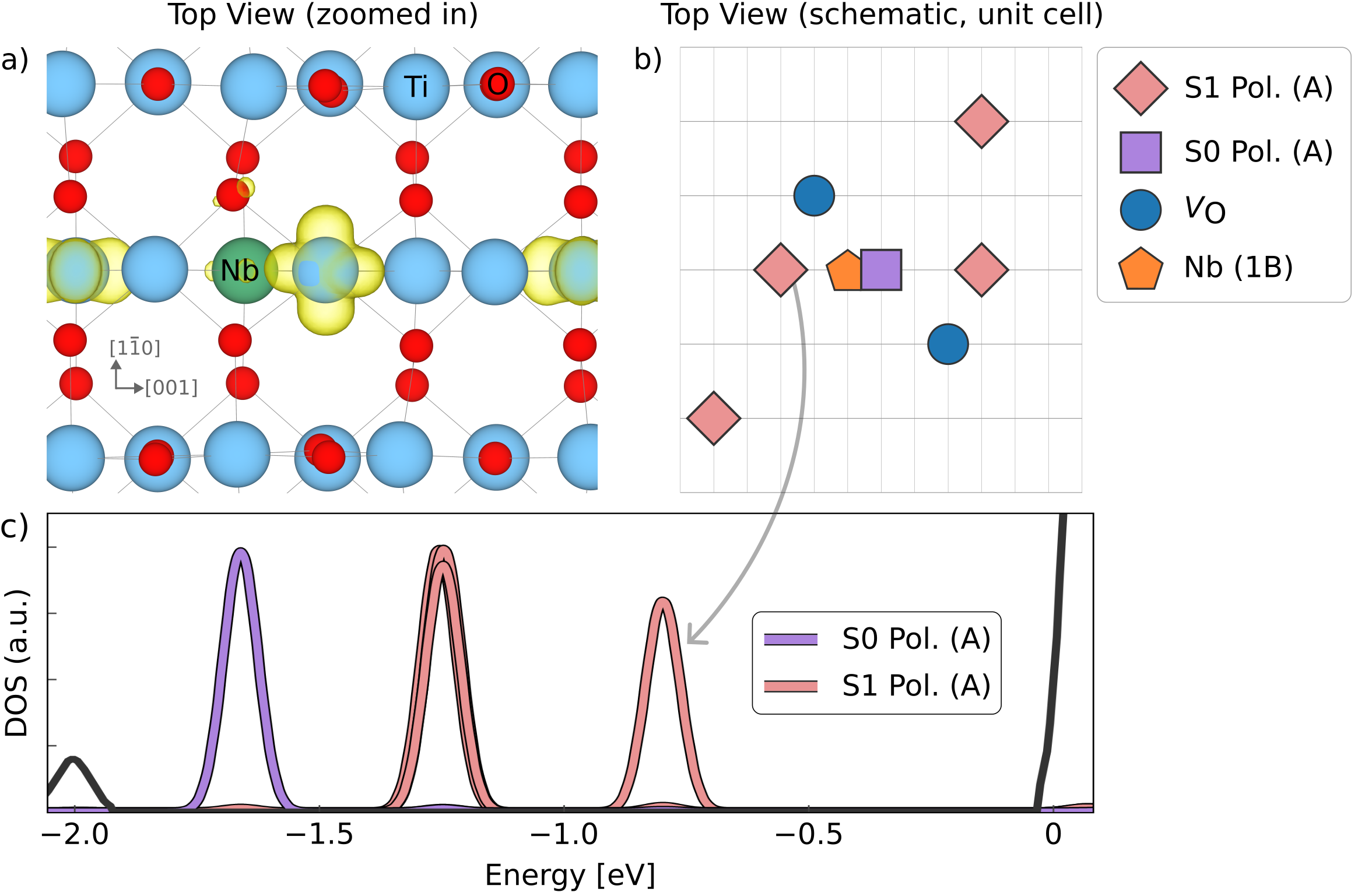}
\caption{Analysis of a configuration with an Nb dopant on the S1 subsurface layer. (a) Top-down, zoomed-in view of the relaxed atomic structure, showing the Nb atom and a nearby S1 polaron (yellow isosurface). (b) Schematic representation of the defect positions within the unit cell. (c) Projected density of states (DOS) comparing the S0 and S1 polarons. The DOS reveals that the S1 polaron in proximity to the S1 Nb dopant has a significantly higher energy eigenvalue compared to the S0 polaron, explaining the energetic instability of this configuration.}
\label{fig_NbOnS1}
\end{figure}

Our configuration space exploration revealed a clear preference for Nb dopants to occupy sites on the S2 layer rather than positions on the S1 layer. To understand the electronic origin of this preference and its implications for surface reactivity, we can analyze the specific configuration shown in Figure~\ref{fig_NbOnS1}, which features an Nb dopant on the S1 subsurface layer. This structure, which corresponds to the configuration represented by a square symbol in Figure 5 of the main text, is particularly informative for understanding both the instability of S1 Nb dopants and their direct influence on CO adsorption.

The projected density of states (DOS) in Figure~\ref{fig_NbOnS1}c provides a clear electronic fingerprint for the instability of this configuration. The eigenvalue for the S1 polaron located near the dopant is exceptionally high, indicating a destabilization of the polaronic state. This unfavorable electronic structure likely stems from the local distortion of the atomic structure around the S1 Nb dopant, which creates a less suitable environment for polaron trapping compared to the more stable S2 configurations.

This same structural model is also useful for evaluating the direct role of Nb in surface reactions. The Nb dopant is in close proximity to the surface, positioned directly below a potential CO adsorption site. Despite this, our \eadsF\ (as defined in Eqn. 3 of the main text) analysis showed that the dopant has a minimal direct effect on the adsorption strength of CO. The reason this configuration is unfavorable for the overall adsorption process (as measured by \eadsG) is not because of a weak interaction with CO, but because of the high intrinsic energy of having an Nb atom at an unstable S1 site. This observation provides strong evidence that the primary role of Nb dopants is to donate excess electrons to the system, rather than to directly participate in surface chemical reactions.

\clearpage

\section{CO Adsorption on Nb-free Surfaces}
To further isolate and quantify the effect of Nb dopants on surface reactivity, we performed additional calculations on Nb-free surfaces and compared the resulting adsorption energies with those obtained for Nb-doped systems.
\begin{figure}[ht]
\centering
\includegraphics[width=0.60\textwidth]{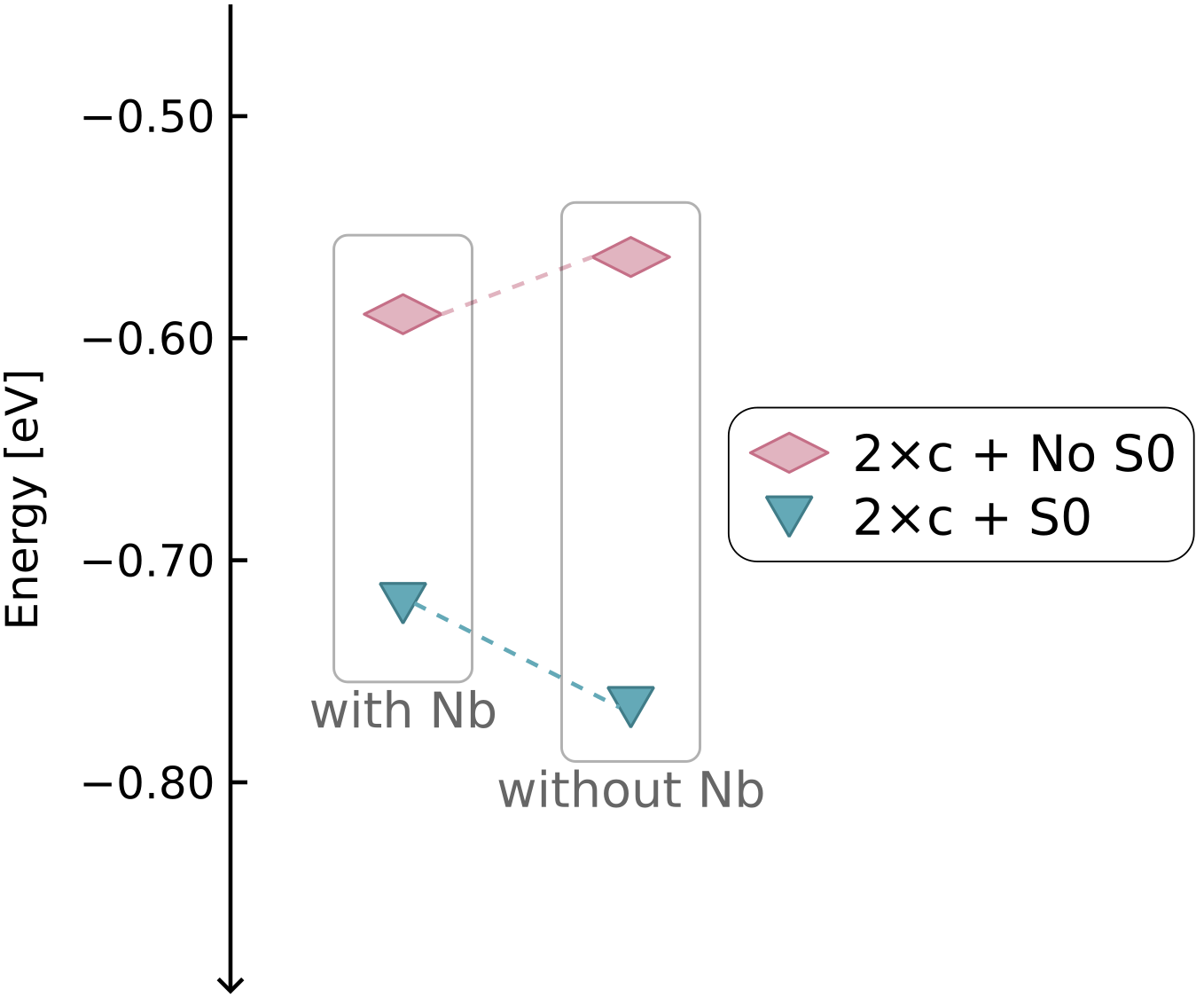}
\caption{Comparison of \eadsF\ adsorption energies for CO on \tios\ surfaces with and without Nb dopants. The figure shows similar adsorption energies in both cases: for S0-polaron driven adsorption, the energy changes from $-720$ meV with Nb to $-760$ meV without Nb, while for surfaces without S0 polarons, it changes from $-590$ meV to $-560$ meV. These results confirm the minimal direct impact of Nb dopants on CO adsorption.}
\label{fig_NbEffectOnAdsorption}
\end{figure}
Figure~\ref{fig_NbEffectOnAdsorption} compares the \eadsF\ adsorption energies for CO on \tios\ surfaces with and without Nb dopants. The data clearly demonstrates that removing the Nb dopant has minimal effect on the CO adsorption energy. For configurations with S0 polarons, the adsorption energy changes from $-720$ meV with Nb to $-760$ meV without Nb—a relatively small difference of 40 meV. Similarly, for configurations without surface polarons, the adsorption energy changes from $-590$ meV to $-560$ meV—an even smaller difference of 30 meV in the opposite direction.

These results provide strong evidence that Nb dopants play a negligible direct role in CO adsorption on \tios\ surfaces. This finding aligns with our overall conclusion that while Nb doping is effective for introducing additional electrons into the system, it does not significantly alter the surface reactivity toward CO molecules. Instead, oxygen vacancies remain the dominant factor in determining surface reactivity, primarily through their role in stabilizing polarons on the surface atomic layer, which then serve as active centers for molecular adsorption.

\clearpage

\bibliography{mrbib,mybib}